     \font\sm=cmbx5    
   \def\({\left(} \def\){\right)}  
   \def\lk{\,\left[ \,} \def\rk{\,\right] \,} 
   \def\rki{\,\right]}  \def\lb{\left\{} \def\rb{\right\}}  
   \def\lw{\left\langle} \def\rw{\right\rangle}
   \def\wu#1{\sqrt{{#1} \,}^{ \hbox to0.2pt{\hss$ 
     \vrule height 2.5pt width 0.6pt depth -.5pt $} \;\! }}
   \font\dick=cmbx12  \font\duenn=cmbx10 
   \def\dop{\duenn \raise0.6pt\hbox to 0.2pt{: \hss}}
   \def\dep{\duenn \raise0.6pt\hbox to 0.3pt{\hss :}}

   \def\cl#1{{\cal #1}}    \def\ct#1{\nz\centerline{#1}}
   \def\schl#1{\widetilde{#1}}  \def\ov#1{\overline{#1}}   
   \def\nz{\par \noindent}   \def\dis{\displaystyle}
   \def\0{\over }  \def\6{\partial }
   \def\ueb#1#2{{\buildrel{#1}\over{#2}}}
   \def\P{ {\mit\Pi} }  
   \def\gll{\; {\dop} =}   \def\glr{= {\dep} \;}
   
   \def\ln{{\rm ln}}   \def\det{{\rm det}}
     \def\Sp{\hbox{Sp}}
   \def\sh{{\rm sh}}    
   \def\th{{\rm th}}   \def\grad{{\rm grad}} 
   \def\div{{\rm div}} \def\rot{{\rm rot}}
   \let\a=\alpha  \let\b=\beta    \let\d=\delta  
     \let\l=\lambda  \let\o=\omega  
   \let\s=\sigma      \let\e=\varepsilon  
   \let\ph=\varphi   \let\ta=\vartheta
   \let\D=\Delta  \let\G=\Gamma  \let\L=\Lambda  
   \let\O=\Omega
   \def\mn{_{\mu\nu}}   \def\omn{^{\mu\nu}}
   \def\eq#1{(\ref{#1})}        \def\nonu{\nonumber}
   \def\be#1{\begin{equation} \label{#1}}
   \def\ben{\begin{equation}}   \def\ee{\end{equation}}
   \def\bea#1{\begin{eqnarray} \label{#1}}
   \def\bean{\begin{eqnarray}}  \def\eea{\end{eqnarray}}
   \let\thq=\theequation
   \def\pfeil{_\rightharpoonup}  \def\leer{\phantom{a}}
   \def\opf{\buildrel \pfeil \over \leer}
   \def\jvv{j \lower0.4pt\hbox to 2pt{\hss $\opf$}} 
   \def\jv{j \lower0.2pt\hbox to 1.4pt{\hss $\opf$}} 
   \def\ivv{i \lower0.4pt\hbox to 2pt{\hss $\opf$}} 
   \def\iv{i \lower0.2pt\hbox to 1.4pt{\hss $\opf$}} 
   \def\hq{h \raise0.2pt\hbox to 0.4pt{\hss $^-$}}   
   \def\vk#1{\hbox{$\buildrel           \pfeil \over #1$}}
   \def\vkk#1{\hbox{$\buildrel   \;     \pfeil \over #1$}}
   \def\vkkk#1{\hbox{$\buildrel  \, \;  \pfeil \over #1$}}
   \def\grpf{\displaystyle  _\rightharpoonup}
   \def\vg#1{\hbox{$\buildrel       \grpf \over #1$}}
   \def\vgg#1{\hbox{$\buildrel  \;  \grpf \over #1$}}
\def\fzz{f} \def\bzz{b} \def\dzz{d} \def\gzz{g} \def\hzz{h} 
\def\jzz{j} \def\kzz{k} \def\lzz{l} \def\mzz{m} \def\wzz{w} 
\def\tzz{t} \def\izz{i} 
\def\bezz{\beta} \def\dezz{\delta} \def\xizz{\xi} 
\def\pszz{\psi}  \def\vthzz{\vartheta}
\def\uph{ \! \mathop{\vphantom{a}} } \def\dph{ \vphantom{a} }
\def\vc#1{\def\tast{\noexpand#1} \def\test{#1}
\ifcat\tast\bzz 
\ifx\test\fzz \vkkk f \uph \else  \ifx\test\bzz \vkk b \uph \else
\ifx\test\dzz \vkkk d \uph \else  \ifx\test\gzz \vkk g \dph \else
\ifx\test\hzz \vkk h \uph \else   \ifx\test\izz \ivv \dph \else 
\ifx\test\jzz \jvv \dph \else     \ifx\test\kzz \vkk k \uph \else
\ifx\test\lzz \vkk l \uph \else   \ifx\test\tzz \vkk t \uph \else
\ifx\test\mzz \vg m \dph \else    \ifx\test\wzz \vg w \dph \else 
\ifnum \lq#1<91 \vgg #1 \uph \else \vk #1 \dph 
  \fi \fi \fi \fi \fi \fi \fi \fi \fi \fi \fi \fi \fi  \else
\ifx\test\bezz \vkk \beta \uph \else  
    \ifx\test\pszz \vkk \psi \dph \else 
\ifx\test\dezz \vkk \delta \uph \else 
    \ifx\test\xizz \vkk \xi \uph \else
\ifx\test\vthzz \vkk \vartheta \uph \else \vk #1 \dph
  \fi \fi \fi \fi \fi \fi }
\def\abst#1{\def\tast{\noexpand#1} \ifcat\tast\bzz 
    \ifnum \lq#1<91 \; \else \, \fi      \else \, \fi}
\def\absv#1{\def\tast{\noexpand#1} \def\test{#1}
\ifcat\tast\bzz \ifx\test\fzz \, \; \else  
    \ifx\test\dzz \, \; \else
\ifx\test\bzz \, \else \ifx\test\gzz \, \else 
       \ifx\test\hzz \, \else
\ifx\test\kzz \, \else \ifx\test\lzz \, \else 
       \ifx\test\tzz \, \else 
\ifnum \lq#1<91 \; \else   
  \fi \fi \fi \fi \fi \fi \fi \fi \fi \else
\ifx\test\bezz \, \else \ifx\test\pszz \, \else 
       \ifx\test\dezz \, \else
\ifx\test\xizz \, \else \ifx\test\vthzz \, \else 
   \fi \fi \fi \fi \fi \fi }
\def\vphan#1{\def\tast{\noexpand#1} \def\test{#1}
\ifcat\tast\bzz \ifx\test\fzz  \uph \else 
      \ifx\test\bzz  \uph \else
\ifx\test\dzz  \uph \else  \ifx\test\hzz  \uph \else
\ifx\test\kzz  \uph \else  \ifx\test\lzz  \uph \else
\ifx\test\tzz  \uph \else  \ifnum \lq#1<91 \uph \else \dph 
  \fi \fi \fi \fi \fi \fi \fi \fi \else
\ifx\test\bezz  \uph \else  \ifx\test\dezz  \uph \else
\ifx\test\xizz  \uph \else  \ifx\test\vthzz \uph \else \dph
  \fi \fi \fi \fi \fi }

  \def\pubox{\dick _{\raise 1pt\hbox{.}} }
  \def\ppubox{\dick _{\raise 1pt\hbox{..}} }
 \def\p#1{{\buildrel \abst #1 \pubox \over #1} \vphan #1 }

  \def\pbox{\dick _{\hbox{.}} } 
  \def\pvc#1{{\buildrel \absv #1 
      \pbox \over {\vc #1}} \vphan #1 }

  \def\vcsm#1{ \def\sm{\raise 1.6pt\hbox to 5pt{\hss $_#1$}} 
               {\buildrel \pfeil \over \sm} \>\!\! }

  \def\pusmbox{\duenn _{\lower 1pt\hbox{.}} }
  \def\ppusmbox{\duenn _{\lower .1pt\hbox{..}} }

   \def\abpfeil{ \raise 1pt\hbox{ $_\vee$ \hskip -6.8pt
           \vrule depth 0.6pt height 4.6pt width 0.3pt} \;\; }
   \def\apf#1{ \buildrel \abpfeil \over {#1} }
   \def\anfu{ $\!\!$\hbox{,$\!\!\;$,} \hskip -.08cm }
   \def\anfo{ \raise .4pt\hbox to .08cm{\hss ''} }

   \font\log=logo10 scaled \magstep2 
   \def\MA{\hbox{\log A}}

\def\dod{\,D\hspace{-.42cm}I\hspace{.2cm}}
\def\sec#1{\let\dq=\thq 
   \renewcommand{\theequation}{\arabic{section}.\dq}    
   \setcounter{equation}{0} \vskip .1cm \section{#1}}         
\def\nz{$ $\\}  \def\ek#1{{\bf [$\,$#1$\,$]}}
\def\ft{\footnotesize}   \def\ou{^{[U]}}
\def\bce{\begin{center}} \def\ece{\end{center}}
\def\unt#1#2#3#4{\hbox{
  \vrule height .2cm width 0.4pt  depth -.5pt
  \vrule height .5pt width #1cm   depth 0pt
  \vrule height .2cm width 0.3pt  depth -.5pt
  \hskip -#1cm \hskip #3cm  \vrule width 0.3pt 
  height 0pt depth #2cm \lower #2cm\hbox{\lower .14cm\hbox{$
  \!\!\!\,\dis = #4$}}} \nonu }
\def\I{\hbox{$\,\cl I \,$}}  \def\2{\hbox{${1\02}$}}
\def\bp{\hbox{\boldmath$\partial$}} \def\AA{\hbox{{\bf A}}}
 \def\fhsp{\hat{\bf S}{\bf p}}
\def\gdw{\,\leftharpoonup \hspace*{-.3cm}\rightharpoondown\,}

\documentstyle[12pt]{article} \parskip .2cm
\begin{document}

\vspace*{-1.4cm}
\leftline{ITP--UH 04/99 \hspace{7cm} \hfill 
{\sl Fr\"uhjahr 1999}} \leftline{hep-ph/9908527}
\thispagestyle{empty}

\vspace*{-.4cm}
\ct{{\Large\bf \ \ \ 2+1 D $\,$Yang--Mills}}  

\bce 
Introductory Lecture Notes\footnote{~hschulz@itp.uni-hannover.de} 
on the Schr\"odinger wave functional and Hamiltonian 
treatment by Karabali, Kim, Nair \cite{kkn} \ece

\nz
Feldtheorie, wenn als Medium am W\"armebad gesehen, gewinnt
mit $T$ (Temperatur) einen n\"utzlichen Parameter. Am oberen
Ende der $T$--Achse ($g$ klein) ist Yang--Mills--Theorie 
(Gluonen--Hohlraumstrahlung $=$ QCD ohne quarks) der 
St\"orungstheorie zugeneigt. Jedoch trifft die entsprechende 
Diagrammatik derzeit bei $g^6$ (Freie Energie, {\sl magnetic 
mass} Skala) bzw. $g^4$ (Selbstenergie, {\sl super soft 
scale}) auf eine \anfu Berechenbarkeits--\hbox{Schwelle\anfo :} 
$\infty\,$viele Diagramme (der Linde--See) tragen 
gleichberechtigt bei. Existiert QCD ? \cite{linde} \ Das 
zugeh\"orige euklidisch 3--dimensionale $T\!=\!0$--System 
wartet auf eine nicht--(oder nicht 
{\sl so})--st\"orungstheoretische Behandlung. 

Der Umstand, da\ss\ diese Schwelle bald \"uberwunden sein 
d\"urfte, ist der eine Grund zur Aufregung, und {\sl wie} 
dies geschieht (und eine \anfu{\sl unification}\anfo ist) 
der andere. Die alte Idee von Feynman \cite{fey} 
aus 1981 kommt zu Ehren und \"ubersetzt sich ins Konkrete. 
Der Eichorbit ist pr\"aparierbar. Die (hier hermitesche) 
Wess--Zumino--Witten--Wirkung findet eine konkrete 
Anwendung. Konforme Feldtheorie macht sich n\"utzlich. Und 
die Thermische Feldtheorie \ --- \ nach ihrer euphorischen 
Zeit um 1990 (Braaten--Pisarski Resummation) etwas 
dahin kr\"ankelnd \ --- \  findet dorthin zur\"uck, wo sie 
schon immer hingeh\"orte, zur grunds\"atzlichen, wiewohl 
Realit\"at verstehen wollenden Feldtheorie. Ein Kreis 
schlie\ss t sich.  
  
Wer alle Details erkl\"art, kommt nicht voran. So 
m\"ochten denn diese Bl\"atter nur einen Einstieg bieten. 
Sie f\"uhren nur bis zum WZW--Term und zum ersten Blick auf 
den {\sl mass gap}. Ist das \anfu Kleine Matterhorn\anfo 
erreicht, finden wahre Bergsteiger schon selbst weiter. 
Die Arbeit \cite{kkn} (kurz KKN) 
\\[8pt] 
\hspace*{1cm} {\small\bf 
   D. Karabali, C. Kim and V. P. Nair, Nucl. Phys. 
   {\bf B 524} (1998) 661 } \\[1pt]
\hspace*{1.8cm}{\small 
   \anfu Planar Yang--Mills theory$\,$: Hamiltonian, 
   regulators and mass gap\anfo} \quad   
\\[8pt] 
ist Leitfaden. Auch der dortige \S 2 ist nur {\sl Outline 
of the main argument}, hat also seine Vorl\"aufer 
\cite{vor}. Der Nachl\"aufer \cite{nach} behandelt 
{\sl confinement}. Auf Gleichungen in \cite{kkn} wird in 
der Form \ek {n.m} verwiesen werden \ --- \ voller Freude, 
sie endlich begriffen zu haben. In der Regel wird aber 
keine Notwendigkeit bestehen, bei KKN nachzuschauen.

\noindent {\ft 
{\bf Deser, Jackiw, Templeton 1982 \cite{deser} :} 
The study of vector and tensor gauge theories in 
three--dimensional space--time is motivated by their 
connection to high temperature be\-ha\-vior of 
four--di\-men\-sio\-nal
models, and is justified by the special properties
which they enjoy.}

\noindent {\ft 
{\bf KKN 1998 \cite{kkn} :}
... there is at least one interesting physical situation, viz., 
the high temperature phase of chromodynamics and associated
magnetic screening effects, to which the (2+1)--dimensional 
theory can be directly applied.}


\sec{Zweidimensionale klassische ED}

\vspace*{-.6cm} \hspace*{2cm} \parbox{6.5cm}{ 
\bea{1.1} \div \vc E &=& \rho  \\  \label{1.2}
        \rot \vc E &=& - \pvc B \eea } \parbox{7cm}{ 
\bea{1.3} \div \vc B &=& 0  \\ \label{1.4}
        \rot \vc B &=& \vc j + \pvc E  \eea } \\
2D Physik ist spezielle dreidimensionale. Die 
\anfu Punkt\anfo$\!\!\!$--Ladungen sind homogen geladene 
Linien parallel z--Achse. Aus ihnen bilden sich die 
Dichten $\,\rho (x,y,t)\,$ und $\;\vc j =$ 
$(\, j_1(x,y,t)\, ,\, $ $ j_2(x,y,t)\, ,\, 0\, )$. 
Unter der Konsistenzannahme, da\ss\ auch $\rot \vc B$ die 
Struktur von $\vc j$ bekommen wird, 
\vspace*{-.6cm}
\bean \hspace{1.2cm} 
      \hbox{folgt aus \eq{1.1}, \eq{1.4}, da\ss\ } 
      \hspace{1cm} \vc E  &=& \Big(\;\, E_1(x,y,t)\; ,
      \; E_2(x,y,t)\; ,\; 0\;\,\Big)  \nonu \\
      \hspace{1.2cm}
      \hbox{Wie nun \eq{1.2}, \eq{1.3} zeigen, hat } 
      \hspace{1cm} \vc B  &=& \Big(\;\, 0\; ,
       \; 0\; ,\; B(x,y,t)\;\,\Big) \nonu
\eea \\[-.8cm] 
nur eine dritte Komponente, womit die Konsistenzannahme 
verifiziert ist. Maxwell hat sich auf \eq{1.1}, \eq{1.2}, 
\eq{1.4} reduziert. Es bedarf auch keines Theorems mehr, 
per  $\, B = \big( \rot \vc A \big)_3 = \6_x A_2(x,y,t) 
- \6_y A_1(x,y,t)\,$ ein 2--komponentiges Vektorpotential 
einzuf\"uhren. Aber \eq{1.2} ist n\"otig, um $\,\vc E 
=  - \pvc A - \grad \,\phi\;$ zu erlauben. Die drei Felder 
$E_1$, $E_2$, $B$ bleiben invariant unter den Umeichungen 
$\vc A \to \vc A + \nabla \chi(x,y,t)\,$ und $\;\phi 
\to \phi - \6_t \chi(x,y,t)\,$. 

Die Eichung $\phi=0$ (temporale Eichung [Muta, S.51], 
Weyl--Eichung, Strahlungs\-eichung) fixiert nur 
unvollst\"andig. Ohne $\vc E =  - \pvc A$ und 
$\,\vc B=\nabla \times \vc A$ zu ver\"andern, 
ist n\"amlich weiterhin die Rest--Umeichung $\vc A \to \vc A 
+ \nabla \chi(x,y)\,$ m\"oglich (note that $\chi$ must not 
depend on time).

Aus Vierer--Notation wird nat\"urlich \anfu 3--Notation\anfo
mit Metrik $+ - -\,$, $\mu=0,1,2\,$, $\big(\, \phi\, , \, 
\vc A \,\big) \glr A^\mu\;$ und $\6^\mu = \big(\,\6_0\, , 
\, -\nabla\,\big)$. Der Zusammenhang Felder--Potentiale
wird folglich zu $E^j=-\6^0A^j+\6^j A^0\,$, 
$B=-\6^1A^2+\6^2A^1 = - \e_{jk}\6^jA^k $ ($\e_{12} \gll 1$). 
Wir definieren wie \"ublich den Feldtensor 
$\, \6^\mu A^\nu -\6^\nu A^\mu \glr F\omn\,$ und finden
die resultierende Matrix--Version ganz h\"ubsch$\,$:
\be{1.5}  F\omn = \( 
     \matrix{  0   & - E_1 & - E_2 \cr
               E_1 &   0   & - B   \cr 
               E_2 &   B   &   0   \cr } \) \qquad , \qquad
    F\mn = \( 
     \matrix{   0   &  E_1  &  E_2  \cr
              - E_1 &   0   & - B   \cr 
              - E_2 &   B   &   0   \cr } \) \quad .
\ee 
Mit dieser rechnet man n\"amlich flugs nach, da\ss\ 
erwartungsgem\"a\ss 
\bea{1.6} 
  \cl L &=&  - {1\04}\, F\omn F\mn \; = \; 
  {1\04}\, \Sp \Bigg(\;\;\; 
\lower 7pt\vbox{\hbox{\scriptsize \ (1.5)}  \vskip -.3cm
                 \hbox{\scriptsize \ linke}  \vskip -.3cm
                 \hbox{\scriptsize Matrix }} \;\,\cdot \;\;
\lower 7pt\vbox{\hbox{\scriptsize \ (1.5)}  \vskip -.3cm
                 \hbox{\scriptsize rechte}  \vskip -.3cm
                 \hbox{\scriptsize Matrix }} \;\; \Bigg)
  \;  = \; {1\02} \( \vc E^2 - B^2 \) \nonu \\
     &=& {1\02} \Big( \, [ - \pvc A - \grad \phi ]^2
     - {1\02} [ ( \rot \vc A )_3 ]^2 \,\Big) \quad 
\eea 
die Lagrange--Dichte ist. Der Vorteil strikter 
Weyl--Eichung $\phi=0$ wird nun bei \"Ubergang zur 
Hamilton--Dichte besonders deutlich$\,$:
\bea{1.7}  
  \cl L &=& {1\02} \pvc A^2 - {1\02} B^2 \qquad\; , 
  \;\;\qquad  \vc \P \; = \;\pvc A \; = \; - \vc E  \\
       \label{8} 
  \cl H &=& \Big[\; \pvc A \vc \P  - \cl L \;\Big]_{\rm 
  ersetze\; ...} \;  = \; {1\02} \big( \vc E^2 + B^2 \big) 
  \; =\; {1\02} \big( \pvc A^2 + B^2 \big) \quad . 
\eea 
Man hat es also (in strenger Weyl--Eichung) nur noch mit 
den zwei reellen Feldern $A_1$ und $A_2$ zu tun.


\sec{Yang--Mills in (2+1) D}

Die Spezialit\"aten nicht--Abelscher Theorie haben fast 
nichts mit der Dimension zu tun. Lediglich l\"auft jetzt 
$\mu$ von 0 bis 2. Als \ g \"a b e \ es in der x-y--Ebene 
auch Teilchen ($\psi$ mit $N$ Farbkomponenten), verlangen 
wir Invarianz jeglicher Physik gegen $\vc r$--$t
$--abh\"angige \"Anderung der $\psi$--Phase. 
Bezeichnungs\"anderungen (um diese geht es hier) r\"utteln
an der Psyche. Wir vergraben uns darum erst einmal in
vertraute Hannover--Notation. Die Kopplung hei\ss e aber
unverz\"uglich $e$ (statt $g\,$)$\,$:
\be{2.1} 
  \! \left. \parbox{14.6cm}{ \vspace*{-.6cm}
  \bean
  U = e^{-ie \L^a (x) T^a } 
  &,&  D_\mu = \6_\mu - i e A_\mu^a T^a   \nonu \\
  \MA_\mu \gll T^a A_\mu^a 
  &,&  \MA_\mu \to \MA_\mu\ou =  U\MA_\mu U^{-1} 
    - {i\0e} \, U_{\prime \mu} U^{-1} \nonu \\ 
   F\mn^{\;\, a} = \6_\mu A_\nu^{\; a}  - \6_\nu 
    A_\mu^{\; a} + e f^{abc} A_\mu^{\; b} A_\mu^{\; c} 
  &,& F\mn = \6_\mu \MA_\nu  - \6_\nu \MA_\mu 
      - i e \lk \MA_\mu , \MA_\nu \rk   \nonu \\
  & & \hspace*{-7cm}  \hbox{\eq{1.5} :} \qquad
  B^a = - F^{12\, a} = - \( \6^1 A^{2\,a}  - 
  \6^2 A^{1\,a} + e f^{abc} A^{1\, b} A^{2\, c}\) \nonu \\
  & & \hspace*{-7cm}  E^{j\, a} = - F^{0j\, a} 
      = - \( \6^0 A^{j\,a} - \6^j A^{0\,a} + e f^{abc} 
      A^{0\, b} A^{j\, c} \) \;\;\; , \;\;\;
      A^{0\, a} \equiv 0 : \; E^{j\, a} 
      = - \p A{}^{j\, a}   \nonu \\  
  & & \hspace*{-6.6cm} \cl L_{\rm streng Weyl} 
    = - {1\04} F^{\mu\nu \, a} F\mn^{\;\, a} = {1\02}
   E^{j\, a} E^{j\, a} - {1\02} F^{12\, a} F^{12\, a}
   = {1\02} \p A{}^{ja} \p A{}^{ja} - {1\02} B^a B^a 
   \quad \nonu 
   \eea \vspace*{-.6cm} } \!\!\right]
\ee 
In gewissen Kreisen um Chern und Simons (aber der Artikel 
von Jackiw \cite{Jack} in den Les Houches von 1983 ist 
trotzdem sehr sch\"on) ist es nun weitgehend \"ublich, die 
Kopplung $e$ in den Feldern zu verstecken und mit 
\ a n t i hermiteschen Feld--Matrizen zu arbeiten. 
Mit den \anfu alten\anfo Feldern in \eq{2.1} 
ist dann folgendes zu veranstalten$\,$:
\bea{2.2}
  \L^a &\gll& e \, \L^{a\; {\bf alt}} \qquad , \qquad 
  A_j^a \gll e \, A_j^{a\; {\bf alt}}  \qquad ,\qquad
  F\mn^a \gll e \, F\mn^{a\; {\bf alt}} \qquad ,
       \;\; \nonu \\
  B^a &\gll& - e \, B^{a\; {\bf alt}} \,\quad ,\qquad
  A_j \gll - i \, e \, \MA_j^{\; {\bf alt}} 
       \quad , \qquad
  F\mn \gll - i \, e \, F\mn^{\; {\bf alt}} 
  \quad . \quad
\eea 
Gleichzeitig mit diesen Ersetzungen nisten wir uns
durchg\"angig in der strengen temporalen Eichung
ein. Es gibt nur noch die $2*n$ Felder $A_j^a (\vc r )$.
Sie leben in der Ebene $\vc r = (x,y)$. Umeichungen
m\"ogen im Endlichen liegen$\,$: $\L^a \( \vc r \to 
\infty \) \to 0\,$, und mit \eq{2.2} ist
\be{2.3}
   U (\vc r ) = \exp{\( -i\L^a(\vc r ) T^a \)} \quad , 
   \quad a = 1, \ldots , N^2-1 \glr n \quad . 
\ee 
KKN geh\"oren nicht zu jenen B\"osewichtern, welche mit
antihermiteschen Generatoren herummachen. Jene Zeile,
welche in \eq{2.1} \anfu vergessen\anfo wurde, gilt 
darum alt wie neu$\,$:
\be{2.4}
   \Sp \( T^a T^b \) = {1\02} \d^{ab} \quad , \quad
   \lk T^a , T^b \rk = i f^{abc} T^c  \quad . \quad
\ee 
Mittels \eq{2.2} wird die kovariante Ableitung h\"ubsch
einfach. $\6_j$ ist antihermitesch, und dies harmoniert
mit dem Umstand, da\ss\ nun auch die Matrixfelder 
(spurfrei und) antihermitesch sind$\,$:
\be{2.5}
  D_j = \6_j + A_j \quad , \quad 
  A_j = - i T^a A_j^a \quad . \quad 
\ee 
Deren Umeichung lautet
\be{2.6}
   A_j \to A_j\ou = U A_j U^{-1} - U_{\prime j} 
   U^{-1}  \qquad , \qquad j=1,2 \quad . \quad
\ee 
Auch Feldtensor und Magnetfeld werfen Ballast ab$\,$:
\be{2.7}
  F_{jk}^{\; a} = \6_j A_k^{\, a} - \6_k A_j^{\, a} 
         + f^{abc} A_j^{\, b} A_k^{\, c} 
   \quad ,\quad   F_{jk} = \6_j A_k - \6_j A_k
       + \lk A_j , A_k \rk \quad , \quad
\ee 
\be{2.8}
  B^a  
   = \6_1 A_2^{\, a} - \6_2 A_1^{\, a} 
   + f^{abc} A_1^{\, b} A_2^{\, c} \quad , \quad
\ee 
und schlie\ss lich bekommt die Lagrangian die Gestalt
\be{2.9}
   \cl L = {1\0 2 e^2} \p A{}_j^a \p A{}_j^a  - 
  {1\0 2 e^2} B^a B^a \;\; \glr \;\; \cl T - \cl V \quad .
\ee 
Wir haben den Fu\ss\ in der T\"ur. \eq{2.8} steht bei KKN 
im Text unter \ek{2.4}. \eq{2.6} ist \ek{2.1}. Aber 
\eq{2.9} ist nicht \ek{2.4}. Nun ja, in letzterer hatte 
ein Tippteufel das $e^2$ in den Z\"ahler gesetzt. Man 
findet das sp\"ater bei KKN heraus, wenn von $\cl L 
= \cl T - \cl V$ per
\be{2.10}
  \P_j^a = \6_{\p A{}_j^a} \cl L = {1\0 e^2}
   \p A{}_j^a \quad , \quad 
   \cl H = \lk \P_j^a \p A{}_j^a - \cl L 
   \rki_{\hbox{\scriptsize eliminiere $\p A$}} 
   \; = {e^2\0 2} \P_j^a\P_j^a \, + \,\cl V
\ee 
zur Hamilton--Dichte \"ubergegangen wird. 

Man kann nun (sollte nicht) hier eine Pause einlegen und 
nachsehen, ob denn nun die Lagrange--Funktion \eq{2.9}
wirklich invariant ist unter den eingeschr\"ankten (nur
$\vc r$--abh\"angigen) $U$--Transformationen. Es \ i s t \
so. Bei entsprechendem Ehrgeiz will man diesen proof mit
endlichen Eichtransformationen
\be{2.11}
  \lb A^a \rb \glr \vc A \;\; , \;\; \lb \L^a \rb \glr
    \vc \L \;\; , \quad {\vc A\ou}_{\!\!\!\!\!\!\!\! j}
    \;\; = \, e^{\vc\L \times } {\vc A}_{\! j}
  \; - \; \int_0^1\!\! ds \,\; e^{s \vc \L \times}\,
  \vc \L _{\,\prime j} \qquad
\ee 
bewerkstelligen und freut sich dar\"uber, da\ss\
$ \pvc A _j\ou = e^{\vc \L \times} \pvc A _j$ ist, da\ss\
tats\"achlich $D \gll e^{\vc \L \times}$ eine
hundsgemeinsnormale Drehmatrix im $n$--dimensionalen 
Farbraum ist und da\ss\ dabei sogar
die alte Weisheit $D\vc a \times D\vc b = D \(\vc a
\times \vc b \)$ Verwendung findet [Bleiphys Gl.(4.9)].
Aber was solls$\,$: proof mit infinitesimalen
Transformationen gen\"ugt nat\"urlich (und nicht einmal
dieser ist n\"otig).


\sec{Matrix--Parametrisierung}

Der erste wichtige Schritt in das KKN--Gesch\"aft braucht
nur eine grobe Begleitphilosophie. Es gibt nur noch die 
2$*n$ Felder $A_j^a$. Sie unterliegen den restlichen 
Umeichungen \eq{2.6}. Quantisierung darf aber nur 
physikalische (nicht durch Umeichung erreichbare) Felder 
betreffen. Jeder bessere Blick in den Raum der Felder $
A_j^a$, jeder Gewinn an Harmonie, ist also gut.

Es bereitet Vergn\"ugen, die Matrix--Parametrisierung 
mit Gem\"ut \anfu selber zu \hbox{erfinden\anfo$\!$.} 
Wir nehmen die Eichtransformation \eq{2.6} zur Hand und 
spielen damit herum, indem wir zun\"achst vor jedem der 
beiden $U^{-1}$ eine Eins einf\"ugen,
\bea{3.1}
   A_j\ou &=& U A_j \, M \, M^{-1} \,U^{-1}
              - U_{\prime j}\, M \, M^{-1} \, U^{-1} 
          \nonu \\ 
          &=& U A_j \, M \, (UM)^{-1} 
              -  U_{\prime j}\, M \, (UM)^{-1} \quad , 
\eea 
woraufhin die zweite Zeile zwanglos in den Sinn kommt. 
Nun starren wir rechts den letzten Term an und w\"urden
statt $U_{\prime j} M$ lieber ein 
$j$--differenziertes Produkt $UM$ dort stehen sehen. 
Also setzen wir $ U_{\prime j} \, M = (UM)_{\prime j} 
- U\, M_{\prime j} $ ein. Das gibt 
\be{3.2}
   A_j\ou = - (UM)_{\prime j} \, (UM)^{-1} \, + \,
   U \lk M_{\prime j} + A_j \,M \rk \, (UM)^{-1} \quad.
\ee 
Bisher wurde nichts \"uber $M$ vorausgesetzt, au\ss er
nat\"urlich, da\ss\ es (wie $U$, seine Ge\-ne\-ra\-toren 
und $A_j$) eine $N\times N$--Matrix zu sein hat. 
Vereinfachen ist die Losung. Die eckige Klammer m\"oge 
verschwinden. Nun handelt es sich bei \eq{3.2} aber um 
\ z w e i \ Gleichungen$\,$: $j=1,2\,$. Ein $M$ zu finden, 
welches (zu beliebigen, spurfreien, antihermiteschen $A_1$, 
$A_2$) beide eckigen Klammern verschwinden l\"a\ss t, das 
d\"urfte (mindestens) problematisch werden. Der Ausweg ist 
wundervoll. Obacht, jetzt wird es h\"ubsch. Die Struktur 
von \eq{3.2} bleibt ja erhalten, wenn man eine beliebige 
Linearkombination der beiden Gleichungen bildet. Vielleicht 
$\;{1\02}\;${\sl erste} $+$ $i$ {\sl mal} ${1\02}\;${\sl 
zweite} ? \ Links entsteht dann die Kombination
\be{3.3}
    A \; \gll \; {1\02} \Big( \; A_1 \, + \, i \, A_2 
    \;\Big) \quad .
\ee 
Das ist \ e i n \ Feld. Man kann sogar $A$ beliebig 
(nur spurfrei) w\"ahlen und sodann $A_1$ und $i\, A_2$
als ihren antihermiteschen und hermiteschen Anteil
identifizieren. Diese Zerlegung geht immer und ist 
eineindeutig. Rechts in \eq {3.3} kombinieren sich die 
Differentiationen zu
\be{3.4}
  \6 \; \gll \; {1\02} \Big( \; \6_1 \, + \, i \, \6_2 
  \;\Big) \quad .
\ee 
Wir erhalten damit
\be{3.5}
   A\ou \; = \; - (\6\; UM) \, (UM)^{-1} \, + \,
           U \lk \6\, M + A \,M \rk \, (UM)^{-1} \quad .
\ee 
Diese \ e i n e \ eckige Klammer in \eq{3.5} verschwindet, 
wenn sich zu jedem spurfreien $A$ eine Matrix $M$ so 
finden l\"a\ss t, da\ss\
\be{3.6}
     A \; = \; - \, \(\6\, M \)  \, M^{-1} \quad .
\ee 
Falls dies m\"oglich ist, l\"auft eine Umeichung \ --- \ 
gem\"a\ss\ \eq{3.5} ohne eckige Klammer \ --- \ schlicht 
auf
\be{3.7}
            M\ou \; = \; U\, M
\ee 
hinaus. Und wir bekommen gratis noch eine Zugabe$\,$:
\be{3.8}
      \( M^\dagger \, M \)\ou =
      (UM)^\dagger \, UM = M^\dagger\, U^\dagger\, U\, M 
      = M^\dagger M\,\glr\, H \qquad \hbox{ist Invariante}
\ee 
unter Eichtransformationen. 

Das obige \anfu Falls\anfo ist noch auszur\"aumen. $A$ ist 
spurfrei (und sonst nichts). Be\-haup\-tung$\,$: \  d i e \ 
Eigenschaft von $M$, welche dies via $A=-(\6M)M^{-1}$ 
liefert (und weiter nichts), ist $\,\det (M) = F \big( 
\,x+iy \glr \ov{z} \,\big)\,$. Und die Funktion $F$ darf  
festgelegt werden$\,$:
\be{3.9}
   \det \( M \) = 1 \qquad , \;\; {\rm d.h.}
   \quad M\; \in \; \hbox{SL(N,C)} \quad .
\ee 
\bea{3.10}  \hskip -.2cm \hbox{Beweis :} \qquad\;
  0 &=& \6 \; \det (M) = \6 \; \e_{j_1 \;\ldots\; j_N} 
        M_{j_1\, 1} \; \ldots \; M_{j_N\, N}  \nonu \\
    &=& \sum _{k=1}^N \sum_{j_k=1}^N \;\, \sum_{{\rm 
        restl.} j's} \;\e_{j_1 \ldots j_k \ldots j_N} \;
        M_{j_1\, 1} \ldots \hbox{\ft (ohne $j_k\, k$)}
        \ldots  \; M_{j_N\, N} \;\; \6 M_{j_k\, k} 
        \nonu \\[-.45cm]
    & & \hskip 2.6cm  \unt{1.5}{.3}{.8}{ 
        \; \e_{j_k \; j_1 \;\ldots\; \hbox{\ft (ohne)}
        \; \ldots \; j_N} \; (-1)^k \;\; , 
        \;\; j_k\glr \ell }   \nonu \\
    & & \hskip 3.4 cm 
      = (-1)^{k+\ell} \e_{j_1 \;\ldots\; \hbox{\ft (ohne)}
        \; \ldots \; j_N}  \;\;\; ,  \nonu \\
    &=&  (\6 M )_{\ell\, k } \, (-1)^{k+\ell}
        \; (\hbox{\ft Unterdeterminante})_{\ell\, k} \nonu \\
    &=& \det (M) \;\Sp \Big( \; (\6 M ) M^{-1} \; \Big)  
        \qquad , \quad \hbox{{ q. e. d.}}   
\eea 

Von gegebenem $A$ zu den $2*n$ reellen Feldern $A_1^a$ und 
$A_2^a\,$ zur\"uckzukehren, ist mittels \eq{2.4} ohne 
weiteres m\"oglich$\,$:
\be{3.11}
 A = -{i\02}\, T^a \( A_1^a + i\,A_2^a \) \quad
     \Rightarrow \quad 2\, i\,\Sp\( T^a A \) = {1\02}\,
     \( A_1^a + i\, A_2^a \) \quad .
\ee 
Wir sind \"ubrigens mit \eq{3.6} bei \ek {2.6} angekommen. 
Und \eq{3.7} ist \ek{2.9}.

Selbstverst\"andlich kann etwas, was von $x$, $y$ 
abh\"angt, auch als Funktion von $\; z \gll x-iy\;$ und 
$\;\ov{z} \gll x+iy\;$ angesehen werden. F\"ur den Fall, 
da\ss\ nur eine dieser Variablen vorkommt, trifft der 
Merkvers
\bea{3.12}
   \hbox{\ft $z \gll x - iy \;\; , \;\; 
         \6 \,\gll {1\02} \(\6_1 + i \6_2 \)$} 
   \quad &,& \quad   
   \6 f(z) = f^\prime (z) \quad , \quad  
           \ov{\6} f(z) = 0   \nonu \\ 
   \hbox{\ft $\ov{z} \gll x + i y \;\; , \;\; 
         \ov{\6} \,\gll {1\02} \(\6_1 - i \6_2 \)$} 
   \quad &,& \quad
   \ov{\6} f(\ov{z}) = f^\prime(\ov{z}) \quad , \quad  
           \6 f(\ov{z}) = 0  \quad  \quad 
\eea 
zu$\,$: Null bei Ableiten nach der \anfu falschen\anfo
Variablen.


\sec{Aufl\"osung von \boldmath$A=-(\6M)M^{-1}$ 
     nach \boldmath$M$}

Wir erinnern uns, wie die Bornsche N\"aherung zu Papier
kommt$\,$: $(\D + k^2 )\,\psi = V \psi\,$, rechte Seite als
bekannte Inhomogenit\"at ansehen, Aufl\"osung nach $\psi$
mittels Greenscher Funktion des Helmholtz--Operators und
schlie\ss lich Iteration mit physikalischem Start--$\psi$.
Der Operator ist hier $\6\,$. Und die Dgl ist $\6 \, 
M = - AM\,$.

Um $\,\6\, G(\vc r) = \d (\vc r )\,$ (2D Delta--Funktion) 
zu l\"osen setzen wir $\,G(\vc r) = 2 (x-iy)\, f(r)\,$, 
so da\ss\
\be{4.1}
  \6\, G\; = \; (\6_x+i\6_y)\, (x-iy)\,f(r) = (2+r\6_r)f(r)
         = {1\0r}\,\6_r\, r^2 f(r) \;\; 
        \ueb{!}{=}\;\; \d(\vc r) \quad . \quad
\ee 
Die Null abseits Ursprung braucht also ein $f(r) \sim 1/r^2$,
dies eine Einbettung von der physikalischen Seite her, und 
die enstehende Delta--Darstellung eine Normierung$\,$:
\bea{4.2}
  f(r) &=& {\a \0 r^2 + \e^2} \quad , \quad  
  1 \;\ueb{!}{=}  2\pi \int_0^\infty \! dr \; r \; \( 
       {1\0r}\, \6_r \, r^2\) {\a \0 r^2 + \e^2} \quad 
   \Rightarrow \quad \a = {1\02\pi} \nonu \\ 
  G(\vc r ) \;& = &\; {1\0 \pi}\, {x-iy \0 r^2 + \e^2 }
  \;\; = \;\; {1\0 \pi}\, {z \0 z \ov{z} + \e^2 }
\eea 
Nat\"urlich ist $\e \to +0$ gemeint \ --- \ aber bitte 
nicht diesen limes ausf\"uhren, \hbox{\anfu $G
$}\hbox{$=1/(\pi \ov{z})$\anfo} schreiben und dann in 
die Falle $\6G=0$ tappen. Wegen Translationsinvarianz von 
$\6$ kann
\be{4.3}
   \6 \, G (\vc r-\vc r^\prime) = \d (\vc r-\vc r^\prime)
   \;\;\;\Rightarrow \;\;\;  
  \6 \int \! d^2r^\prime \; G(\vc r-\vc r^\prime) \( - 
   A (\vc r^\prime)  M(\vc r^\prime) \) 
   = - A(\vc r ) M(\vc r ) \quad
\ee 
geschrieben und eine spezielle L\"osung $M_{\rm spez}$ der
inhomogenen Dgl $\6\, M = - AM$ abgelesen werden. Also ist
\be{4.4}
  M=M_{\rm hom} - \int^\prime G A M \quad , \quad
  \6 M_{\rm hom} = 0 \quad , \quad 
   M = 1 - \int^\prime G A M \quad . \quad
\ee 
Die homogene Dgl wird durch eine beliebige Matrix
$M_{\rm hom}(\ov{z})$ gel\"ost. Mehr hierzu steht in 
\S~11.1$\,$. Die gew\"unschte Aufl\"osung ist also 
nicht eindeutig. $\, M_{\rm hom}=1\,$ ist \ e i n e \ 
erlaubte einfache Wahl.

Mit \eq{4.4} rechts gehen wir jetzt in Matrixsprache
unter Weglassen der Integrale (Summenkonvention). $M$
ist (kontinuierlich) mit $\vc r$ indizierter Vektor, 
ebenso die 1 in \eq{4.4}. $G$ ist doppeltindizierte 
Matrix, und $A(\vc r^\prime )$ d\"urfen wir in 
\eq{4.4} durch die Matrix $\AA (\vc r^\prime , 
\vc r^{\prime\prime}) \gll A(\vc r^\prime )\,
\d (\vc r^\prime - \vc r^{\prime\prime})$ ersetzen.
\hbox{\bf 1} stehe f\"ur $\d (\vc r - \vc r^\prime )\,$. 
Das in der n\"achsten Gleichung jeweils rechts 
stehende $A$ ist wieder nur Vektor. Hiermit sieht 
die Iteration von \eq{4.4} folgenderma\ss en aus$\,$:
\bea{4.5}
  M &=& 1\, -\, GA\, +\, G\AA\, GA\,  - \, G\AA\, 
     G\AA\, GA\, + G\AA\, G\AA\, G\AA\, GA\, 
     - \, \ldots \nonu \\[2pt]
  &=& 1 - { 1 \0 \hbox{\bf 1} + G \AA }\, G A
        \, = \, 1 - {1\0 1/G + \AA\,}\, A
        \, = \, 1 - {1\0 \,\bp + \AA\,} \, A \quad , 
\eea 
wobei wir $\,1/G=\bp\,$ aus $\,\bp\, G = \hbox{\bf 1}$ 
abgelesen haben \ --- \ $\bp$ ist Matrix (!!), 
n\"amlich $\6_{\vcsm r , \vcsm r^\prime} = {1\02} 
\d^\prime (x-x^\prime) + {i\02} \d^\prime (y-y^\prime)\,$.
Aufl\"osung gelungen (und \ek{2.7}, \ek{2.8} verstanden). 

Warnung$\,$: \eq{4.4} kann man als $(\hbox{\bf 1} + G\AA) 
M = 1$ schreiben. $\bp$--Anwenden gibt $(\bp + \AA ) M 
= 0\,$: auch richtig, denn das ist \eq{3.6}. Aber 
Aufleiten per $G$ von links $\big(\,\leadsto\; 
(\hbox{\bf 1} + G\AA ) M = 0\,\big)$ holt die 
Inhomogenit\"at nicht zur\"uck (\anfu $\;\bp 1 = 0 
\;\leadsto\; 1=0\;$\anfo$\!$).


\sec{Eichinvariante Freiheitsgrade : \boldmath$M=V\rho$}

In den vorigen beiden Abschnitten wurden die Verh\"altnisse
im Raum der Eichfelder $A_j^a$ (\anfu Oberwelt\anfo) 
umformuliert. Dabei war eine eineindeutige Abbildung in 
den Raum der SL(N,C)--Matrizen $M\,$ (\anfu Unterwelt\anfo)
entstanden \ --- \ jedenfalls dank der 
Unterwelt--Einschr\"ankung \eq{4.4}. Jedes $A$--Feld 
wei\ss\ von seinem Partner $M$ und umgekehrt$\,$:
\be{5.1}  
    M \;\; \ueb{ A = -(\6 M)\, M}{ \vrule depth -2.6pt 
    height 3pt width 2cm\!\!\longrightarrow} 
   \;\; A  \qquad , \qquad 
   A \;\; \ueb{M = 1 - {1\0 \6+A} \, A}{
   \vrule depth -2.6pt height 3pt width 2cm\!\!
   \longrightarrow}  \;\; M \quad .
\ee 
Jeder Weisheit in der Unterwelt der $M$'s entspricht eine 
solche in der realen Oberwelt der $A$--Felder. Vorteil
der Unterwelt ist die besonders einfache Formulierung 
$M \to M\ou = UM$ der Umeichung. Den Wunsch--Raum der 
physikalischen Felder k\"onnen wir ebensogut in der 
Unterwelt zu konstruieren versuchen. Und diese Aufgabe 
ist ein Halbzeiler. Wenn sich generell zeigen l\"a\ss t, 
da\ss\ sich eine beliebige Matrix $M$ aus SL(N,C) als das 
folgende \ P r o d u k t \ schreiben l\"a\ss t,
\be{5.2}
  M \; = \; V \; \rho \quad\qquad \vtop{\hbox{mit \quad
  $V V^\dagger =1 \; , \;\; \det(V)=1$}
  \hbox{und \qquad $ \rho^\dagger 
        = \rho \; , \;\; \det(\rho) = 1$ \quad , }}
\ee 
dann ist klar, da\ss\ der Vorfaktor $V$ (in jedem $M$) nur 
eine Ein--St\"uck--weit--Umeichung von $M=\rho$ ist und 
da\ss\ die hermiteschen $\rho$--Matrizen (mit 
Determinante eins) bereits {\bf den} Wunsch--Raum in der 
Unterwelt darstellen und ausf\"ullen. KKN unter \ek{2.9} 
sagen es so$\,$: {\sl $\rho$ represents the 
gauge--invariant degrees of freedom}. \ {WOW$\,$!}

Wir haben jetzt etwas zum Festhalten. \eq{5.2} ist 
zu beweisen. Wenn in der Unterwelt schon alles klar ist,
dann wird es irgendwie auch m\"oglich sein, die 
Integration ($\,d\mu (\cl C)\,$) \"uber physikalische 
Oberwelt--Freiheitsgrade explizit zu machen und zu 
Quantenmechanik \"uberzugehen$\,$: $\psi(\rho )$. KKN 
behaupten, man k\"onne (statt mit $\rho$) den 
Wunsch--Raum ebensogut auch mit den Elementen
\be{5.3}
  \rho^2 \; = \;\rho^\dagger \,\rho 
   = M^\dagger V \, V^\dagger M  = M^\dagger M 
   \; = \; H \quad , \quad \det(H)=1 \qquad 
\ee 
ausstaffieren. Im ersten Moment mag man hier erschrecken, 
weil ja auch $(-\rho)\, (-\rho)$ zu $H$ f\"uhrt. Jedoch 
wird alsbald an \eq{5.6} klar, da\ss\ $\rho$ positiv 
definit ist$\,$: $\rho^2$--Bilden verliert keine 
Information. Fazit, die hermiteschen, 
eins--determinantischen $N\times N$--Matrizen $H$ bilden 
den Unterwelt--Wunsch--Raum. F\"ur Asylanten und andere 
Fremdsprachler ist dieser der \anfu SL(N,C) / SU(N)\anfo 
\ --- \ und Fernziel ist $\psi(H)$.
      \nz[.4cm] \hspace*{2cm} 
{\bf Beweis von \boldmath$\; M=V\rho$ :} 
      \nz[2pt]
Man jammere ein wenig herum bis Erbarmen eintritt, etwa 
in Form einer einschl\"agigen e--mail von York Schr\"oder 
(DESY). Und dann geht das wie folgt. 
\begin{enumerate}
\item
$M^\dagger M$ ist hermitesch und somit diagonalisierbar$\,$: 
\be{5.4}
      U \; M^\dagger\; M\; U^\dagger\; = \,\; \hbox{diag}
      \,(\l_1, \ldots, \l_N ) \;\;\glr\; {\rm diag} \quad . 
\ee 
\item
Wie $M^\dagger M \ph = \l \ph \;\; \Rightarrow \;\; 
\int | M \ph |^2 = \l$ zeigt, sind die Diagonalelemente 
$\l_j$ nicht--negativ. Sie sind sogar echt positiv, weil           
\be{5.5}
  1\; = \; \det (UM^\dagger M U^\dagger)
   \; = \; \det ({\rm diag})\quad 
\ee 
Null--Eigenwerte verbietet. 
\item  
Jetzt \ d e f i n i e r e n \ wir die hermitesche Matrix 
\be{5.6}
   \rho \; \gll \; U^\dagger \,\wu {\rm diag}\, U 
   \qquad \Rightarrow \quad  \det(\rho) \, 
   = \, 1 \;\; , \;\;\;  {1\0 \rho} \, 
   = \, U^\dagger\, {1\0 \wu {\rm diag} }\, U \quad , 
\ee 
wobei die Eins--Determinante mittels \eq{5.5} folgte 
und $\wu {\phantom{aa}} \gll + \wu {\phantom{aa}}$. 
Ist $M$ gegeben, so liegt die Matrix $\rho$ eindeutig
fest (weil einerseits wegen \eq{5.4} $\rho^2=M^\dagger M$ 
ist und andererseits $\rho$ nur positive Eigenwerte hat,
die Elemente von $\wu {\rm diag}$n\"amlich). F\"ur $U$ 
gilt das \"ubrigens keineswegs. Es ist anders als bei 
reellen Drehmatrizen. Selbst bei Nicht--Entartung 
und Ordnung der diag--Elemente nach Gr\"o\ss e kann aus 
$U$ eine Diagonalmatrix $U_{\rm ph}$ aus Phasenfaktoren 
nach links abgespalten werden. Diese rekombinieren in der 
Bildung $U_{\rm ph}^\dagger\, {\rm diag}\, U_{\rm ph}$. 
$U$ liegt nicht fest, aber $M^\dagger M$ und $\rho$ 
sehr wohl. \\
Parameter--Z\"ahlen \ --- \ siehe unten um \eq{5.8} herum 
\ --- \ gibt \"ubrigens in \eq{5.6} auf der linken Seite 
$N^2\!-\!1$ (in $\rho $) und auf der rechten Seite 
$N\!-\!1$ ($N$ in diag$\,$; $-1$ wegen $\det=1$) plus 
$N^2$ (in $U$) minus $N$ (in $U_{\rm ph}$). Es pa\ss t.
\item  
So nun also $\rho$ ein Inverses hat, d\"urfen wir getrost 
$M \glr V \rho$ nach $V$ aufl\"osen und unverz\"uglich
dach Determinante und Unitarit\"at fragen$\,$: 
\bea{5.7}
  V &=& M \; {1\0\rho} \;\;\qquad \Rightarrow \qquad 
  \det(V)\, = \, 1 \quad \hbox{und} \quad \cr
  V^\dagger \, V \; &=& \; {1\0\rho}\, M^\dagger\, M\, 
   {1\0 \rho} \; = \; U^\dagger\, {1\0 \wu {\rm diag} }\, 
   U\, M^\dagger\, M\, U^\dagger\, {1\0 \wu {\rm diag} }\, 
   U \;\; = \;\; 1 \quad .
\eea  
Was man sogar aufschreiben kann (mit beiden gew\"unschten
Eigenschaften), das existiert. Qed, fertig und dankesch\"on.
\end{enumerate}
          
Zu begreifen, weshalb \eq{5.2} gilt, bereitete anfangs 
einige M\"uhe. Z\"ahlen wir doch erst einmal reelle 
Paramter (sagte Ketov), um zu sehen, ob $M=V\rho$ 
m\"oglich ist. Und auf Zusammenhang mit 
Lorentz--Transformation verwiesen beide, Prof.~Lechtenfeld 
und Dr.~Ketov. Unabh\"angig vom obigen Beweis ist 
Parameter--Z\"ahlen eigentlich ganz am\"usant. Eine
komplexe $N\!\times\!N$--Matrix hat $2N^2$ Elemente, die
Forderung $\,\det (M)\,\ueb{!}{=}\,1$ reduziert auf
$2N^2\!-\!2 = 2n\,$. $V$ ist SU(N)--Element und
hat $n$ reelle Parameter$\,$: $V=e^{i \vcsm \L \vcsm T}$.
Der interessantere Faktor ist $\rho$ mit $\det(\rho)=1$. 
Weil positiv definit, ist folgendes erlaubt$\,$:
\be{5.8}
  0 = \ln \lk \det \( \rho \) \rk 
  = \Sp \lk \ln \( \rho \) \rk \quad \Rightarrow \quad
  \ln \( \rho \) = \o^a T^a \quad , \quad
  \rho  = e^{\vcsm \o \vcsm T} \quad .
\ee 
Auch $\rho$ hat $n$ reelle Parameter. Und man kann sie
ebenfalls im Exponenten ansiedeln. $2n=n+n$. Genug 
gez\"ahlt. $M=V\rho$ ist aus dieser Sicht nichts weiter 
als eine spezielle Aufschreibungsart der SL(N,C)--Elemente. 


\sec{Die R\"aume \quad \lower .6cm\vbox{
     \hbox{\boldmath$\;\; \cl A \,\quad | \quad \cl C 
                     \;\quad |$ } \vskip -.4cm
     \hbox{--------------------------} \vskip -.4cm
     \hbox{ \boldmath$\cl M \quad | \quad \cl H 
                     \quad |\quad \cl G_* $ }} }

Etwas Besinnung ist angezeigt, Strategie und Ausblick.
Die Horizontale trennt Oberwelt und Unterwelt. Alle diese 
R\"aume, welche in der \"Uberschrift Namen bekommen haben, 
kennen wir eigentlich schon$\,$:  
\\[4pt] 
$\cl A \;$: \parbox[t]{15cm}{
   der Raum \ a l l e r \  Eichfelder (egal, ob wir 
   dabei an die reellen $A_j^a(\vc r)$ denken, oder an 
   die elegante Zusammenfassung zu einem komplexen 
   spurfreien Matrix--Feld $A$).}
\\[8pt] 
$\cl M \;$: \parbox[t]{15cm}{ 
       der Raum aller ($\vc r$--abh\"angigen)
       Matrizen aus SL(N,C).}
\\[4pt] 
$\cl H \;$: \parbox[t]{15cm}{
    der Unterwelt--Wunschraum $=$
    Raum aller ($\vc r$--abh\"angigen) hermiteschen 
    $N\times N$--Matrizen $H$ mit Determinante 1.}
\\[8pt] 
$\cl G_* \;$: \parbox[t]{15cm}{
    die {\sl gauge group} $=$ der Raum aller ($\vc r
    $--abh\"angigen) unit\"aren Matrizen $U$ aus SU(N).}
\\[4pt] 
$\cl C \; $: \parbox[t]{15cm}{der Raum nur aus solchen 
      Eichfeldern, welche nicht mehr durch Umeichung 
      aus\-einander hervorgehen k\"onnen $=$ {\bf der} 
      interessierende physikalische Unterraum, in 
      welchem zu Quantenmechanik \"ubergegangen werden 
      kann.}
\\[.6cm]
Kein graues Haar m\"oge beim Anblick der
folgenden Gleichung wachsen,
\be{6.1}
 \lower 10pt\hbox{
 \vbox{\hbox{\ft \ \ Raum aller 2-komponentigen 
                     Eichfelder} \vskip -.2cm
       \hbox{\ft \ \ \quad b i s \ a u f \ 
                     Umeichungen $=$} \vskip -.2cm
       \hbox{\ft {\sl space of gauge--invariant 
                   field configurations} $\cl C$ } } }
    \;\; = \quad { 
 \hbox{ 
 \vbox{\hbox{\ft Raum \ a l l e r \ 
                 Eichfelder $=$} \vskip -.2cm
       \hbox{\ft {\sl set of all gauge potentials} 
                  $\cl A\;$}}}   \0 
       \;\hbox{\ft {\sl gauge group} 
       $\cl G_*\;$} } \quad , \;\;
\ee 
denn \eq{6.1} \ d e f i n i e r t \ lediglich, was ein 
Bruchstrich im Gruppenchinesisch bedeutet. Bei KKN 
(Zitat [4] dort) ist allerlei Literatur zur Geometrie 
des Raumes $\cl C$ angegeben.

Jemand k\"onnte fragen, weshalb nicht $\cl H$ und $\cl C$ 
identisch seien. Nun, die $H$'s sind Unterwelt--Bewohner. 
Und wenn Integration \"uber $\cl A$ nicht jene \"uber 
$\cl M$ ist, weil eine Jacobi--Determinante dazwischen 
sitzt, dann wird wohl auch bei $\cl H$ und $\cl C$ etwas
\anfu dazwi\-schen sitzen\anfo$\!$.

Im Schr\"odinger--Bild einer Feldtheorie sind an jedem der 
$\infty$ vielen Punkte $\vc r$ des diskretisierten Raumes 
einige (hier $2*n$) \anfu Feldachsen\anfo angebracht. Die 
reellen Werte $A_j^a$ auf diesen sind die Variablen$\,$: 
$\psi \!\lk A_j^a(\vc r) \rk$. Jaja, wir vergessen jetzt 
kurzzeitig die Eichfreiheit. $\mu$, $a$, $\vc r$ z\"ahlen 
nur die $2*n*\infty$ Variablen ab. Ein Skalarprodukt 
$\lw 1 | 2 \rw$ zwischen zwei $\psi$--Zustandsvektoren 
enth\"alt also $2*n*\infty$ Integrale$\,$:
\be{6.2}
  \int \psi_1^* \psi_2 \; = \; \int \! dA_1^1(1) 
  \, \ldots \, dA_2^n(\infty)\bigg|_? \;\; 
  \psi_1^* \!\lk A_j^a(\vc r) \rk \,
  \psi_2 \!\lk A_j^a(\vc r) \rk \,  \glr \, 
  \int\! d\mu (\cl A)\bigg|_? \;\psi_1^* \psi_2
\ee 
Dem Produkt der Differentiale wurde rechts in \eq{6.2}
ein Name gegeben$\,$: Vo\-lu\-menelement $d\mu (\cl A)$ 
im Raum $\cl A$. 

Der Fragezeichen--Index verweist auf das Problem. \eq{6.2} 
gibt nur Sinn, wenn zuvor andere Variable (physikalische 
kontra umeichische) so einge\-f\"uhrt werden, da\ss\ sich 
die Integrationen in \eq{6.2} auf physikalische 
beschr\"anken lassen. Nur von diesen h\"angt $\psi$ ab$\,$: 
$\psi\!\lk H(\vc r )\rk$. $\vc r$ ist nur Variable 
abz\"ahlender Index. Damit zeichnet sich eine Stategie ab, 
genau jene, welche auch bei Anwendungen der 
Fourier--Transformation zutrifft$\,$: Marsch hinab in die 
Unterwelt ($M$'s), dort den physikalischen Raum ($H$'s) 
abspalten, Umeichvo\-lumen wegwerfen (Fad\-de\-ev und 
Popov lassen gr\"u\ss en) und zur\"uck nach 
oben$\,$: 
\bea{6.3}  
   & & \nonu \\[-.4cm]
   & & \lower .4cm\vbox{
  \hbox{$d\mu \(\cl A \) \hspace*{2cm}  d\mu \(\cl C \)$}
  \vskip -.1cm
  \hbox{$\;\;\; \downarrow \hspace*{3cm} \uparrow$}
  \vskip -.1cm
  \hbox{$ d\mu \(\cl M \) \hspace*{.5cm}
         \longrightarrow \hspace*{.5cm}
          d\mu \(\cl H \) \;\; \cdot \;\; \lb \; 
          d\mu (\cl G_*) \; \rb$} }
\eea 
Die n\"achsten vier Abschnitte werden in den Details 
der Volumenelemente ertrinken. Aber dann werden sich die
$\psi$'s melden und nach einem Hamilton--Operator rufen.
Die kinetische Energie $\int\! d^2r\;\cl T$ besteht nach 
\eq{2.9} aus $2*n*\infty$ $\p A$--Quadraten. Sie wird
quantenmechanisch zu einem multidimensionalen 
Nabla--Quadrat. Wenn beschr\"ankt auf $\cl C$ wird man 
es den Laplace--Operator auf $\cl C$ nennen k\"onnen.

Wir erwarten, da\ss\ ein neues Volumenelement $d\mu$ das 
Produkt der Differentiale neuer Variabler ist {\bf mal} 
eine Jacobi--Determinante. Letztere ist Betrag der 
Determinante der \anfu Jacobi--Matrix\anfo $\Im \gll 
\6({\rm alte Variable})/\6({\rm neue})\,$. Am Beispiel 
Kugelkoordinaten werde klar, da\ss\ und wie $\Im$ aus der 
Metrik $ds^2$ erhalten werden kann. So dienen denn die 
Zeilen 1 und 2 der nachfolgenden Tabelle zum Aufw\"armen. 
Bei $ds^2$ in Zeile 2 teile man im Geiste durch $dt^2$ und 
denke an die kinetische Energie ($v^2$) eines Teilchens. 
Wer mit allgemeiner Relativistik zu tun hat, wird hier 
wohl milde l\"acheln ($g\omn$ aus $ds^2$ und $\wu {\rm 
det}$ in der Wirkung). Ausgangspunkt ist der Raum $\cl A$. 
Seine Metrik in Zeile 3 ist Euklidisch und harmlos (lies 
$\int\! d^2r$ als $\sum_{\vcsm r}$). Alles weitere gehe 
auf Konto Ausblick. 
\bea{6.4} 
   & &  \nonu  \\[-.2cm]  
   & & {\ft  \hspace{-.8cm}
   \begin{tabular}{ r | c | c | l | l | }\hline
  & Raum & Elemente & \qquad Metrik & 
           \ \ \ Volumenelement \\ \hline\hline
{\ft 1} & $ R^3 $  &  $\vc r $  
        &  $ ds^2 = dx^2 + dy^2 + dz^2 $ 
        &  $ d^3r = dx\, dy\, dz $  \\ \hline 
{\ft 2} & $ R^3 $  &  $r$, $\ta$, $\ph$  
        & $ ds^2 = dr^2 + r^2 d\ta^2 + 
            r^2 \sin^2 (\ta ) d\ph^2 \!$ 
        & $ d^3r = dr r^2 \, d\ta \sin (\ta )\, d \ph $ \\  
        & &  $ \matrix{ \Im\,\gll\, \cr
               {\6 (x,y,z) \0 \6 (r,\ta ,\ph )} \cr } $  
        &  $ ds^2 = \hbox{\ft $\( \matrix{ dr \cr d\ta 
             \cr d\ph \cr} \)$} \, \Im^{T} \; \Im 
   \hbox{\ft $\( \matrix{ dr \cr d\ta \cr d\ph \cr} \)$} $ 
   & $ d^3r = dr \, d\ta \, d \ph \, |\det (\Im )| $ \\ \hline 
{\ft 3} &  $\cl A$   &  $A$  
        &  $ ds^2 = \int\! d^2r \;\d A_j^a \d A_j^a $ 
        & $ d\mu \( \cl A \) = d\mu \( \cl M \) \, 
           \det ( D^\dagger D ) $ \\ \hline
{\ft 4} & $\cl M $  &  $M$  
        & $ ds^2 = 8 \int\!d^2r \, \hspace{2.9cm} $ 
        & $ d\mu \( \cl M \) = \hspace{.8cm} $ \\ 
   & {\scriptsize SL(N,C) }    & 
        & $ \quad  \Sp \lk ( \d M M^{-1} )
            ( M^{\dagger\, -1} \d M^\dagger ) \rk $ 
        & $ \qquad  d\mu(\cl H ) \, vol(\cl G_*)$ \\ \hline
{\ft 5} & $\!\cl H = {{\rm SL(N,C)} \0 {\rm SU(N)} }\!$ 
           \rule[-.3cm]{0pt}{.9cm} &  $H$  
        &  $ds^2 = 2 \int\! d^2r \; 
            \Sp (H^{-1} \d H H^{-1} \d H) $
        & $ d\mu \( \cl H \)$  \\ \hline 
{\ft 6} &  $\cl G_* =${\ft SU(N) }  &  $U$  
        & & $ d\mu \( \cl G_* \)$  \\ \hline
{\ft 7} & $\cl C$   &  $A_{\rm phys}$  
        & & $ d\mu \( \cl C \) = 
        d\mu (\cl H )\, \det ( D^\dagger D) $ \\ \hline
\end{tabular}  } \hspace{1cm}
\eea 

\noindent
Zum Stichwort Besinnung haben vielleicht die vier Sorten
$A$'s eine \"Ubersicht verdient$\,$:
\bea{6.5}
     A_j = - iT^a A_j^a \quad \longleftarrow 
   & \hbox{\ft $2n$ reelle} \; A_j^a & 
     \longrightarrow \quad A^a = {1\02} \( A^a_1 
         + i A^a_2 \) \nonu \\[-.2cm]
     \downarrow \hspace*{3.4cm} 
   & & \hspace*{1.3cm} \downarrow \nonu \\[-.15cm]
     {1\02}\(A_1+iA_2 \) \;\; = \; 
   &\hbox{\ft ein spurfreies}\, A& = \; - i T^a A^a
   \quad . \quad
\eea 


\sec{\boldmath$ d\mu(\cl A) \to d\mu(\cl M) $ : 
     Jacobi--Determinante}

An einer Stelle im Raum $\cl A$ werde eine kleine 
\"Anderung $\d A_j^a (x,y)$ des Feldes vorgenommen. Es 
versteht sich, da\ss\ dabei die linearen 
in--Matrix--Umwandlungs--Relationen \eq{6.5}
mitziehen$\,$:
\bea{7.1}  
  \d A_1^a \d A_1^a + \d A_2^a \d A_2^a  
  & = & (\d A_1^a + i \d A_2^a)\, (\d A_1^a - i \d A_2^a) 
  \; = \; 4\; \d A^a\; \d A^{a*} \nonu \\
  & = & 8\,\Sp \( T^a \d A^a \, T^b \d A^{b*} \) 
  \; = \; 8\,\Sp \( \d A \,\d A^\dagger \) \quad .\quad
\eea 
\pagebreak[3]
     \nz[.5cm]  \hspace*{2cm} 
{\bf 7.1 \quad \boldmath$ d s_{\cl A}^2$ und $\d M$ }
     \nz[.1cm] 
\nopagebreak[4]
Da $A$ eindeutig mit $M$ zusammenh\"angt, wird sich 
\eq{7.1} auf $\d M$ umschreiben lassen. Wir haben 
lediglich $A=-(\6 M) M^{-1}$ zu bedienen und flei\ss ig
von 
$$ M\, M^{-1} = 1 \;\; \Rightarrow \;\; 
   \d M^{-1} = - M^{-1} \d M M^{-1} \;\; \hbox{und} \;\;
   \6 M^{-1} = - M^{-1} (\6 M) M^{-1}\, $$
Gebrauch zu machen$\,$:
\bea{7.2}
 \d A &=& - (\6\d M ) M^{-1} - (\6 M) \d M^{-1} \nonu \\
      &=& \; - \6 \lk \d M M^{-1} \rk + \d M \6 M^{-1}
           + (\6 M) M^{-1} \d M M^{-1} \nonu \\
      &=& \; - \6 \lk \d M M^{-1} \rk - \d M M^{-1} (\6 M ) 
          M^{-1} + (\6 M) M^{-1} \d M M^{-1}  \nonu \\
      &=& \; - \lb \; \6 \lk \d M M^{-1} \rk \; + \; 
          \lk A \, , \,\d M M^{-1} \rk \; \rb  \nonu \\
      &=& \; - \; \dod \;\,\d M M^{-1} \qquad 
     \hbox{mit} \quad \dod \,\gll 
                \6 + \lk A \, , \, \;\; \rk
\eea 
Vermutlich (Sinn oder Willk\"ur bei KKN$\,$?) kann statt 
$\d$ ebensogut auch $d$ geschrieben werden, wobei dann 
aber Wirkungsweise--Begrenzer n\"otig w\"urden. $\d$ 
beziehe sich stets nur auf die unmittelbar folgende Matrix. 
Die kovariante Ableitung wird in allerlei Varianten
auftauchen. In fundamentaler Darstellung haben wir $D_j 
= \6_j + A_j$ und k\"onnen die beiden zu $D \gll {1\02} 
( D_1 + i D_2) = \6 + A$ kombinieren. $\dod$ in \eq{7.2} 
ist die Kommutator--Variante der Ableitung in adjungierter 
Darstellung. Deren Index--Variante $D^{ab}$ kommt ins 
Spiel, wenn man $\dod$ auf ein Matrixfeld $\,\L^a\, T^a\,$ 
anwendet:
\bea{7.3}
  \dod\, \L^a T^a &=& T^a \6 \L^a 
      - i A^b \lk T^b\, ,\, T^c \rk \L^c  \nonu \\
      &=& T^a D^{ac} \L^c \qquad \hbox{mit} \qquad
      D^{ac} \, \gll \, \d^{ac} \6 + f^{abc} A^b \quad .
\eea 
Mit $A^b$ ist nat\"urlich ${1\02}\( A_1^b + i A_2^b\)$ 
gemeint. In \eq{7.1} wird auch $\d A^\dagger$ ben\"otigt.
Ausgehend von $A^\dagger = - M^{\dagger\, -1} \,\ov{\6} 
M^\dagger$ mit $\ov{\6} ={1\02} \( \6_1 - i\6_2 \)$
entsteht in jedem Schritt das Gekreuzte von \eq{7.2}$\,$:
\be{7.4}
  \d A^\dagger = \;\ldots\; = - \; \ov{\dod} \; 
  M^{\dagger\, -1} \d M^\dagger \quad \hbox{mit} \quad 
  \ov{\dod} \,\gll \ov{\6} - 
       \lk A^\dagger \, , \, \;\; \rk \quad .
\ee 
Nach Einsetzen in \eq{7.1} und Summation $\int\! d^2r$
\"uber Raumpunkte entsteht als Zwischenresultat
\be{7.5}
   ds^2_{\cl A} = \int\! d^2r \; \d A_j^a \,\d A_j^a  
   \; = \; 8 \int\! d^2r \;\Sp \( \lk \dod \; 
  \d M M^{-1} \rk \lk \ov{\dod} \; 
  M^{\dagger\, -1} \d M^\dagger \rk \) \quad ,
\ee 
Dies war nur der erste von drei Schritten. Im zweiten
Schritt interessiert $\d M$, und wir finden \ --- \
unabh\"angig von \eq{7.5} \ --- \ das Volumenelement 
$d\mu (\cl M )\,$. Erst danach (3.~Schritt) wird der 
Zusammenhang zwischen $d\mu (\cl A )$ und $d \mu (\cl 
M )$ hergestellt. 
\pagebreak[2]
     \nz[.5cm]  \hspace*{2cm} 
{\bf 7.2 \quad \boldmath$d \mu (\cl M )$ }
     \nz[.1cm] 
\nopagebreak[3]
Gruppenelemente (hier $M$'s) gehen durch Multiplikationen
auseinander hervor. Die Me\-trik, sagt Prof. Dragon, 
erschlie\ss e sich ausgehend vom 1--Element.  
Infini\-te\-si\-ma\-le Ab\-wei\-chung von der Eins kann 
per $ 1 + {1\02} \vc \e \vc T $ parametrisiert werden.
Die Komponenten $\e^a$ von $\vc \e$ sind komplex. Linear
in $\vc \e$ ist $\det (M)=1$ gew\"ahrleistet. Nun 
setzen wir uns mitten in die Gruppe hinein und wollen einen 
kleinen Unterschied zwischen $M$ und Nachbar--Element $M + 
\d M$ dingfest machen \ --- \ durch Multiplikation$\,$:
\bea{7.6}
  M + \d M  &=& \big( \, 1 + \2 \vc \e \vc T \, \Big) \; 
  M \quad \Rightarrow \quad \d M = \2 \vc \e \vc T M 
      \nonu \\
  \d M \, M^{-1} &=& \2 \vc \e \vc T \quad , \quad 
  \e^a \; = \;  4 \;\Sp \( T^a \,\d M \, M^{-1} \)\quad .
\eea 
Der unn\"otig erscheinende Vorfaktor \2 wird sogleich f\"ur 
einfache Resultate sorgen. $d \mu (\cl M )$ folgt aus 
Metrik, und diese braucht quadratisch infinitesimale
Gr\"o\ss en. Die erste naheliegende Idee funktioniert$\,$:
\be{7.7}
  8\; \Sp \( \d M M^{-1} \, M^{\dagger\, -1} \d M^\dagger \)
  \; = \; 2\; \Sp \( \e^a T^a \; \e^{b\, *} T^b \)
  \; = \;  \e^a \e^{a\, *} \; = \; \e_1^2 + \e_2^2 \quad ,
\ee 
wobei nat\"urlich \"uber $a$ summiert wird, und $\e_1^a 
\gll \Re e \(\e^a\)\,$, $\e_2^a \gll \Im m \(\e^a\)\,$. 
Die Metrik im Raum der $\vc r$--abh\"angigen $M$--Felder
ist somit
\be{7.8} 
     ds^2_{\cl M} = 8 \int\! d^2r \;
     \Sp \( \d M M^{-1} \, M^{\dagger\, -1} \d M^\dagger \)
      \; = \; \int\! d^2r \;  \e^a \e^{a\, *} \quad .
\ee 
Die erste Gleichung in \eq{7.8} ist \ek{2.12} und in der
4. Zeile der Tabelle \eq{6.4} verzeichnet. Die Epsilontik 
ist Eigenbr\"au. Es steht keine Jacobi--Matrix rechts 
in \eq{7.8}. In unserer epsilontischen Notation ist
folglich
\be{7.9}
  d\mu (\cl M ) \; = \;\prod_{\vcsm r}\;
   \prod_{a,\, j} \,\; \e_j^a  \; = \;
   \prod_{\vcsm r}\; \prod_a \,\; \e_1^a \e_2^a \quad
\ee 
das Volumenelement ({\sl Haar measure}) im Raum $\cl M\,$. 
Mit \eq{7.9} startet \S~8. 
\pagebreak[2]
     \nz[.5cm]  \hspace*{2cm} 
{\bf 7.3 \quad \boldmath$d \mu (\cl A )\;= \; d \mu 
     (\cl M )\;$ mal Jacobi }
     \nz[.1cm] 
\nopagebreak[3]
Jetzt erst steht eine ordentliche Jacobi--Matrix ins Haus.
Wir beginnen wieder mit der Metrik, n\"amlich mit \eq{7.5},
setzen dort die Epsilontik \eq{7.6} ein und greifen auf 
\eq{7.3}, \eq{7.4} zur\"uck$\,$: \bea{7.10}
   ds^2_{\cl A} &=&  2 \int\! d^2r \;\Sp \( 
       \lk \dod \;\e^a T^a \rk 
       \lk \ov{\dod} \;\e^{b\, *} T^b \rk \) 
   \; = \;  2 \int\! d^2r \; \Sp \Big( 
      \lk T^a \, D^{ac} \e^c \rk \lk T^d \, D^{db \,*} 
      \e^{b\, *} \rk \Big) \nonu \\
   &=&  \int\! d^2r \; 
     \lk D^{ac} \e^c \rk \lk D^{ab\, *} \e^{b\, *} \rk
     \qquad \hbox{wobei} \quad D^{ab\, *} \gll \,\lb \, 
      \d^{ab} \, \ov{\6} + A^{\bullet\,*} f^{a\bullet b} 
      \,\rb    \nonu \\
   &=& \int\! d^2r \;\lk D \vc \e \rk \cdot 
       \lk D \vc \e \rki^*  \; = \; 
       \int\! d^2r \; \( \lk \Re e ( D \vc \e ) \rki^2
	       \, + \, \lk \Im m ( D \vc \e ) \rki^2 \) \quad . 
       \nonu \\ 
  & &  \qquad\quad  D \glr D_1 + i D_2 \;\; , \;\;
       \vc \e = \vc \e_1 + i \vc \e_2 \;\; : \nonu \\
  &=&  \int\! d^2r \; \( 
       \lk D_1 \vc \e_1 \rki^2 + \lk D_2 \vc \e_2 \rki^2 
        - 2 \lk D_1 \vc \e_1 \rk \lk D_2 \vc \e_2 \rk  
        \right. \nonu \\[-9pt]
   & & \hspace*{2cm} \left.  + \;
       \lk D_1 \vc \e_2 \rki^2 + \lk D_2 \vc \e_1 \rki^2 
       + 2 \lk D_1 \vc \e_2 \rk \lk D_2 \vc \e_1 \rk 
        \) \nonu \\
   &=&  \int\! d^2r \;  \lk 
   \( \matrix{ D_1 & -D_2 \cr D_2 & D_1 \cr } \)
   \( \matrix{\vc \e_1 \cr \vc \e_2 \cr} \) \rk \cdot \lk
   \( \matrix{ D_1 & -D_2 \cr D_2 & D_1 \cr } \)
   \( \matrix{\vc \e_1 \cr \vc \e_2 \cr} \) \rk \;\glr\; 
    \ueb{\rightharpoonup}{\hbox{\small\bf [\ ]}} \cdot 
    \ueb{\rightharpoonup}{\hbox{\small\bf [\ ]}} \quad .
\eea 
In der dritten Zeile ist nat\"urlich \anfu $D$\anfo nur die 
Abk\"urzung f\"ur die Matrix $D^{ab}$ (Verzeihung$\,$! Zu 
viele $D$'s. Die alte Bedeutung als $\6+A$ ziehen wir jetzt 
kurzerhand aus dem Verkehr). Der Operator $D$ ist nicht nur 
$n\times n$--Matrix sondern enth\"alt auch Differentiation. 
Wie weit diese wirkt, ist in \eq{7.10} konsequent durch 
eckige Klammern abgegrenzt. 

Rechts unten in \eq{7.10} haben wir auch noch $\int\! d^2r$ 
als Skalarprodukt--Summation interpretiert. Vorsicht Falle. 
In diesem Moment ist nun auch $\ueb{\rightharpoonup}
{\hbox{\small\bf [\ ]}}$ als $\vc r$--indizierter, riesiger 
Vektor zu lesen. Schlie\ss lich werden die $D$'s zu 
Riesenmatrizen ${\bf D}$, welche das Indexpaar $\vc r$, 
$\vc r^\prime$ tragen und per $\int^\prime \!\! 
D_{r r^\prime} \e_{r^\prime}$ anzuwenden sind. Wir hatten 
solcherlei bereits in \eq{4.5} und \"ubernehmen darum die 
boldface--Notation. All dies war zu sagen, um aus 
\eq{7.10} die Jacobi--Matrix $\Im$ richtig abzulesen. 
Das Volumenelement bekommt daraufhin die Gestalt
\be{7.11}
  d \mu (\cl A ) \; = \; d \mu (\cl M ) \; \left| 
      \det (\Im ) \right| \; = \;\; d \mu (\cl M ) \;
  \left| \det \( \matrix{ {\bf D}_1 & -{\bf D}_2 
   \cr {\bf D}_2 & {\bf D}_1 \cr } \) \right| \quad .
\ee 
Die Matrix $\Im$ ist reell. Aber $\det (\Im )$ hat noch 
nicht die gew\"unschte Form $\det ( {\bf D}^\dagger 
{\bf D} )\,$. Diese wird mit der folgenden sch\"onen 
Herleitung (diagonalisiere \I, sagt Dr.~J.~Schulze) 
erreicht$\,$: 
\bea{7.12}
  \Im &=& \( \matrix{ {\bf D}_1 & -{\bf D}_2 \cr 
   {\bf D}_2 & {\bf D}_1 \cr } \) \; = \; 
   {\bf D}_1 \, 1 + {\bf D}_2 \, \I \quad , \quad
   \I = \( \matrix{ 0 & -1 \cr 1 & 0 \cr } \) \quad ,
     \nonu \\
  W &=& {1\0 \wu 2 } \( \matrix{ 1 & i \cr i &
       1 \cr } \) \;\; ,  \;\;
   W \I W^\dagger \; = \; \( \matrix{ i & 0 
      \cr 0 & -i \cr } \) \;\; , \;\; \det(W) 
      = \det (W^\dagger) = 1 \quad , \nonu \\
  \det ( \Im ) &=& \det \( W \Im W^\dagger \) \; = \; 
  \det \( \matrix{ {\bf D}_1 + i {\bf D}_2 & 0 \cr 
          0 & {\bf D}_1 -i {\bf D}_2 \cr } \) \; = \;
  \det \( {\bf D}\, {\bf D}^* \) \quad .
\eea 
Das $\cl A$--Raum--Volumenelement hat nun die Gestalt 
$\, d\mu (\cl A ) = d \mu(\cl M ) \left| \det ({\bf D} 
{\bf D}^* ) \right|$. Sind wir zufrieden$\,$? Die 
Betragsstriche st\"oren noch. 

Was mag wohl beim Vorzeichen solcher Determinanten schon
ges\"undigt worden sein (nur wir nicht$\,$: die 
Betragsstriche stehen da). Nach Blick auf \eq{7.3} 
schreiben wir die Matrix ${\bf D}$ mit allen Indizes auf,
\be{7.13}  \;\;
  {\bf D}\, : \quad
  D^{ab}_{r r^\prime} = \d^{ab} \6_{r r^\prime} + 
  A^\bullet (\vc r ) f^{a \bullet b} \d_{r r^\prime}
  \quad , \hspace*{5.1cm}
\ee 
und erkennen, da\ss\ 
\be{7.14}
  {\bf D}^T\, : \quad  
  (D^T)^{ab}_{r r^\prime} = \d^{ab} \( - \6_{r r^\prime}\)
   +  A^\bullet (\vc r ) f^{b \bullet a} \d_{r r^\prime}
  \qquad \hbox{ergo} \quad {\bf D}^T = - {\bf D} \quad
\ee 
ist und folglich $\,{\bf D}^* = - {\bf D}^\dagger\,$. 
Zur Erinnerung$\,$: $\,A^\bullet = \2 \( A_1^\bullet 
+ i A_2^\bullet \)\,$, $\,A_j^\bullet$ reell, und 
$\6_{r r^\prime} = -\6_{r^\prime r}$ stand unter \eq{4.5}. 
Vektorpfeile \"uber $r$--Indizes sind bitte hinzu zu 
denken. Falls die hermitesche Matrix $\,{\bf D}^\dagger 
{\bf D}\,$ nicht gerade einen Null Eigenwert hat (das 
Risiko gehen wir ein), dann haben wir
\be{7.15} 
  - {\bf D} {\bf D}^* = {\bf D} {\bf D}^\dagger  \quad
  \lower 4pt\vbox{\hbox{{\ft ist positiv}} \vskip -.26cm
                  \hbox{{\ft \ \ definit}}} \qquad
  \Rightarrow \qquad  \big| \;\det \( {\bf D} {\bf D}^* 
  \)\,\big| \; = \; \det \( {\bf D}^\dagger  {\bf D} \) 
  \;\; \glr \; e^\G \quad ,
\ee 
und
\be{7.16}
  d \mu \( \cl A \) \; = \; d \mu \( \cl M \) 
  \; \det \( {\bf D}^\dagger  {\bf D} \) \quad
\ee
ist das Resultat dieses Abschnitts. Es vervollst\"andigt 
Zeile 3 der Tabelle \eq{6.4}. Jenes $\G$ in \eq{7.15} ist 
reell. Mit \eq{7.10} bis \eq{7.15} sind wir \"ubrigens 
bei KKN um zwei Zeilen Text unter \ek{2.12} voran gekommen.


\sec{\boldmath$ d\mu(\cl M) \to d\mu(\cl H) $ : Abspalten 
     des \anfu volume\anfo}

Einen \anfu Fu\ss weg\anfo gehen ist eine Sache, ihn 
anzulegen (wo entlang?) eine andere. Es war sehr 
hilfreich, da\ss\ nun, mit Beginn dieses Abschnitts, Herr 
Jens Reinbach voll mitgearbeitet hat. Man frage nicht nach 
\anfu wer--was\anfo, denn wirklich getan haben es ja KKN. 

Zum Abkoppeln unphysikalischer Freiheitsgrade steht aus 
Abschnitt 5 der \anfu Separationsansatz\anfo
\be{8.1}
    M \; = \; U \, \rho \qquad \Rightarrow \qquad
    \d M = \d U \,\rho \, + \, U \,\d\rho
\ee 
bereit. Die \"Anderung der Bezeichnung von $V$ zu $U$ 
signalisiere eine solche der Philosophie$\,$: von jedem
physikalischen \anfu Punkt\anfo $\rho$ starten s\"amtliche 
Umeichungen $U\,$. Die Grobstruktur der jetzt anstehenden 
Rechnungen l\"a\ss t sich in Worte fassen. Und das Resultat 
ist so einfach, da\ss\ man es in \eq{8.2} vorweg verstehen 
kann.

Ausgangspunkt ist $d \mu (\cl M )$, d.h. das Produkt 
\eq{7.9} von Differentialen. Der Zusammenhang der Epsilons 
mit $\d M$ steht in \eq{7.6}. Mittels \eq{8.1} wird sich 
$\e_1^a$ als rein $\rho$--ische Bildung erweisen (nur von 
$\rho$ und $\d \rho$ abh\"angend). Hingegen setzt sich 
$\e_2^a$ additiv zusammen aus einem rein $U$--ischen und 
einem rein $\rho$--ischen Ausdruck. Der letztere kann nur 
Linearkombination der $\e_1^a$'s sein. Hiernach sind neue 
Variable $h$ und $u$ nahegelegt$\,$: 
\bea{8.2}
  & & \e_1^a \; = \; d h^a \quad , \quad \e_2^a \; = \; 
  Q^{ab} d h^b \; + \; d u^a \quad : \nonu \\[6pt]
  \Big[ \prod_a \e_1^a \e_2^a \Big]
  \!\! &=&  \Big[ \prod_a dh^a du^a \Big] \;
  \left| \, \det \( \matrix{ 1 & 0 \cr Q^{ab} & 1 \cr } \) 
  \,\right| =  \Big[ \prod_a d h^a \Big] 
  \Big[ \prod_a du^a \Big] \quad . \quad
\eea 
Ganz h\"ubsch, nicht wahr, denn rechts in \eq{8.2} steht 
bereits die gew\"unschte Zerlegung. \\
\parbox[t]{8.8cm}{Integration \"uber 
$du$ gibt den Umeichraum $\cl G_*$, und aus den $d h$'s 
wird sich $d \mu (\cl H )$ stricken lassen. Anders als in 
\ek{2.13}, \ek{2.14} sind weder {\sl wedge} Produkte noch 
$\backslash$propto Zeichen erforderlich. Im \"ubrigen 
mu\ss\ auch einmal eine Figur zu sehen sein. \vskip .2cm } 
\hspace*{.3cm} \parbox[t]{5.3cm}{\unitlength .96cm 
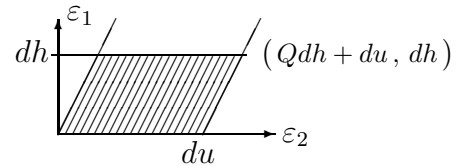
\begin{picture}(5,0)
   \put(1,-1.8){\vector(1,0){3}} 
   \put(1,-1.8){\vector(0,1){1.6}}
   \put(.9,-.7){\line(1,0){2.7}}  
   \put(1,-1.8){\line(1,2){.8}} 
   \put(3,-1.8){\line(1,2){.8}}
   \put(1.12,-.22){$\e_1$} \put(4.1,-1.9){$\e_2$}
   \put(.4,-.8){$dh$}     \put(2.7,-2.2){$du$}
   \put(0,-5.3){$ $}  
   \put(3.8,-.79){\ft $\big(\, Q dh + du\, ,\, dh\, \big)$}
   \multiput(1,-1.8)(.1,0){21}{\line(1,2){.54}}
\end{picture}} 

Der Reihe nach. Mit \eq{7.6}, \eq{8.1} und $\lk \Sp(A) 
\rk^* =\Sp (A^\dagger)$ haben wir zun\"achst
\bea{8.3}
  \e^a &=& 4 \, \Sp \( T^a \lk \d U \rho + U \d\rho \rk 
  \rho^{-1} U^\dagger \) 
  \; = \; 4\,\Sp \( \ov{T}^a U^\dagger \d U \)
  \; + \; 4\,\Sp \( \ov{T}^a \d \rho \,\rho^{-1} \) 
     \qquad \nonu \\
  \e^{a\, *} &=& {} - 4\,\Sp \( \ov{T}^a U^\dagger \d U \)
  \; + \; 4\,\Sp \( \ov{T}^a \rho^{-1} \d\rho \) \qquad ,
  \qquad \ov{T}^a \gll U^\dagger T^a U \quad ,
\eea 
wobei $\d U^\dagger U = - U^\dagger \d U$ zu Ehren kam.
Man sieht bereits, da\ss\ sich $U$--ische Terme im
Realteil kompensieren werden$\,$:
\bea{8.4}
  \e_1^a = {\e^a + \e^{a\,*} \0 2} 
  &=& 2 \,\Sp \( \ov{T}^a \rho^{-1} \lk \rho \,\d\rho 
   + \d\rho \,\rho \rk  \rho^{-1} \) \nonu \\
  &=& 2\, \Sp \( \ov{T}^a H^{-1/2} \d H H^{-1/2} \) 
     \;\; \glr \;\; dh^a \quad .
\eea 
Aber der Imagin\"arteil beh\"alt zwei Terme$\,$:
\bea{8.5}
  \e_2^a = {\e^a - \e^{a\,*} \0 2 \, i } &=& - 2\, i 
  \,\Sp \( \ov{T}^a \lk \d\rho \,\rho^{-1} 
   - \rho^{-1} \,\d\rho \rk \) \; + \; du^a \nonu \\
  \hbox{mit} \qquad  du^a &\gll& - \; 4\, i \, 
  \Sp \(\ov{T}^a U^\dagger \,\d U \)  \; = \; - \; 4\, 
  i \, \Sp \( T^a \d U\, U^\dagger \) \quad . \quad
\eea 
Unter \eq{5.8} war klar geworden, da\ss\ $\rho$ durch
$n$ reelle Parameter festliegt. Jede 
einfach--infinitesimale rein $\rho$--ische Bildung \ --- 
\  so auch der entsprechende Term in \eq{8.5} \ --- \ 
kann also aus $dh^a$ linearkombiniert werden$\,$:
\be{8.6}  - 2\, i \,
  \Sp \( \ov{T}^a \lk \d\rho \,\rho^{-1} - \rho^{-1} 
  \,\d\rho \rk \) \;\glr\; Q^ {ab} dh^b \quad .
\ee 
Damit haben wir $\e_2^a = Q^{ab} dh^b + du^a$ und sind 
schon bei \eq{8.2} angekommen \ --- \ es sei denn,
jemand will zu \eq{8.6} die absolute Sicherheit erlangen.
Wir kommen darauf unter \eq{8.13} zur\"uck. Man darf 
einwenden, da\ss\ sich ja wegen $\ov{T}^a \gll 
U^\dagger T^a U$ in \eq{8.4}, \eq{8.5} und somit in
$Q$ noch $U$--Abh\"anigkeit verberge. Richtig. Aber in
\eq{8.2} f\"allt die gesamte Matrix $Q$ heraus$\,$!

Inwiefern die unit\"are Drehung von $\vc T$ auch in 
$d\mu (\cl H )$ irrelevant ist, das sehen wir uns als 
n\"achstes an. Die Matrizen $H$ sind hermitesch und 
haben bei infinitesimaler \"Anderung hermitesch
zu bleiben$\,$:
\be{8.7}
  H + \d H = H^{1/2} \( 1 + \eta^a \ov{T}^a \) H^{1/2} 
  \quad , \quad \eta^a \;\; {\rm reell} \quad .
\ee 
Wegen $\Sp(T^a)=0$ bleibt linear in $\eta$ auch die
Determinante eins. Die Aufl\"osung nach $\eta^a$ 
enth\"ullt, da\ss\
\be{8.8}
  \eta^a = 2\, \Sp \( \ov{T}^a H^{-1/2} \d H 
   H^{-1/2} \) \; = \; dh^a \quad 
\ee 
ist, vgl.~\eq{8.4}. Im Raum $\vc r$--abh\"angiger $H$'s
ist somit 
\be{8.9}
    d \mu (\cl H ) = \prod_{\vcsm r} \prod_a d h^a \quad ,
    \quad d \mu ( \cl M ) = \prod_{\vcsm r} 
    \prod_a d u^a \,\cdot\, d \mu ( \cl H )  \quad ,
\ee 
wobei wir in jede eckige Klammer in \eq{8.2} auch noch
das Produkt \"uber Raumpunkte aufgenommen haben. Die 
Metrik im Raum $\cl H$ ist \"ubrigens
\be{8.10}
  ds^2_{\cl H} = \int\! d^2 r \; \eta^a \eta^a 
  = \int\! d^2 r \; 2 \,\Sp \(\eta^a \ov{T}^a 
    \ov{T}^b \eta^b \) \, = \, 2 \int\! d^2 r \; 
    \Sp \( H^{-1} \d H H^{-1} \d H \) \quad 
\ee 
(wieder) in \"Ubereinstimmung mit \ek{2.17}.

Nun. Das ist ja alles schlau gemacht, aber via \eq{8.8}
steht doch noch immer $U$ im $dh^a\;$! Ja. Aber eine
$U^\dagger \vc T U$--Transformation enspricht einer 
normalen reellen Drehmatrix ${\cl D}$, angewandt auf 
$\vc \eta$, und die Determinante von ${\cl D}$ ist eins. 
\ --- \ \anfu Will sehen\anfo:
\be{8.11}
  \eta^a \glr {\cl D}^{ab} \,\eta^b_{\bf un} \;\; , \;\; 
  \eta^a_{\bf un} \eta^a_{\bf un} \, = \,\eta^a \eta^a
   = \eta^b_{\bf un} \, ({\cl D}^T)^{ba} \,
  {\cl D}^{ac} \, \eta^c_{\bf un} \quad
  \Rightarrow \quad {\cl D}^T {\cl D} = 1 \quad .
\ee 
$\eta^a_{\bf un}$ ist hierbei \eq{8.8} mit 
{\bf un}verdrehten $T^a$. Und bei $\,\eta^a_{\bf un} 
\eta^a_{\bf un} = \eta^a \eta^a\,$ haben wir auf die 
Glei\-chung \eq{8.10} geblickt, welche auch unverdreht 
gilt. Man schreibe zuerst alle $u$--Integrale nach links 
und begradige dann den $\cl H$--Raum bis er von $U$ nichts 
mehr wei\ss . Jetzt k\"onnen die $u$--Integrale wieder 
nach rechts$\,$:
\be{8.12}
  \int d \mu (\cl M ) = \int d\mu (\cl H ) \;\,
  \prod_{\vcsm r} \prod_a d u^a \;\; \glr \; \int 
   d \mu (\cl H ) \cdot d \mu ( \cl G_* ) \quad .
\ee 
Die $du^a$ h\"atten wir \"ubrigens vorweg als {\sl Haar 
measure} in $\cl G_*$ per $U+ \d U = ( 1 + {i\02} du^a 
T^a ) U$ einf\"uhren k\"onnen. Aber nun ist sie doch
schon weg, die gesamte Eichfreiheit. 

Wir ziehen Bilanz aus drei Abschnitten und kombinieren
die Gleichungen \eq{6.1} ($\cl C$ ist der von Umeichungen
befreite Raum $\cl A$), \eq{7.16} ($d\mu (\cl A ) \to 
d\mu (\cl M )$) und \eq{8.12}$\,$:
\be{8.13}
  \fbox{\qquad $\dis
  d \mu (\cl C )  \;\; = \;\; { d\mu (\cl A ) \0 
  d \mu ( \cl G_* ) } \;\; = \;\; { d \mu (\cl M ) \, 
  \det \( {\bf D}^\dagger  {\bf D} \) \0 
  d \mu ( \cl G_* )} \;\; = \;\; d\mu ( \cl H ) \;
  \det \( {\bf D}^\dagger  {\bf D} \)  \;\;\quad . $ 
  \quad}
\ee 
\eq{8.13} ist \ek{2.19}$\,$: {\sl the problem is thus
reduced to the calculation of the determinant of the
two--dimensional operator $\;{\bf D}^\dagger {\bf D}\,$.}

Es gab da noch die unter \eq{8.6} angesprochene 
Sicherheitsfrage um die Existenz der Matrix $Q^{ab}$.
Um sie explizit anzugeben, machen wir von der
$\rho$--Darstellung \eq{5.8} Gebrauch$\,$:
\bea{8.14}
 \rho \; = \; e^{\vcsm \o \vcsm T} \; &,& \;
  \d\rho = d\o^a \,\6_{\o^a} \, e^{\vcsm \o \vcsm T} 
    = d\o^a \int_0^1\! ds \; e^{s\,\vcsm \o \vcsm T} 
     \big[ \6_{\o^a} \;\vc \o \vc T^a \big]
      e^{(1-s)\,\vcsm \o \vcsm T} 
      \quad \nonu \\
  \d\rho\, \rho^{-1} 
  &=& d\o^a \int_0^1 \! ds\; \tau^a(s) \qquad , \qquad  
    \tau^a(s) \;\gll\;  e^{s\,\vcsm \o \vcsm T} 
    \, T^a \, e^{-s\,\vcsm \o \vcsm T} \nonu \\
  \rho^{-1} \d\rho &=& \big[ \d\rho\, \rho^{-1} 
    \big]^\dagger \; = \; d\o^a \int_0^1 \! ds \;
    \tau^{a\, \dagger} ( s ) \quad . \quad
\eea 
Mit diesen Details ($\ov{T}^a$ ist jetzt einfach $T^a$
genannt) werden \eq{8.4} und \eq{8.6} zu
\be{8.15}
  \matrix{ \,\; dh^a = 2 \, S^{ab}\, d\o^b \cr
   Q^{ab} dh^b = - 2 \, R^{ab}\, d\o^b\cr} \qquad 
  \hbox{mit} \qquad 
 \matrix{S^{ab} \cr R^{ab} \cr } \Bigg\} \;\gll\; 
 \int_0^1 \! ds\; \Sp \( T^a  \lb
   \matrix{ [\tau^b + \tau^{b\,\dagger} ] \cr
   i \, [ \tau^b - \tau^{b\,\dagger} ] \cr} \rb\) \quad ,
 \ee 
und $d\o^b$ l\"a\ss t sich eliminieren$\,$:
\be{8.16}
  2 \, d \vc \o = S^{-1}\, d\vc h \quad : \qquad 
  Q \; = \; - \, R \; S^{-1} \quad . \quad
\ee 
Es ist recht vergn\"uglich (aber hier allzu unwichtig),
weitere Details zu ergr\"unden. $S$, $R$ sind reell.
$S$ ist symmetrisch, $R$ antisymmetrisch. $R(\vc \o =0)$
verschwindet. Da $R$ \anfu nur von $\vc \o$ wei\ss 
\anfo$\!$, mu\ss\ $R= \beta \,\vc \o \times$ gelten
mit $\b (\vc \o =0 ) = - \2\,$. Schlie\ss lich kommt
man bei
\be{8.17}
   Q  = - \b \, \rho^{-1} \, \vc \o \times
   \;\;\;\qquad \Big[ \, N=2 : \; \b = {-2 \0 \o^2}\,
   \sh^2 \( {\o \0 2} \)  \; , \;
   Q={1\0 \o}\,\th\( {\o \0 2} \) \vc \o \times \,\Big]
   \quad
\ee 
an. Alles fein, Sicherheitsstandard erreicht$\,$: 
$\rho$ legt $\vc \o$ fest, beide gemeinsam $Q$.


\sec{Funktionale Dgln f\"ur \boldmath$S(H)$}

Wollen wir dem KKN--Leitfaden \cite{kkn} folgen, so 
ist jetzt die eigenwillige Beziehung \ek{2.20} 
\be{9.1}
  e^\G \, = \,\det \( {\bf D}^\dagger {\bf D} \) \,
  = \,\s^n \, e^{2N\, S}  \qquad \hbox{mit} \quad 
  \s \; = \; {\det^\prime \( - \ov{\6} 
          \6 \) \0 \int \! d^2r } \quad
\ee 
an der Reihe. Hieran interessiert uns aber (vorerst) nur
der folgende Umstand. Der Vorfaktor $\s$ enth\"alt
(d\"urfte enthalten\footnote{Der Nenner im Faktor $\s$
    ist noch nicht recht verstanden. Entweder verweist er 
    nur auf eine spezielle $\backslash$prime--Festlegung, 
    oder \eq{9.1} f\"uhrt in h\"ohere Sp\"aren 
    \cite{bn,gaw}. Die Determinante eines positiv 
    definiten Operators ist das Produkt seiner 
    Eigenwerte$\,$: $\det(P) =  \exp\(\Sp[\ln(P)]\)
    = \exp\(\sum \ln (\l )\) = \prod \l\,$. Speziell 
    $ - \ov{\6} \6 = - \D/4$ hat $\l = \vc k^2/4$ und 
    $\vc k =0$ ist auszuschlie\ss en (bei $A=0$ haben 
    r\"aumlich konstante $\e^a$ bereits zu \eq{7.10} 
    keinen Beitrag). Es ist auch richtig, da\ss\ 
    $\det^\prime$ eine gro\ss --$k$--Regularisierung
    braucht. Die Potenz $n$ ist ebenfalls klar \ -- \ 
    nur die Fl\"ache im Nenner von $\s$ nicht.}) 
alles, was \"ubrig bleibt, wenn man $A$ Null werden
l\"a\ss t. Nichttriviale $A$--Abh\"angigkeit verbleibt
in $S$ und wir gewinnen eine Nebenbedingung an das durch 
\eq{9.1} definierte Funktional $S$, n\"amlich
\be{9.2}
   \lim_{A\to 0} S \; = \;\lim_{H \to {\rm const}} 
   S\lk H \rk \; = \; 0 \quad .
\ee 
Da\ss\ $S$ wegen Eichinvarianz nur von den physikalischen
Freiheitsgraden $H$ abh\"angen d\"urfte, ist plausibel
und wird uns noch besch\"aftigen. $A=-(\6 M)M^{-1}$ 
verschwindet bereits bei r\"aumlich konstanter Matrix $M$.
Dann ist auch $H = M^\dagger M$ konstant. Das wiederum
wird f\"ur $S=0$ ausreichen.

Was jetzt anhebt, um das Funktional $S$ zu ergr\"unden,
folgt dem Motto \anfu ableiten, regularisieren, wieder 
aufleiten\anfo und scheint bei Anomalie--Rechnungen
\"ublich zu sein \cite{powi} (Anomalie: 
Noether--Conti--Verletzung bei Quantisierung). 
Unter Variation nach $A$--Feldern ist der $\s$--Faktor
in \eq{9.1} irrelevant
\be{9.3}
  \G = \ln \lk \det \( {\bf D}^\dagger {\bf D} \) \rk
     = n \ln (\s ) + 2NS \quad \Rightarrow \quad
    \d\,\G = 2N \,\d\,S \quad .
\ee 
     \nz[-.1cm] \hspace*{2cm} 
{\bf\boldmath 9.1 \quad $\d \,$ ln ( det ) }
     \nz[.1cm]
Wer ein rechter Fu\ss g\"anger ist, der blickt kurz 
auf \eq{7.13}, und l\"auft dann mal einfach los$\,$: 
\bea{9.4}
  \G &=& \ln \lk \det \( \Diamond \) \rk 
     = \fhsp \lk \ln ( \Diamond ) \rk \quad , \quad
       {\bf D}^\dagger {\bf D}\,\glr\,\Diamond \;\; , \;\;
       1-\Diamond \,\glr \,\mho  \nonu \\
   &=& \fhsp \( - \sum_{n=1}^\infty \,{1\0n} \;\mho^{\; n}\) 
   \; = \; - \sum_{n=1}^\infty \,{1\0n} \; \( \;\;\;
   \unitlength 1cm \parbox[t]{3cm}{\begin{picture}(2,1.1)
    \put(1,.6){$\mho_{12}$}     \put(2,.3){$\mho_{23}$}
    \put(-.1,.3){$\mho_{n1}$}   \put(2,-.3){$\mho_{34}$}
    \put(-.1,-.3){$\mho_{n-1\, n}$}
    \put(1,-.08){\vbox{\hbox{\tiny $\;\;\; n$} 
       \vskip -.36cm \hbox{\tiny St\"uck}}}
    \put(.7,-.8){$\bullet$} \put(1.2,-.92){$\bullet$}
    \put(1.7,-.8){$\bullet$} \end{picture}}\;\) \quad.
\eea 
Spur in boldface verweise auf $r r^\prime$--indizierte 
Riesenmatrizen, und das Dach auf deren $ab$--Indizes. 
So steht der Index 1 z.B. f\"ur $b, \vc r^\prime$, 2 f\"ur 
$c, \vc r^{\prime\prime}$ und so fort. Eine $\d$--Variation 
greift in jeden der $n$ gleichberechtigten Eimer $\mho\;$:
\bea{9.5}
  \d\,\G &=& - \sum_{n=1}^\infty \lk \d\,\mho_{12} \rk 
          \mho_{23} \ldots \mho_{n1} \; = \;
  - \lk \d\,\mho_{12} \rk \( {1\0 1 - \mho} \)_{21} 
  \; = \; \lk \d\,\Diamond \rki_{12} \( {1\0 \Diamond} 
  \)_{21} \nonu \\
  &=& \fhsp \( \lk \d ( {\bf D}^\dagger {\bf D} ) \rk
      {\bf D}^{-1} {\bf D}^{\dagger -1} \)
  \; = \; \fhsp \( {\bf D}^{-1} {\bf D}^{\dagger -1}
          \d ( {\bf D}^\dagger {\bf D} ) \) \quad .
\eea 
Erfa\ss t die Variation nur ein $D$ (z.B ${\bf D}$ 
selbst), dann f\"allt das jeweils andere (n\"amlich 
${\bf D}^\dagger\,$) in \eq{9.5} heraus. Wenn wir also 
speziell nach $A^a (\vcsm r )$ ableiten, dann entsteht
\bea{9.6}
  \d_{A^a (\vcsm r )} \G 
  &=& \fhsp \( {\bf D}^{-1} \d\, {\bf D} \) 
    \; = \; (D^{-1} )_{12} (\d\, D)_{21}
    \; = \; \int^\prime \int^{\prime\prime} 
    (D^{-1})^{bc}_{r^\prime r^{\prime\prime}} 
    \d_{A^a (\vcsm r )} D^{cb}_{r^{\prime\prime} 
    r^\prime}  \nonu \\
  &=& \int^\prime \int^{\prime\prime} 
    (D^{-1} )^{bc}_{r^\prime r^{\prime\prime}} \;
    \d (\vc r - \vc r ^{\prime\prime} ) \; f^{abc} \; 
    \d (\vc r ^{\prime\prime} - \vc r ^\prime ) \nonu \\
 \d_{A^a (\vcsm r )} \G
 &=& f^{abc} \, (D^{* -1} )^{bc}_{r^\prime \, r} 
    \;\hbox{\Large \boldmath$|$}_{\vcsm r^\prime \to 
    \vcsm r} \quad . \quad
\eea 
In der zweiten Zeile stehen zwei Deltafunktionen. Die erste 
ergab sich beim funktionalen Ableiten nach dem $A$--Feld 
in $D^{cb}_{r^{\prime\prime} r^\prime} = \d^{cb} 
\6_{r^{\prime\prime} r^\prime} + A^\bullet (\vc r^{\prime
\prime}) f^{c \bullet b} \d_{r^{\prime\prime} r^\prime}$ 
($\equiv$\eq{7.13}). Sie ist harmlos$\,$: Integration 
\"uber $\vc r^{\prime\prime}$ erlaubt. Die zweite ist jene, 
welche bereits vorher rechts in $D^{cb}_{r^{\prime\prime} 
r^\prime}$ steht. Sie m\"ochte die r\"aumlichen Indizes 
gleich werden lassen. In der dritten Zeile z\"ogern wir
(KKN folgend) diese Koinzidenz, welche Regularisierung 
erfordert, noch ein Weilchen hinaus und be\-trach\-ten die 
Indizes an $D^{-1}$ vorerst als verschiedene. Da $\G$ 
reell, konnte die zweite Dgl in der dritten Zeile als c.c. 
der ersten notiert werden. Diese c.c.--Version ist 
\ek{2.22}.
\pagebreak[2]
     \nz[.5cm]  \hspace*{2cm} 
{\bf 9.2 \quad Das Inverse von {\bf D} }
     \nz[.1cm] 
\nopagebreak[3]
Das in \eq{9.6} ben\"otigte Inverse {\boldmath$D^{-1}$}
ist \ \ d i e \ \ e i n e \ L\"osung der $n^2$ Gleichungen
\be{9.7}
  D^{ac}_{r r^{\prime\prime}} \; 
  (\,\hbox{\bf ?}\, )^{cb}_{r^{\prime\prime} r^\prime}
  \; = \; \Big(\d^{ac} \6 + A^{ac}(\vc r )\Big) \;
  (\,\hbox{\bf ?}\, )^{cb}_{r r^\prime} 
  \;  = \; \d^{ab} \,\d (\vc r - \vc r^\prime ) 
  \quad \hbox{mit} \;\; 
  A^{ac} \gll A^\bullet f^{a \bullet c} \quad . \quad
\ee 
Wir erfinden gern selber. Das Fragezeichen sollte wegen
\eq{4.3}, n\"amlich $\6 G_{r r^\prime} = \d (\vc r - 
\vc r^\prime )$, etwas mit $G$ zu tun haben. Setzen wir 
$\d^{cb} G_{r r^\prime}$ f\"ur ({\bf ?}) ein, entsteht 
rechts zwar $\d \d$ aber auch ein Zusatzterm $A G$. Um
zu dessen Kompensation einen weiteren Term zu erhalten, 
versuchen wir es mit $M^{cb} (\vc r ) G_{r r^\prime}\,$,
und das gibt rechts $(\6 M^{ab} ) G + M^{ab} \d \d
+ A^{ac} M^{cb} G\,$. Aha$\,$! Ein weiteres $(M^{-1})^{db}
(\vc r^\prime )$ von rechts k\"onnte die Dinge ins Lot
bringen$\,$:
\be{9.8}
  \Big(\d^{ac} \6 + A^{ac}(\vc r )\Big) \;
  M^{cd}_r G_{r r^\prime} (M^{-1})^{db}_{r^\prime}
  = \lk (\6 M)_r^{ad} + A^{ac}_r M^{cd}_r \rk
    G_{r r^\prime} (M^{-1})^{db}_{r^\prime} 
    + \d^{ab} \d_{r r^\prime} \quad  
\ee 
Wenn wir die \anfu adjungierte\anfo Matrix $\hat{M}$ mit 
Elementen $M^{ab}$ finden, welche die eckige Klammer 
verschwinden l\"a\ss t, d.h. wenn wir $\hat{A} \hat{M} 
= - \6 \hat{M}$ l\"osen k\"onnen, dann ist auch das
Invertierungsproblem gel\"ost$\,$:
\be{9.9}
  (D^{-1})^{ab}_{r r^\prime} =  M^{ac}_r G_{r r^\prime} 
  (M^{-1})^{cb}_{r^\prime} \quad : \quad {\bf D}^{-1} 
   = \hat{M} \, G \, \hat{M}^{-1} \quad .
\ee 
\"Uber jene Paare von r\"aumlichen Indizes in \eq{9.8}, 
\eq{9.8} ist \"ubrigens {\sl nicht} zu integrieren (das 
letzte zu summierende Paar gab es ganz links in \eq{9.7}).

$\hat{A}$ habe die rechts in \eq{9.7} angegebenen 
Elemente. Bei $\hat{A} = - (\6 \hat{M})\hat{M}^{-1}$ 
handelt es sich offenbar um eine adjungierte Version der 
vertrauten Beziehung $A= -(\6 M) M^{-1}\,$.  
     \nz[.5cm] \hspace*{2cm} 
{\bf 9.3 \quad Die adjungierte Matrix \boldmath$\hat{M}$ }
     \nz[.1cm] 
Der Ausgangspunkt $A= - i T^a A^a = - (\6 M) M^{-1}$ ist in
fundamentaler Darstellung angesiedelt. Wir nehmen ihn im
Kommutator $[ T^b\, , \,  \ldots ]$ und erhalten
\be{9.10}
  A^{bc} T^c = - T^b (\6 M ) M^{-1} + (\6 M) M^{-1} T^b
             = - T^b (\6 M ) M^{-1} - M (\6 M^{-1} ) T^b
\ee 
Weil $\hat{A} \hat{M} = - \6 \hat{M}$ rechts/links 
das gleiche unbekannte Objekt $\hat{M}$ braucht, 
multiplizieren wir von links mit $M^{-1}$ und von rechts mit 
$M\;$:
\be{9.11}
 A^{bc} M^{-1} T^c M \, = \; - M^{-1} T^b \6 M 
 - (\6 M^{-1} ) T^b M \; = \; - \6 \; M^{-1} T^b M \quad .
\ee 
Fast. Fundamentale Elemente lassen sich per Spurbildung 
beseitigen. Aber die Nor\-mie\-rung von $\hat{M}$ wird 
durch \eq{9.11} nicht festgelegt. Dazu lassen wir uns 
von \eq{4.4}, \eq{4.5} $M\to 1$ bei $A\to0$ erz\"ahlen 
und machen es per Forderung $M^{ab} \to \d^{ab}$ 
adjungiert ebenso. Fazit$\,$:
\be{9.12}
 \hat{M} \, : \;\,  M^{ab} = 2 \, \Sp 
 \Big( T^a M T^b M^{-1} \Big) \qquad \hbox{erf\"ullt} 
 \quad \hat{A} \hat{M} = - \6 \hat{M} \quad . \qquad
\ee 
Wegen $(\hat{M}^\dagger )^{ab} = M^{ba\; *} = 
2\, \Sp ( M^{\dagger -1} T^a M^\dagger T^b )$
folgt nun
\be{9.13}   \hspace*{.5cm}
 \hat{M}^\dagger \, : \;\,  (M^\dagger )^{ab} = 2\,\Sp 
 \Big( T^a M^\dagger T^b M^{\dagger -1} \Big) \quad 
 \hbox{erf\"ullt} \quad \hat{M}^\dagger \hat{A}^\dagger  
  = - \ov{\6} \hat{M}^\dagger \quad . \quad
\ee 
Besteht $\hat A$ aus $A^{ab} = A^\bullet f^{a\bullet b}$, 
so hat $\hat{A}^\dagger$ die Elemente $(A^\dagger)^{ab} 
=  A^{\bullet *} f^{b\bullet a} = - A^{\bullet *} 
f^{a\bullet b}\,$.
\pagebreak[2]
     \nz[.5cm] \hspace*{2cm} 
{\bf 9.4 \quad Sprung in die regularisierte Version}
     \nz[.1cm] 
\nopagebreak[3]
Da\ss\ wir uns nicht sch\"amen. Aber die KKN--Leitlinie
erlaubt es hier, das Resultat der Regularisierungs--M\"uhen
vorwegzunehmen. Es kann nur die Greensche Funktion 
$G_{r r^\prime}$ sein, welche in \eq{9.9} unter 
$\vc r^\prime \to \vc r$ leidet. Sie wird durch ihre
regularisierte Version $\cl G_{r r^\prime}$ zu ersetzen
sein. $\vc r^\prime \to \vc r$ ist dann m\"oglich und
f\"uhrt auf ({\bf Zitat} aus \S~11, Gleichung \eq{11.28}$\,$)
\be{9.14}
  \lk D^{-1}_{r^\prime r} \rki_{\vtop{\hbox{\ft \ reg}
  \vskip -.24cm \hbox{\ft $\vcsm r^\prime\!\to\!\vcsm r $}}}
  \, = \, - \, { \hat{A}^\dagger - ( \ov{\6} \hat{M} ) \, 
     \hat{M}^{-1}  \0 \pi } \;\;\; , \;\;\; 
  \lk D^{* -1}_{r^\prime r} \rki_{\vtop{\hbox{\ft \ reg}
  \vskip -.24cm \hbox{\ft $\vcsm r^\prime\!\to\!\vcsm r $}}}
  \, = \; { \hat{A} - \hat{M}^{\dagger -1} \, 
    \6 \, \hat{M}^\dagger \0 \pi } \;\;\; . \quad
\ee 
Hiermit werden nun die Gleichungen \eq{9.6} zu 
anst\"andigen (lokalen, noch adjungiert 
ge\-schrie\-be\-nen) funktionalen Dgln$\,$:
\be{9.15}
  \d^a \G = {- 1\0\pi}\, f^{abc} 
  \( \hat{A}^\dagger - ( \ov{\6} \hat{M} ) \, 
   \, \hat{M}^{-1} \)^{bc} \quad , \quad 
  \d^{a*} \G = {1\0\pi}\, f^{abc} 
   \( \hat{A} - \hat{M}^{\dagger -1} \, 
    \6 \, \hat{M}^\dagger \)^{bc} \quad , \quad
\ee 
worin nat\"urlich $\d^a$ f\"ur $\d_{A^{a} (\vcsm r )}$
steht und $\d^{a*}$ f\"ur $\d_{A^{a*} (\vcsm r )}\,$.
Die rechte Dgl ist das c.c. der linken. Es geht jetzt 
nur noch um R\"uckkehr zu den liebenswerten 
$N\times N$--Matrizen $M\,$. Aus den adjungierten
Spuren in \eq{9.15} sollen fundamentale werden. 
\pagebreak[2]
     \nz[.5cm]  \hspace*{2cm} 
{\bf 9.5 \quad Spur mal Spur \ -- \ zur\"uck zum 
               Fundamentalen}
     \nz[.1cm] 
\nopagebreak[3]
Ein neckisches technisches Detail kommt bereits ins Spiel, 
wenn das Inverse von $\hat M$ zu Papier soll. Es hat sich 
aus $(M^{-1})^{ac} M^{cb} = \d^{ab}$ zu ergeben. Wir 
\ r a t e n \ das Resultat,
\be{9.16}
   ( M^{-1} )^{ac} = 2 \, \Sp \( T^a M^{-1} T^c M \)
   \; = \;  M^{ca} \quad ,
\ee 
und pr\"ufen nach. Zwei fundamentale Spuren stehen da und
sollen zu einer einzigen verkettet werden$\,$: {\sl 
concatenation} w\"urde Miss Maple sagen$\,$:
\bea{9.17}
  ( M^{-1} )^{ac} M^{cb} &=& 4 \, \Sp \( T^a 
   M^{-1} T^c M \) \; \( T^c M T^b M^{-1} \) \nonu \\
  &=&  4 \, \Sp \( \lk M T^a M^{-1}\rk T^c \) \; \( T^c 
      \lb M T^b M^{-1}\rb \)  \nonu \\
  &=& 2 \lk \qquad \rki_{\ell\, m} \; 2 \, T^c_{m\, \ell}   
      \, T^c_{p\, q} \; \lb \qquad \rb_{q\, p} \nonu \\[-.4cm]
  & & \hspace*{2cm} \unt{1.8}{.4}{.3}{\d_{m\, q} \; 
      \d_{\ell \, p} \, - \, {1\0N}\; \d_{m\, \ell}\; 
      \d_{p\, q} }   \nonu \\
  &=& 2 \,\Sp\( \lk \qquad \rk \; \lb \qquad \rb_{q\, p} \)
      \; = \; \d^{ab} \quad \hbox{q.e.d.} \quad
\eea 
Sieh an, das unterklammerte Detail aus SU(N)--Weisheiten
leistet also {\sl concatenations}. Der $1/N$--Term gab
keinen Beitrag wegen $\Sp ( \lk \quad \rk ) = \Sp ( \lb 
\quad \rb ) = 0\,$. 

Unverz\"uglich wenden wir uns mit Blick auf \eq{9.15}
der n\"achsten Verkettung zu$\,$:
\bea{9.18}
  (\ov{\6} M^{bd} ) \, (M^{-1})^{dc} 
  &=&  4\, \Sp \( T^b (\ov{\6} M)  T^d M^{-1} + T^b M 
       T^d \ov{\6} M^{-1} \)  \, \Sp \( T^d M^{-1} 
       T^c M \)   \nonu \\
  &=& 4\, \Sp \( M^{-1} \lk T^b \ov{\6} M - 
      (\ov{\6} M ) M^{-1} T^b M \rk T^d \) \,  
      \Sp \( T^d M^{-1} T^c M \)  \nonu \\
  &=& 2 \, \Sp \( \lk T^b (\ov{\6} M ) M^{-1} 
        - (\ov{\6} M ) M^{-1} T^b \rk T^c \) \nonu \\
  &=& 2\,\Sp \( \lk T^c , T^b \rk (\ov{\6} M ) M^{-1} \) 
      \; = \; - \, 2 \, i \, f^{abc} \,\Sp \( T^a 
      (\ov{\6} M) M^{-1} \) \quad , \quad \nonu \\
  (M^{\dagger -1} )^{bd}\, \6 \, (\hat{M}^\dagger )^{dc}   
  &=& \quad \ldots \quad = \; - \, 2 \, i \, f^{abc} \,\Sp 
      \( T^a M^{\dagger -1} \6 M^\dagger \)  \quad . 
\eea 
Es ist hingegen recht trivial, wie die $A$--Terme in 
\eq{9.15} zu einer Spur werden$\,$: 
\be{9.19}
   A^{bc} =  A^a f^{bac} =  - 2i f^{abc} \,\Sp 
   \( T^a A \) \;\; , \;\; (A^\dagger )^{bc}  
   = - A^{\bullet *} f^{b \bullet c} =  - 2i 
     f^{abc} \,\Sp \( T^a A^\dagger \) \quad .
\ee 
An dieser und der vorigen Gleichung gefallen zwei Dinge.
Sie enthalten nur noch eine fundamentale Spur. Und endlich 
ist das zweite $f$ zu sehen, welches in \eq{9.15} per 
$f^{abc} f^{bc\bullet}=N\d^{a\bullet}$ aus $\G$ einen 
Vorfaktor $N$ abspalten wird. Wir setzen \eq{9.19}, 
\eq{9.18} in \eq{9.15} ein und erhalten
\be{9.20}
 \d^a \G = 2N {i\0\pi} \,\Sp \( T^a \big[ A^\dagger 
     - (\ov{\6} M ) M^{-1} \big] \) \; , \;
 \d^{a*} \G = - 2N {i\0\pi} \,\Sp \( T^a \big[ A 
  - M^{\dagger -1} \6 M^\dagger \big] \) \; . \;
\ee 
Die Dgln f\"ur das in \eq{9.1} eingef\"uhrte Funktional 
$S$ sind wegen \eq{9.3} ($\d\G = 2N\d S$) einfach die 
obigen beiden ohne $2N$--Vorfaktor. Da $\G$ reell, ist
erneut die zweite Dgl das c.c. der ersten. Erst am Ende
dieses langen Abschnittes wird klar, weshalb (s. 
9.8--\"Uberschrift) wir artig diese Spiegelbilder mit 
uns herumschleppen. \eq{9.20} rechts ist \ek{2.23}.
\pagebreak[2]
     \nz[.5cm]  \hspace*{2cm} 
{\bf 9.6 \quad \boldmath$\, S\, =\, S_1 + S_2\;$: $\;$die 
                Dgln f\"ur $\,S_2$}
     \nz[.1cm] 
\nopagebreak[3]
Umformulieren, jetzt noch einmal, dann noch einmal$\,$: 
mit welchem Ziel$\,$? Der Leser mag ruhig einmal 
vorausschauen auf die erste Gleichung in \S~10. $S(H)$ ist 
das Ziel. Er mag auch bitte etwas von der M\"uhsal 
ermessen, den hier dokumentierten (idealen?) Weg zu 
finden. Es war wie bei einer Schachpartie$\,$: drei gute 
Z\"uge, auf jeden gibt es drei weitere und so fort. Die 
n\"achsten Schritte bringen zwei neue Buchstaben ($F$, 
$\O$), beide nur vor\"ubergehend. Nein, Protest 
unzul\"assig, Abk\"urzungen machen Strukturen sichtbar. 

Neben den $A$'s gibt es in \eq{9.20} Bildungen mit der 
\anfu falschen\anfo Differentiation. Sie bekommen den 
Namen $F\,$: 
\be{9.21}
   \Big(\; A = - \( \6 M \) M^{-1} \;\; : \;\Big) \qquad  
   F \,\gll\, - \( \ov{\6} M \) M^{-1} \quad , \quad  
   F^\dagger \, = \, - M^{\dagger -1} \6 M^\dagger 
   \quad . \quad
\ee 
F\"ur das folgende ist es recht gescheit, die Generatoren 
$T^a$ in \eq{9.20} als Ableitungen zu schreiben, n\"amlich 
mittels $\, T^a = i \int \d^a A\,$ und $\,T^a = - i \int 
\d^{a*} A^\dagger\,$. Damit haben wir
\be{9.22}
  \d^a S  =  - {1\0\pi} \int \Sp \Big( \lk A^\dagger 
             + F \rk \d^a A \Big)  \quad , \quad
  \d^{a*} S  = - {1\0\pi} \int \Sp \Big( \lk A 
  + F^\dagger \rk \d^{a*} A^\dagger \Big) \quad . \quad
\ee 
Das sind $2*n$ funktionale Dgln f\"ur $S$. $\int$ steht 
f\"ur $\int\! d^2r$ \"uber die ganze 2D Ebene. $\vc r$ ist 
jetzt Integrationsvariable. Also lesen wir $\d^a$ als 
$\d_{A^a (\vcsm r_0)}\,$. In \eq{9.22} darf $\d^a A$ 
durch $\d^a [ A + F^\dagger ]$ ersetzt werden, und 
$\d^{a*} A^\dagger$ durch $\d^{a*} [ A^\dagger + F ]$,
weil
$$  A \; , \; F \quad \hbox{nur von} \quad A^a \qquad 
   \;\; \hbox{und} \qquad \;\; A^\dagger \; , \; 
   F^\dagger \quad \hbox{nur von} \quad A^{a*} \quad $$
abh\"angen. Wer hieran zweifelt, blicke auf \eq{4.5}
zur\"uck, wonach nicht nur $M$ sondern auch $\ov{\6} M$
eine Entwicklung (nur) nach $A^a$--Feldern hat. Die 
n\"achste Definition, n\"amlich $\,\O \gll A+F^\dagger\,$,
macht \eq{9.22} so kurz,
\be{9.23}  \hspace*{-.2cm}
  \d^a (2\pi S) = \int \Sp \( - 2\,\O^\dagger \d^a \O \)
  \quad , \quad  \d^{a *} (2\pi S) = \int \Sp \( - 2\,\O 
  \d^{a*} \O^\dagger \) \quad , \quad
\ee 
da\ss\ man bereits zu raten beginnen m\"ochte. Aber ein 
$\O \O^\dagger$ unter $\int\Sp$ taugt nur f\"ur eine 
Abspaltung$\,$:
\be{9.24} 
  2\pi S \; = \; 2\pi S_1 + 2\pi S_2 \qquad , \qquad 
       2\pi S_1 \,\gll\, \int \Sp \( - \O \O^\dagger 
  \) \quad . \quad
\ee 
Ausblick$\,$: $S_1$ wird der \anfu harmlose\anfo erste 
Term in der gesuchten Wirkung bleiben und $S_2$ zur
WZW--Wirkung werden. Um die funktionalen Dgln separat 
f\"ur den Anteil $\,S_2 = S - S_1\,$ zu ergr\"unden, sind 
lediglich die Dgln $\d (2\pi S_1) = \int \Sp \( - 
\O^\dagger \d \O - \O  \d \O^\dagger \)$ von \eq{9.23} zu 
subtrahieren$\,$:
\bea{9.25}
 \d^a (2\pi\, S_2) = \int \Sp \Big( \O \d^a \O^\dagger
  - \O^\dagger  \d^a \O  \Big) \quad , \quad
 \d^{a*} (2\pi\, S_2) = \int \Sp \Big( \O^\dagger 
 \d^{a*} \O - \O \d^{a*} \O^\dagger  \Big) \quad . \quad
\eea 
\pagebreak[2]
     \nz[.1cm]  \hspace*{2cm} 
{\bf 9.7 \quad Nur noch Variable \boldmath$H$}
     \nz[.1cm] 
\nopagebreak[3]
Es folgt eine landschaftlich sch\"one Strecke. Wenn $\G$ 
und $S$ als eichinvariante Bildungen nur von $H = 
M^\dagger M$ abh\"angen (sollten), dann erwarten wir, 
da\ss\ die Dgln neben $H$ nur $\d^a H$ bzw. $\d^{a*} H$ 
enthalten$\,$:
\bea{9.26}
  \O &=& A + F^\dagger = - \lk (\6 M) M^{-1} 
       + M^{\dagger -1} (\6 M^\dagger ) \rk \nonu \\
     &=& - M^{\dagger -1} \lk M^\dagger \6 M  + 
         ( \6 M^\dagger ) M  \rk M^{-1} = - M^{\dagger -1} 
         (\6 H) M^{-1} \quad , \quad \nonu \\
  \O^\dagger &=&  - M^{\dagger -1} (\ov{\6} H) M^{-1} 
  \; = \; M (\ov{\6} H^{-1} ) M^\dagger \quad . \quad
\eea 
Unverz\"uglich wird der Anteil $S_1$, \eq{9.24}, zu einem 
rein $H$--ischen Objekt$\,$:
\be{9.27}
  2\pi\, S_1 = \int \Sp \Big( (\6 H) \, \ov{\6} 
               H^{-1} \Big) \quad . \quad
\ee 
F\"ur $S_2$ haben wir nur Dgln. Wie sie sich mittels 
\eq{9.26} $H$--ifizieren, sei an der linken Dgl 
\eq{9.25} vorgef\"uhrt$\,$:
\bea{9.28}
  \d^a (2\pi\, S_2) &=& \int \Sp \( M^{\dagger -1}
  (\6 H) M^{-1} \d^a \lk M^{\dagger -1} (\ov{\6} H)
  M^{-1} \rk \; - \; \hbox{dito}_{\6 \gdw \ov{\6}} \;\)
  \nonu \\
  &=& \int \Sp \( (\6 H) H^{-1} \d^a \lk (\ov{\6} H) 
      H^{-1} \rk \; - \; \hbox{dito}_{\6 \gdw 
      \ov{\6}} \;\)  \nonu \\
  &=& {i\02} \int \Sp \( H_{\prime 2} H^{-1} \d^a 
      \lk H_{\prime 1} H^{-1} \rk \; - \; 
      \hbox{dito}_{1 \gdw 2} \;\)  \nonu \\
  &=& {i\02} \int \Sp \( H^{-1} H_{\prime 2} H^{-1} 
      \d^a H_{\prime 1} - H^{-1} H_{\prime 2}
      H^{-1} H_{\prime 1} H^{-1} \d^a H \; - \; 
      \hbox{dito}_{1 \gdw 2} \;\)  \nonu \\
  &=& {i\02} \int \Sp \Big( \ph^a \lk X_1 X_2 - X_2 X_1 
      \rk \, + \6_1 (X_2 \ph^a) - \6_2 (X_1 \ph^a) \Big) 
      \quad . \quad
\eea 
Das Ziel ist erreicht$\,$: nur $H$ und $\d^a H\,$. In der 
ersten Zeile durften wir $M^{\dagger -1}$ an $\d^a$ vorbei 
schieben. Die dritte Zeile entstand per \anfu $\,\6\ov{\6} 
- \ov{\6}\6 = {1\04} ( \6_1 + i \6_2) ( \6_1 - i \6_2) 
- {1\04} ( \6_1 - i \6_2) ( \6_1 + i \6_2) = {i\02} \lk 
\6_2\,\6_1 - \6_1\,\6_2 \rk\,$\anfo. Die letzte Zeile in
\eq{9.28} wurde erreicht mit naheliegenden Abk\"urzungen 
f\"ur die zwei rein $H$--ischen Bildungen
\be{9.29}
   X_j \gll H^{-1} H_{\prime j} \quad (j=1,2)
   \quad \hbox{und} \quad
   \ph^a \gll H^{-1} \d^a H \quad 
\ee 
einerseits, und zum anderen mit der Umformung
\bean
 X_2 H^{-1} \d^a H_{\prime 1}  
 &=&  X_2 H^{-1} \6_1 \d^a H 
      \; = \; \6_1 ( X_2 \ph^a ) - ( H^{-1} 
      H_{\prime 2} H^{-1})_{\prime 1} \d^a H \nonu \\
 &=&  \6_1 (X_2 \ph^a) - H^{-1} H_{\prime 2 \prime 1} 
      \,\ph^a + \lk X_1 X_2 + X_2 X_1 \rk \ph^a 
      \quad , \qquad \nonu
\eea
wonach 1-2--symmetrische Terme unter dito entfallen.  
\pagebreak[2]
     \nz[.5cm]  \hspace*{2cm} 
{\bf 9.8 \quad Stokes \ --- \ ein gemeinsames 
         \boldmath$\d$ }
     \nz[.1cm] 
\nopagebreak[3]
Das Resultat \eq{9.28} enth\"alt die Drittkomponente
einer Rotation. Die aus analoger Rechnung folgende \anfu
Spiegelbild--Dgl\anfo f\"ugen wir jetzt hinzu$\,$:
\bea{9.30}
  \d^a S_2 
&=& {i\04\pi} \int \Sp\,\bigg( \,\ph^a [ X_1 X_2 
       - X_2 X_1 ] + \left[ \nabla \times 
        \hbox{\Large\bf (}  X_1 \ph^a 
        \hbox{\Large\bf\ , } X_2 \ph^a  
        \hbox{\Large\bf\ , } \ldots  
        \hbox{\Large\bf\ )} \right]_3 \;\bigg) \nonu \\
  \d^{a*} S_2 
&=& {i\04\pi} \int \Sp\,\bigg( \,\psi^a [ X_1 X_2 
       - X_2 X_1 ] - \left[ \nabla \times 
        \hbox{\Large\bf (}  X_1 \psi^a 
        \hbox{\Large\bf\ , } X_2 \psi^a  
        \hbox{\Large\bf\ , } \ldots  
        \hbox{\Large\bf\ )} \right]_3 \;\bigg) \qquad 
\eea 
mit $\psi^a \gll H^{-1} \d^{a*} H\,$. Ist da sonst noch
ein kleiner Unterschied$\,$?! Im je ersten Term hat sich
lediglich $\d^a$ durch $\d^{a*}$ ersetzt. Hier d\"urfen
wir also ein \anfu gemeinsames $\d$\anfo einf\"uhren$\,$:
$\d$ ist wahlweise $\d_{A^a (\vcsm r_0)}$ oder $\d_{A^{a*}
(\vcsm r_0)}\,$. Statt $\ph^a$, $\psi^a\,$ gen\"ugt ein
$\ph \gll H^{-1} \d H\,$. Aber in den Rotationstermen 
wird solcherlei durch den Vorzei\-chen\-unterschied 
unterbunden. 

Rotation ruft nach Stokes (wiewohl hier nur f\"ur ebene
Fl\"ache). Es ist eine delikate Frage, ob nach Integration 
die Randterme entfallen d\"urfen. $\ph$ ist Funktion von
zwei Orten, von $\vc r$ (wovon $H$ abh\"angt und worauf 
$\nabla\times$ wirkt) und von $\vc r_0$, die Marke an 
$\d^a =\d_{A^a(\vcsm r_0)}\,$. Wie \eq{4.4}, \eq{4.5} 
zeigt, sind die $A$--Felder, von denen $H$ (neben $\vc r$) 
funktional (auch noch) abh\"angt, s\"amtlich 
\"uberintegriert$\,$: sie sitzen bei $\vc r^\prime\,$. 
$\d^a$ zwingt ein $\vc r^\prime$ auf $\vc r_0$. Und wenn 
sich nun $\vc r$ weit von $\vc r_0$ entfernt, dann sollten 
die Greensfunktionen $G$ in \eq{4.5} f\"ur Abfall nach 
au\ss en sorgen. \\[.3cm]
\hspace*{.6cm}
\vrule depth -3pt height 3.5pt width 1.2cm 
\ \ $\vc r$ \ Rand \ 
\vrule depth -3pt height 3.5pt width 2cm 
\ \ $A(\vc r^\prime )$ \
\vrule depth -3pt height 3.5pt width .6cm 
\ \ $\d_{A^a(\vcsm r_0)}$ \ 
\vrule depth -3pt height 3.5pt width 2cm 
\ \ Rand \ $\vc r$ \
\vrule depth -3pt height 3.5pt width 1.2cm \\[.3cm]
Fallen die $A$--Felder selbst nach au\ss en ab ($H$ und 
$M\to$ const), dann helfen auch die $X_j$--Faktoren
mit. Unter solcherlei Unsicherheiten und Vorbehalten 
k\"onnen wir schreiben$\,$: 
\bea{9.31}
  \d\, S_2 &=& {i\04\pi} \int \Sp \Big( \, \ph \,
           [ X_1 X_2 - X_2 X_1 ] \,\Big) 
      \quad , \quad \ph = H^{-1} \d H  \\
  &=& {i\04\pi} \int \Sp \Big( \, H^{-1} \d H \lk H^{-1} 
  H_{\prime 1} H^{-1} H_{\prime 2} - H^{-1} H_{\prime 2}   
  H^{-1} H_{\prime 1} \rk \Big) \quad . \quad \nonu 
\eea 

Weil wir $S_2$ noch nicht kennen, ist \eq{9.31} \ d a s \
Resultat des gesamten langen \S~9$\,$. F\"ur sp\"ateren
Gebrauch, n\"amlich in \S~12.2, sollten wir hier aber 
auch noch $\d S_1$ und $\d S$ H--isch notieren. Dazu nehmen
wir \eq{9.23} zur Hand, setzen den H--isierer \eq{9.26}
ein und erhalten
\be{9.32}
  \d S_1 = {1\04\pi} \int \Big\{ \; \Sp \Big( \,\ph 
     \lk \6_1 X_1  + \6_2 X_2 \rk \Big)\,  
     + \6_1 \,\Sp \( \ph X_1 \)  + \6_2 \,\Sp \( 
     \ph X_2 \) \;\Big\} \quad . \quad
\ee 
Noch vor partieller Integration hat also $\d S_1$ bereits
ein gemeinsames $\d\,$. Statt Stokes beseitigt diesesmal 
der ebene Gau\ss\ die reinen Ableitungen$\,$:
\bea{9.33}
  \d S_1 &=& {1\0 2\pi} \int \Sp 
           \Big( \, \ph \lk \6 \ov{X} + \ov{\6} X \rk \big)
     \qquad ,\qquad X \gll H^{-1} \6 H \;\; , \;\; 
     \ov{X} \gll H^{-1} \ov{\6} H \nonu \\
  \d S_2 &=& {1\0 2\pi} \int \Sp \Big( \, \ph \lk \ov{X} X 
             - X \ov{X} \rk \Big) \nonu \\
  & & \hspace*{-2cm} 
      \vrule depth -2pt height 2.2pt width 9.3cm \nonu \\
  \d S &=& {1\0 2\pi } \int \Sp \Big( \, \ph \lk \6 \ov{X} 
         + \ov{\6} X + \ov{X} X - X \ov{X} \rk \Big) 
    \; = \; {1\0 \pi } \int \Sp \Big( \;\ph  \;\6 
         \ov{X} \,\Big) \quad , \quad
\eea 
denn 
\be{9.34} 
   \ov{\6} X + \ov{X} X = \ov{\6} \lk H^{-1} \6 H \rk 
   + \ov{X}\, X \, = \, H^{-1}  \6 \ov{\6} H \, 
    = \, \6 \ov{X} + X \ov{X} \quad . \quad 
\ee 
\eq{9.33} zeigt einmal mehr, da\ss\ der WZW--Term $S_2$ 
im gegenw\"artigen Kontext eine ganz bestimmte St\"arke 
braucht, um die Dgln zu erf\"ullen, n\"amlich genau den 
in \eq{10.1} angegebenen Vorfaktor.


\sec{Die L\"osung \boldmath$S(H)$ der Dgln}

Selber finden ist zwar sch\"oner, aber der Blick auf 
das angebliche Resultat \ek{2.21} ist jetzt allzu 
unwiderstehlich. $S$ ist die hermitesche WZW--Wirkung 
\be{10.1}  \!\!   
    \fbox{\rule[-.6cm]{0pt}{1.4cm} \  $\dis      
    S = {1\0 2\pi} \int \Sp 
           \Big( ( \6 H ) \ov{\6} H^{-1} \Big)
     + {i\0 12\pi} \int_V \epsilon^{j k \ell} 
       \Sp \Big( H^{-1} ( \6_j H ) H^{-1} ( \6_k H )
           H^{-1} ( \6_\ell H ) \Big) \; . \;  $ } 
\ee 
Der erste Term ist ein alter Bekannter, $S_1$ 
n\"amlich$\,$: \eq{9.17}. Der zweite Term \ --- \ ebenfalls 
rein $H$--isch \ --- \ ist also $S_2$ und hat die Dgln 
\eq{9.31} zu erf\"ullen. Beide Terme stehen \"ubrigens mit 
unserer Nebenbedingung \eq{9.2} in Einklang$\,$: $H\to$ 
const gibt Null (vorbehaltlich Integrale--Existenz).  

Obiges $S_2$ ist ein Volumenintegral. Wie bitte$\,$?! $H$ 
ist auf xy--Ebene definiert. Auf dieser leben wir und 
k\"onnen nicht anders. Offenbar hat irgendein B\"osewicht 
unseren $H$'s eine zus\"atzliche Abh\"angigkeit von der 
Variablen $z$ verpa\ss t. So schreibt denn auch Dimitra 
Karabali in ihrem TFT'98--Proceedings--Artikel
\cite{dika}  da\ss\     \nz
{\sl \anfu the integrand thus requires an extension of the 
matrix field $H$ into the interior of $V$, but physical 
results do not depend on how this extension is done. 
Actually for the special case of hermitian matrices, the 
second term can also be written as an integral over} [2D] 
{\sl spatial coordinates only.\anfo} \ --- \ Aha \ --- 
\ ? \ --- . 

An dieser Stelle steht eine b\"ose Falle bereit. Das 
$V$--Integral macht nur Sinn, so meint man wohl 
leichtfertig, wenn $\int_V$ unverz\"uglich mittels Gau\ss\ 
in ein Oberfl\"achen\-inte\-gral $\int_{\6V}$ verwandelt 
wird (ein skalarer Integrand l\"a\ss t sich stets als 
$\nabla\cdot \vc C$ schreiben). Auf dieser Oberfl\"ache 
sollten sodann die skalaren Werte $S_2$ liegen, welche die 
Dgln \eq{9.31} erf\"ullen. Es hat lange gedauert, ehe die 
soeben genannte Leichtfertigkeit \"uberwunden war. Man soll 
n\"amlich (zumindest geht es dann) \ e r s t \ funktional 
ableiten und \ d a n a c h \ Gau\ss\ bedienen. Hinter
dieser Anmerkung verbirgt sich die (unvermindert 
bohrende) Frage nach einem anderen Weg. Wir gehen jetzt 
da entlang, wo es nicht weiter anstrengt. 

Die hermiteschen Matrizen $H$ m\"ogen also auch 
(irgendwie) von $z$ abh\"angen. Neben $X_1$, $X_2$ gibt 
es nun auch noch ein $X_3 = H^{-1} H_{\prime 3}$. In 
der epsilon--Tensor--Sprache von \eq{10.1} stehen sechs 
Terme unter der Spur. Aber zyklische Vertauschungen unter 
derselben reduzieren auf zwei$\,$: 
\be{10.2} 
   S_2 = {i\0 4\pi} \int_V \Sp \Big( \, X_3 \lk X_1 X_2 
         - X_2 X_1 \rk \Big) \quad . \quad 
\ee 
Es gibt nur noch ein $\d$, und dessen Marke $\vc r_0$
liege irgendwo in $V$. Wir sehen nach, ob \eq{10.2} die
Dgln \eq{9.31} erf\"ullt (Erinnerung$\,$: $\ph 
= H^{-1} \d H\,$)$\,$:
\bea{10.3}
  \d\, S_2 
  &=& {i\0 4\pi} \int_V \Sp \Big(\,\lk X_1 X_2 - X_2 X_1\rk 
      \d X_3  \;\; + \;\; \hbox{zyklisch}
      \,\;\Big) \nonu \\[2pt]
  & & \hspace*{1.4cm}
      \d X_3 = \d ( H^{-1} H_{\prime 3} ) \, = \, X_3 \ph 
       - \ph X_3 + \6_3 \ph \quad , \quad \nonu \\[2pt]
  & & \hspace*{1.4cm}
      \Big[\, [ X_1 , X_2 ]\, ,\, X_3\, \Big] \; + \; 
      \hbox{zyklisch} \; = \; 0  \quad
      \hbox{(Jacobi--Identit\"at)} \nonu \\[2pt]
  &=& {i\0 4\pi} \int_V \Sp \Big(\,\lk X_1 , X_2 \rk \6_3 
      \ph \;\; + \;\; \hbox{zyklisch}\;\Big) \nonu \\[2pt]
  &=& {i\0 4\pi} \int_V \Sp \Big(\,\6_3\, \ph \lk X_1 , 
       X_2 \rk  \;\; + \;\; \hbox{zyklisch} \; - \; \ph 
       \,\hbox{\large\boldmath$\{\,$} \6_3 \lk X_1 , X_2 
       \rk \; + \; \hbox{zyklisch} 
       \,\hbox{\large\boldmath$\,\}$}\;\Big) \nonu \\[2pt]
  & & \hspace*{1.4cm}
       X_1 X_2 - X_2 X_1 \, = \,\6_2 X_1 
       - \6_1 X_2 \qquad \Rightarrow \quad 
      \hbox{\large\boldmath$\{ \qquad \}$}\; = \; 0  
      \nonu \\[2pt]
  &=& {i\0 4\pi} \int_V \nabla \cdot \,
      \hbox{\Large\bf ( }  \Sp \( \ph [ X_2 , X_3 ] \) 
      \hbox{\Large\bf\ , } \Sp \( \ph [ X_3 , X_1 ] \) 
      \hbox{\Large\bf\ , } \Sp \( \ph [ X_1 , X_2 ] \) 
      \hbox{\Large\bf\ )} \nonu \\[4pt]
  & & \hspace*{3cm} \hbox{{\bf jetzt erst Gau\ss\ :}} 
      \nonu \\[3pt]
  &=& {i\0 4\pi} \int_{\6 V} d\vc f \cdot \; 
      \hbox{\Large\bf ( }  \Sp \( \ph\,[ X_2 , X_3 ] \)
      \hbox{\Large\bf\ , } \Sp \( \ph\,[ X_3 , X_1 ] \)
      \hbox{\Large\bf\ , } \Sp \( \ph\,[ X_1 , X_2 ] \)
      \hbox{\Large\bf\ )} \nonu \\[4pt]
  \d\, S_2 
  &=& {i\0 4\pi} \int \Sp \Big(\,\ph \lk X_1 X_2 
       - X_2 X_1 \rk \Big) \quad \hbox{auf Deckfl\"ache ,} 
      \;\;\qquad \hbox{q.e.d.} \qquad
\eea 
Die Dgln sind erf\"ullt. Dieser \S\ hat damit seiner
\"Uberschrift schon gen\"uge getan. Das Fragezeichen hinter 
obigem Karabali--Zitat bezieht sich auf die Behauptung, 
$S_2$ lasse sich explizit als \ \hbox{e b e n e s} \
Integral aufschreiben. Aber das will uns partout nicht 
gelingen.
\pagebreak[2]
     \nz[.6cm]  \hspace*{2cm} 
{\bf \quad R\"uckblick und Fazit}
     \nz[.1cm] 
\nopagebreak[3]
Es ging darum \ --- \  \"uber f\"unf Abschnitte \ --- \ 
die Wellenfunktionen $\psi$ der 
Schr\"odinger--funk\-tio\-na\-len Quantenmechanik richtig 
normieren zu k\"onnen. Die reellen Variablen leben auf den 
\hbox{\anfu Feld}\-achsen\anfo$\!$. Schon in \eq{6.2} 
hatten wir einmal das Skalarprodukt notiert und mit
Fragezeichen das Problem markiert. Da\ss\ das dortige 
Ma\ss\ $d\mu (\cl A)$ durch $d\mu (\cl C )$ zu ersetzen 
ist, war bereits in \S~6 verstanden$\,$: Integration \"uber 
unphysika\-li\-sche Freiheitsgrade einfach weg\-las\-sen. 
\S\S~7 und 8 bescherten uns dann die eingerahmte Glei\-chung 
\eq{8.13}. Und von da an haben wir uns nur noch mit der 
dortigen Determinante herumgeschlagen. Fazit$\,$:
\be{10.4}
  \int \psi_1^* \psi_2 \; = \; \int\! d\mu (\cl C)
  \;\,\psi_1^*(H) \, \psi_2 (H) 
  \; = \; \int\! d\mu (\cl H) \; e^{2NS(H)} \;
  \,\psi_1^*(H) \, \psi_2 (H) \quad .
\ee 
\eq{10.4} ist \ek {2.25}. KKN$\,$: {\sl This formula shows 
that all matrix elements in (2+1)--dimensio\-nal SU(N) gauge 
theory can be evaluated as correlators} des hermiteschen 
WZW--Modells. Es gibt nur noch $n$ \anfu Feldachsen\anfo$
\!\!$, n\"amlich eine f\"ur jeden reellen Parameter $\o^a$
(nichts gegen Blick auf \eq{8.14} und $H=\rho^2\,$). Eine 
Vorgabe aller $\o^a$ legt n\"amlich eine be\-stimm\-te 
hermitesche eins--determinantische Matrix $H$ fest. Ab 
sofort k\"onnen wir Wellenfunktionen $\psi$ normieren, 
mittels \eq{10.4} n\"amlich. Unter Funktionen $\psi$ erlaubt 
die Quantenmechanik bekanntlich nur die normierbaren. Und 
dann gibt es da noch die Bewegungsgleichung $i \hbar \p 
\psi = {\bf H} \psi\;$. Mindestens auch noch zum 
Hamilton--Operator ${\bf H}$ im Schr\"odinger--Bild sollte 
dieses Traktat etwas sagen$\,$: \S~12$\,$.  
\pagebreak[2]
     \nz[.5cm]  \hspace*{2cm} 
{\bf \quad Polyakov--Wiegmann--Identit\"at}
     \nz[.1cm] 
\nopagebreak[3]
Die Wirkung $S(H)$, \eq{10.1}, erf\"ullt eine neckische 
Beziehung f\"ur Produkt--Argument. In \cite{powi2} wird sie 
eine {\sl remarkable identity} genannt. Ihre dreizeilige
Begr\"undung ist anstrengend, und das in \cite{powi2} 
angebene Resultat ist falsch. Manche Autoren sollten das
Vordiplom zu wiederholen haben. Als \ek{3.2} ist die
Identit\"at auch bei KKN falsch angegeben. Wir stellen
richtig$\,$:
\be{10.5}
  S(AB) = S(A) + S(B) - {1\0\pi} \int \Sp \Big( \lk 
  \6 B \rk B^{-1} A^{-1} \,\ov{\6} A \,\Big) \quad .\quad
\ee 
Um \eq{10.5} herzuleiten, schreiben wir wieder $S = 
S_1 + S_2$ und erhalten zun\"achst
\bea{10.6}
  S_1(AB) &=& {1\0 2\pi} \int \Sp \Big( \lk \6 A B \rk 
              \ov{\6}\, B^{-1} A^{-1} \Big) \nonu \\
    &=& {1\0 2\pi} \int \Sp \Big( \lk (\6A) B + A\, \6 B \rk 
       \lk  (\ov{\6} B^{-1} ) A^{-1} + B^{-1} \ov{\6} 
        A^{-1} \rk \Big) \nonu \\
    &=& S_1(A) + S_1(B) - {1\0 2\pi} \int \Sp \Big( \;
        (\ov{\6} B ) B^{-1} A^{-1} \6 A + (\6 B) B^{-1} 
        A^{-1} \ov{\6} A \;\Big) \nonu \\[3pt]
    &=& S_1(A) + S_1(B) - {1\0 4\pi} \int \Sp 
        \Big( \, b_1 a_1 + b_2 a_2 \,\Big) \quad ,\quad 
   \lower 10pt\vbox{\hbox{$a_j \gll A^{-1} A_{\prime j}$}
     \hbox{$b_j \gll B_{\prime j}\,B^{-1}$}} \quad . \quad
\eea 
Auch im Term $S_2$ ersetzen wir $H$ durch $AB\,$. In der
Version \eq{10.2} ist folglich $X_j = B^{-1}A^{-1} 
(AB)_{\prime j} = B^{-1} \big( a_j + b_j \big) B\,$ 
einzusetzen$\,$:
\pagebreak[3]
\bea{10.7}
 S_2(AB) &=& {i\0 4\pi} \int_V \Sp \Big( \, X_3 \lk X_1 
             X_2 - X_2 X_1 \rk \Big) \nonu \\ 
  &=&  {i\0 4\pi} \int_V \Sp \lk \, (a_3 + b_3 ) 
       (a_1+b_1)(a_2+b_2) \;\; - \;\; 
       \hbox{dito}_{\rm antizyklisch} \rk  \\
  &=&  S_2(A) + S_2(B) + {i\04\pi} \int_V \Sp 
       \,\Big[\; \hbox{Mischterme}\; \Big] \quad . 
       \quad \nonu 
\eea 
Die Mischterme gruppieren sich in solche mit einem $b$ 
mal Kommutator zweier $a$'s und solche mit einem $a$ und 
$b$--Kommutator. Die drei $a$--Kommutatoren bilden die 
Komponenten von $\vc a \times \vc a\,$. Unter der Spur
steht schlie\ss lich $[${\small Mischterme}$] = \vc b 
(\vc a \times \vc a ) + ( \vc b \times \vc b ) \vc a\,$.
Hiermit lassen sich nun \ --- \ darum hatten wir gebangt 
\ --- \ die Mischterme als Divergenz schreiben$\,$:
\bea{10.8}
 \nabla \times \vc b = \nabla \times \lk  (\nabla B ) 
    B^{-1} \rk &=& \vc b \times \vc b  \quad , \quad 
    \nabla \times \vc a = \nabla \times \lk A^{-1} \nabla A 
    \rk = - \, \vc a \times \vc a \quad , \nonu \\
   \,\Big[\; \hbox{Mischterme}\; \Big] 
   &=& (\nabla \times \vc b ) \vc a - \vc b (\nabla \times 
   \vc a ) = \nabla (\vc b \times \,\vc a ) \quad , \quad
\eea 
und das macht ihn gl\"ucklich, den Meister Gau\ss . Auf der
Deckfl\"ache des Volumens $V$ entsteht $ ( \vc b \times 
\,\vc a )_3 = b_1 a_2 - b_2 a_1 $ und insgesamt
\be{10.9}
  S_2(AB) = S_2(A) + S_2(B) - {1\0 4\pi} \int \Sp
  \Big( \, i \, b_2 a_1 - i \, b_1 a_2 \, \Big) \quad . \quad
\ee 
Addition von \eq{10.9} zu \eq{10.6} gibt
\be{10.10}
 S(AB) = S(A) + S(B) - {1\0 4\pi} \int \Sp \Big(\; 
   (b_1 + i b_2 ) \, (a_1 - i a_2 ) \; \Big) \;\; 
   \equiv \;\; \hbox{\eq{10.5} \quad ,\quad q.e.d.} \quad
\ee 
Sagt doch der Meister Ketov so nebenbei, jaja, als 
Fu\ss g\"anger brauche man Gau\ss , aber die elegante
Herleitung sei jene (schreckliche) in \cite{powi2}.
Im n\"achsten Abschnitt werden wir auf $S(ABC)$ zu blicken
haben. Mittels \eq{10.5} folgt
\bea{10.11}
  S(ABC)&=& S(A) + S(B) + S(C) - {1\0\pi} \int \Sp 
            \bigg( \nonu \\ 
  & & \hspace{-2.8cm} (\6 C) C^{-1}B^{-1}A^{-1} 
      (\ov{\6} A ) B \; + \; (\6 C) C^{-1}B^{-1} 
       \,\ov{\6} B   \; + \; (\6 B) B^{-1} 
    A^{-1}\,\ov{\6} A  \;\;\bigg) \quad . \quad
\eea 
Um dies (insbesondere den Vorfaktor $-1/\pi\,$) irgendwie 
zu testen, setzen wir $\, A B C = H H^{-1} H\,$, 
verifizieren (\"Ubung) die Eigent\"umlichkeiten
\be{10.12}
  S_1 \big( H^{-1} \big)\, = \, S_1 \big( H \big) 
  \qquad , \qquad   
  S_2 \big( H^{-1} \big)\, =\, -\, S_2 \big( H \big)
  \qquad 
\ee 
und erhalten wunschgem\"a\ss\ $\, 3 S_1 + S_2 - 2 S_1 = S\,$
auf der rechten Seite von \eq{10.11}.


\sec{Regularisierung}

Im Abschnitt 9.4 hatten wir den Schritt zur 
regularisierten inversen $D$--Matrix \eq{9.14} per 
Zitat erledigt. Es ist an der Zeit, diesen Versto\ss\ gegen 
die guten Sitten endlich auszuwetzen. In den Dgln f\"ur 
$\G$, \eq{9.6}, war $\vc r^\prime \to \vc r$ auszuf\"uhren.
Aber dieser limes \ \hbox{g e h t} \ s o \ 
\hbox{n i c h t }, weil gem\"a\ss\ \eq{9.9} die Greensche 
Funktion an ihr pathologisches Argument Null ger\"at.
 
An die Regularisierung sind mehrere Anspr\"uche zu stellen. 
Sie hat zum einen eichinvariant zu sein, wird also $H$'s 
gegen\"uber $M$'s favorisieren. Es gibt aber noch eine
andere \ \hbox{R e d u n d a n z} , welche sie respektieren 
sollte, und dieser sei der n\"achste Unterabschnitt 
gewidmet. 
\pagebreak[2]
     \nz[.5cm]  \hspace*{2cm} 
{\bf 11.1 \quad Holomorphe Invarianz}
     \nz[.1cm] 
\nopagebreak[3]
Dieser Terminus d\"urfte von KKN erfunden sein ({\sl
we shall refer to ... as ...}). Der zugeh\"ori\-gen
Freiheit waren wir schon einmal begegnet, als die 
Inhomogenit\"at in \eq{4.4} auf 1 festgelegt wurde. 
Stattdessen war dort eine beliebige von $\ov{z}=x+iy$ 
abh\"angige Matrix $\ov{V}$ erlaubt. L\"a\ss t man 
s\"amtliche solche Inhomogenit\"aten zu, dann ist einem 
Element $A$ aus $\cl A$ ein ganzer Unterraum 
$M\,\ov{V}$ zugeordnet ($M$ in SL(N,C), aber $M\ov{V}$ 
nicht mehr), denn
\be{11.1}  
  M \to M\, \ov{V} \qquad \hbox{l\"a\ss t} \qquad
  A=-(\6M)\, M^{-1} \;\to\; - \(\6 M \ov{V} \) \; 
  \ov{V}^{-1}  M^{-1} = A \quad  \;\; , \quad
\ee 
unver\"andert, weil $\,\6\ov{V}(\ov{z})=0\,$, s.~\eq{3.12}.
Keine Physik sollte davon abh\"angen, mit welchem $\ov{V}$
auf Erde, Mond oder Neptun gearbeitet wird. 
\\[12pt]
{\bf Die Wirkung {\boldmath$S(H)$} ist holomorph invariant}.
Mit $M\to M\ov{V}$ geht $H=M^\dagger M$ in $H \to VH\ov{V}$
\"uber, wobei $V\gll \ov{V}^\dagger$ nur von $z=x-iy$
abh\"angt$\,$: $\ov{\6} V(z)=0\,$. Die Frage ist somit,
ob $S(VH\ov{V})$ mit $S(H)$ \"ubereinstimmt \ --- \ womit
bereits klar ist, weshalb wir im vorigen Abschnitt die 
Beziehung \eq{10.11} vorbereitet hatten. Die drei dort unter 
Spur ausgeschriebenen Terme verschwinden wegen $\6 C = \6 
\ov{V} = 0$ und/oder $\ov{\6} A = \ov{\6} V = 0\,$. Jetzt 
st\"oren nur noch $S(V)$ und $S(\ov{V})\,$. $S_1(V)$ und 
$S_1(\ov{V})$ verschwinden, weil darin $V$ bzw. $\ov{V}$ 
je einmal die \anfu falsche Differentiation\anfo erleiden. 
Zu $S_2(V)$ sehen wir uns in \eq{10.2} die Bildung
\bea{11.2}
   \lk X_1 X_2 - X_2 X_1 \rk &=& \lk V^{-1} V_{\prime 1}
   V^{-1}V_{\prime 2} - V^{-1} V_{\prime 2}
   V^{-1}V_{\prime 1} \rk \nonu \\
   &=& - i \lk V^{-1} V^\prime V^{-1} V^\prime 
   - V^{-1} V^\prime V^{-1} V^\prime \rk = 0 \quad 
\eea 
an. Analog verschwindet auch $S_2(\ov{V})\,$. Ergo
$S(VH\ov{V})= S(H)\,$, q.e.d.
\\[12pt]
{\bf Die Greensche Funktion wird Matrix.} Die Leute auf
dem Mond (L wie Luna) stellen das Feld $A$ per
$A= - (\6 L ) L^{-1}$ dar und benutzen zur Aufl\"osung
die Inhomogenit\"at $\ov{V}$ statt 1, d.h. sie schreiben
($G_{r r^\prime} = - G_{r^\prime r}$ ausnutzend)
\be{11.3}
  L = \ov{V} + \int^\prime (AL)_{r^\prime} G_{r^\prime r}
  \quad \hbox{statt} \quad
  M = 1 + \int^\prime (AM)_{r^\prime} G_{r^\prime r}
\ee
auf. Wir multiplizieren unsere Igl (die irdische
in \eq{11.3} rechts)  mit $\ov{V}\,$ von rechts,
$$ M \ov{V} = \ov{V} + \int^\prime (AM\ov{V})_{r\prime}
   \;\;\ov{V}_{r^\prime}^{-1} G_{r^\prime r} \ov{V}_r 
   \quad , \quad $$
und sehen, da\ss\ uns die beiden Handgriffe
\bea{11.4}
  M \;\to\; M\,\ov{V} \quad  &\hbox{und zugleich}& 
  \quad G_{r r^\prime} \;\to\; \ov{V}^{-1}_r 
  G_{r r^\prime} \ov{V}_{r^\prime} \;\glr\, 
     \schl{G}_{r r^\prime} \hspace{1cm} \nonu \\
  \hbox{bzw.} \quad M^\dagger \;\to\; V\, M^\dagger\quad 
  &\hbox{und zugleich}& \quad \ov{G}_{r r^\prime} \;\to\; 
  V_r \;\ov{G}_{r r^\prime} V_{r^\prime}^{-1}
\eea 
in das Reich der Lunasen beamen (KKN verwirren hier, weil
sie $\ov{G}\to$ richtig, aber $G \to$ falsch angeben).
\"Ubrigens ist auch $\schl{G}_{r r^\prime}$ eine Greensche 
Funktion, denn $\6 \schl{G}_{r r^\prime} = \d(\vc r 
- \vc r^\prime)\,$. Aus Spa\ss\ an der Freude gehen wir 
noch auf den Neptun, $A = -(\6N)N^{-1}$, wo die Leute
mit Inhomogenit\"at $\ov{V}$ arbeiten und mit 
$\ov{U}_r G_{r r^\prime} \ov{U}_{r^\prime}^{-1}\,$ 
als Green$\,$:
$$ N = \ov{V} + \int^\prime (AN)_{r^\prime} \, 
   \ov{U}_{r^\prime} G_{r^\prime r} \ov{U}_r^{-1} 
   \quad . \quad $$ 
Jetzt f\"uhrt Multiplikation mit $\ov{U}$ von rechts
auf die Luna--Gleichung zur\"uck, n\"amlich mit $L=N\ov{U}$
und Inhomogenit\"at $\ov{V}\ov{U}$. Alles in Ordnung.

Wir tragen noch nach, wie $\schl{G}_{r r^\prime}$ in 
adjungierter Darstellung aussieht, d.h. als 
$ab$--Matrix$\,$:
\be{11.5}
 G_{r r\prime} \,\d^{ab} \;\to\; ( \ov{V}_r^{-1} )^{ac}
 G_{r r\prime} (\ov{V}_{r^\prime})^{cb} \; \glr\, 
  ( \schl{G}_{r r\prime})^{ab} \quad . \quad
\ee 
Um sich dies herzuleiten, wiederholt man am besten die 
Schritte von \eq{11.3} bis \eq{11.4}, aber in adjungierter 
Darstellung.
\\[12pt]
{\bf Das Integrations--Volumenelement $d\mu(\cl H)$ ist
holomorph invariant}. Da\ss\ dem so ist, sei {\sl easily 
checked}, sagen KKN. Wir blicken dazu auf \eq{8.10}, d.h.
auf $ds_{\cl H}^2 = 2 \int \Sp ( H^{-1} \d H H^{-1} 
\d H)\,$. Vermutlich ist nun zu sagen, da\ss\ man 
mit einer \ \hbox{f e s t e n } \  Inhomogenit\"at 
$\ov{V}$ (bzw $V$) zu arbeiten habe (man ist entweder auf
Neptun, Mond, oder hinieden). Wenn aber die Variationen
kein $\ov{V}$ oder $V$ erfassen, dann fallen sie in der 
Tat aus \eq{8.10} heraus \ --- \ und aus $d\mu (\cl H)$ 
ebenso.
\\[12pt]
{\bf Die Operatoren \boldmath$p^a$ und $\ov{p}^a$ spalten
eine Matrix ab}. Zum einen, um den Laplace--Operator
auf $\cl C$ zu konstruieren$\,$\footnote{\ 
      Auf dieses Detail (Laplace on $\cl C$) werden wir 
      uns hier nicht mehr einlassen. Es wird im \S~12 
      nicht ben\"otigt. Wir hadern aber auch mit dem 
      Umstand, da\ss\ KKN in \ek{2.39} (und Text davor) 
      rechts-- und links--Translationen von $H$ betrachten, 
      welche aus dem Raum hermitescher Matrizen
      herausf\"uhren.},
und zum anderen weil hilfreich beim anstehenden Gesch\"aft 
der Regularisierung \ \hbox{d e f i n i e r e n} \ KKN in
\ek{2.34} die
folgenden beiden Operatoren $p\;$:
\be{11.6}
  \d^a \,\glr\, M_r^{ab} \int^\prime G_{r r^\prime} 
   p^b_{r^\prime} \quad , \quad
  \d^{a*} \,\glr\, - (M_r^\dagger)^{ba} \int^\prime 
   \ov{G}_{r r^\prime} \ov{p}^b_{r^\prime} \quad . \quad
\ee 
(Zur Erinnerung$\,$: $\d^a=\d_{A^a(\vcsm r)}$.) Wir l\"osen
diese Gleichungen nach $p$ auf (z.B. die erste per Anwenden
von $\hat{M}^{-1}$ und sodann $\6_r\,$)$\,$:
\be{11.7}
  p_r^a = \6_r \, (M_r^{-1})^{ac} \; \d_r^c \quad , 
  \quad  \ov{p}_r^a = - \ov{\6}_r \, (M_r^\dagger)^{ac} 
  \; \d_r^{c*}  \quad . \quad
\ee 
Die Orts--Marken sind angegeben, um deutlich zu machen, auf 
welche zwei Abh\"angigkeiten $\6$ bzw. $\ov{\6}$ wirken.
Beim Umsteigen nach Luna widerf\"ahrt den $p$'s 
folgendes$\,$:
\be{11.8}
  p^a \;\to \; (\ov{V}^{-1})^{ab} p^b \qquad , \qquad
  \ov{p}^a \;\to \; V^{ab} \ov{p}^b \quad . \qquad
\ee 
Wir leiten dies f\"ur $p^a$ her (wof\"ur KKN's Angaben
nicht richtig sind). $A^a$ kennt keinen 
Erde--Mond--Unterschied und folglich auch $\d^a$ nicht$\,$:
\bea{11.9}
  p^a \;\to\; \6 \,\; 2 \,\Sp \Big( T^a \ov{V}^{-1} 
     M^{-1} T^c M \ov{V} \Big) \; \d^c  
  &=& \6 \; 2\Sp \Big( \ov{V} T^a \ov{V}^{-1} T^b \Big)\, 
   2\Sp \Big( T^b M^{-1} T^c M \Big)\; \d^c \qquad \nonu \\
     = \,\6 \; ( \ov{V}^{-1})^{ab} (M^{-1})^{bc} \,\d^c 
  &=&  (\ov{V}^{-1} )^{ab} p^b \quad . \quad
\eea 
Der Schritt in der ersten Zeile, das war {\sl concatenation} 
r\"uckw\"arts.
\pagebreak[3]
\\[12pt]
{\bf Die Hamilton--Dichte ist holomorph invariant}. 
Der Potentialterm enth\"alt nur die holomorph 
unempfindlichen $A^a$ selbst. Man kann sich grob 
vorstellen, da\ss\ die 
kinetische Energie $\d^{a*}\d^a$ enth\"alt und da\ss\ 
hier via \eq{11.6} die $p$--Ope\-ra\-to\-ren ins Spiel 
gebracht werden k\"onnen. Der Vorgriff auf
\be{11.10}
 \cl T = - {e^2\02} \d^{a*} \d^a \; = \; 
    {e^2\02} H^{ab} \,\big( \ov{G} \ov{p} 
     \big)^a\, \big( G p \big)^b  \qquad
\ee 
ist also nicht tragisch. Wir wollen hier lediglich
nachsehen, ob der Ausdruck rechts in \eq{11.10} holomorph 
invariant ist (was er als $\d^{a*}\d^a$--Bildung 
nat\"urlich zu sein hat). Es versteht sich, da\ss\ 
$H^{ab}=2\Sp(T^a H T^b H^{-1})\,$ ist. Die letzte Klammer 
erlebt 
\be{11.11}
  \big( G p \big)_r^b \,\gll\, \int^\prime 
   G_{r r^\prime} p_{r^\prime}^b \; \to \; 
   \int^\prime \Big( \ov{V}^{-1}_r G_{r r^\prime} 
       \ov{V}_{r^\prime} \Big)^{bc} 
       \( \ov{V}_{r^\prime}^{-1} \)^{cd} p_{r^\prime}^d
   = \big( \ov{V}_r^{-1} \big)^{bc} \big( G p \big)^c_r 
     \quad , \quad
\ee 
und man sieht sch\"on, wie \eq{11.5} und \eq{11.8} 
miteinander harmonieren. Analog folgt $(\ov{G} \ov{p})^a 
\to V^{ab}(\ov{G} \ov{p})^b\,$. Jetzt fehlt uns, um 
\eq{11.10} auf den Mond zu beamen, nur noch $H^{ab} \to 
V^{ac} H^{cd} \,\ov{V}^{db}$. Wer dies aus $H\to VH\ov{V}$ 
herleiten will (\"Ubung), braucht zwei R\"uckw\"arts--{\sl 
concatenations}. Insgesamt verwandelt sich $\cl T$ in
\be{11.12}
  \cl T \;\;\to\;\; {e^2\02} \, V^{ac} \, H^{cd} \, 
   \ov{V}^{db}\, V^{ae} \, \big( \ov{G} \ov{p}\big)^e \,
   (\ov{V}^{-1})^{bf} \, \big( G p \big)^f  \,\; = \,\; 
   \cl T \quad , \quad \hbox{q.e.d.} \quad . \quad
\ee 
\eq{11.12} ist ein Suchbild nach zwei 
Kroneckern (note that $V^{ae}=(V^{-1})^{ea}\,$).
\pagebreak[2]
     \nz[.5cm]  \hspace*{2cm} 
{\bf 11.2 \quad Point splitting}
     \nz[.1cm] 
\nopagebreak[3]
Die Divergenzen einer Feldtheorie werden ja meist (aber 
nicht notwendigerweise) im $K$--Raum detektiert und 
regularisiert ($\L$, $M$, $d-\e$). Wenn dies im Ortsraum 
geschehen soll, werden kurze Abst\"ande zu verwaschen sein. 
KKN's verwaschene Fast--Deltafunktion ist
\be{11.13}
   \s (\vc r ) \gll {1\0\pi\eta} e^{-r^2/\eta}
   \quad , \quad \int \! d^2r \;\s (\vc r ) = {1\0\pi} 
   \int e^{-r^2} = 1
\ee 
Der \anfu reziproke $K$--cutoff\anfo $\eta$ ist klein aber 
nicht Null. Demgegen\"uber ist der kleine Parameter $\e$
in der Greensfunktion \eq{4.2} ein eiskaltes $+0\,$. 

In einer Weise, welche holomorphe Transformationen aufrecht
erh\"alt, bauen KKN die Verwaschung $\s$ zuerst in den 
Operatoren $p$ ein und lesen dann aus $(Gp)^a$ ab, wie $G$
zu verarzten ist. Ob dies \ d e r \ geeignete Weg (oder
ein Umweg) ist, wissen wir nicht. Es folgen Auswertung der 
regularisierten Greens und schlie\ss lich der limes 
$\vc r^\prime \to \vc r\,$.

Es beginnt mit Notation$\,$: $H(\vc r)$, wenn als Funktion
von $z$ und $\ov{z}$ gelesen, hei\ss e $K\,$:
\be{11.14}
  H^{ab}(\vc r) \, \glr\, K^{ab}(z,\ov{z}) \; \to \;
  V^{ac} (z) \, K^{cd}(z,\ov{z}) \,\ov{V}^{db}(\ov{z}) 
  \quad . \quad
\ee 
Mittels \eq{11.14} ist direkt einzusehen, da\ss\ die
beiden Bildungen
\bea{11.15}
  p_{\rm reg}^a \,&\gll& \int^\prime \s_{r r^\prime}
  \lk K^{-1} (z^\prime , \ov{z} ) \, K(z^\prime , 
   \ov{z}^\prime ) \rki^{ab} \, p_{r^\prime}^b
         \nonu \\
  \ov{p}_{\rm reg}^a \,&\gll& \int^\prime \s_{r r^\prime}
  \lk K(z, \ov{z}^\prime ) \, K^{-1}(z^\prime , 
   \ov{z}^\prime ) \rki^{ab} \, \ov{p}_{r^\prime}^b
   \quad
\eea 
das richtige holomorphe Verhalten \eq{11.7} auch dann
zeigen, wenn $\s \neq \d_{r r^\prime}$ bleibt$\,$: was die
$p$'s nach links ausgeben, wird aufgefressen und ebenso das,
was sich zwischen $K$'s ansammeln m\"ochte$\,$:
$\;\big[\, K^{-1}(z^\prime,\ov{z})\,\big]^{ac} \to 
\,\big[\, \ov{V}^{-1}(\ov{z}) K^{-1}(\ldots ) V(z^\prime)
\,\big]^{ac}\;$. Und nach links rutscht je die richtige 
$V$--Matrix bei ungestrichenem Argument heraus. Irgendwie 
schlau$\,$!     

Wie angek\"undigt, wird nun in der Bildung $(Gp)^a$ der
Operator $p^a$ durch seine regularisierte Version 
$p^a_{\rm reg}$ ersetzt$\,$,
\be{11.16}
  (G p)^a = \int^\prime G_{r r^\prime} p_{r^\prime}^a
   \;\to\; \int^\prime G_{r r^\prime} \int^{\prime\prime}
   \s_{r^\prime r^{\prime\prime}} \lk \vphantom{K} \ldots 
   \rki^{ab}_{r^\prime r^{\prime\prime}} \, p_{r^{\prime\prime}}^b
   \; \glr \int^{\prime\prime} \; 
   \cl G^{ab}_{r r^{\prime\prime}} \; p_{r^{\prime\prime}}^b
   \quad , \quad
\ee 
um die regularisierte Greensfunktion $\cl G$ zu 
entnehmen$\,$:
\be{11.17}
  \cl G^{ab}_{r r^{\prime\prime}} = \int^\prime
  G_{r r^\prime} \s_{r^\prime r^{\prime\prime}}
  \lk K^{-1} ( z^{\prime\prime} , \ov{z}^\prime ) \,
  K ( z^{\prime\prime} , \ov{z}^{\prime\prime} ) \rki^{ab}
  \; \glr \; \int^\prime  G_{r r^\prime} 
  \s_{r^\prime r^{\prime\prime}} \; f^{ab}(\ov{z}^\prime )
  \quad . \quad
\ee 
Die uninteressanten \"au\ss eren Variablen 
$z^{\prime\prime}$, $\ov{z}^{\prime\prime}$ wurden rechts
in $f$ unterdr\"uckt. Wir merken uns, da\ss\ $f^{ab}
(\ov{z}^{\prime\prime}) = \d^{ab}$ ist.
\pagebreak[2]
     \nz[.5cm]  \hspace*{2cm} 
{\bf 11.3 \quad Ausintegration von \boldmath$\cl G$}
     \nz[.1cm] 
\nopagebreak[3]
Es klingt abenteuerlich, die $d^2r^\prime$--Integration in 
\eq{11.17} ausf\"uhren zu wollen. Zwar sind $G$ und $\s$ 
konkrete Funktionen, aber wir wollen \ \hbox{n i c h t} \ 
etwa $f(\ov{z})$ spezifizieren. Es geht dennoch \ --- \ 
aufgrund dessen, da\ss\ $G$ eine Greens von $\6$ ist.

Das Resultat wird einen Test zu bestehen haben. Gem\"a\ss\ 
\eq{11.17} gilt n\"amlich
\be{11.18}
  \6 \, \cl G^{ab}_{r r^{\prime\prime}} 
  = \s_{r r^{\prime\prime}}\, f^{ab} (\ov{z}) 
  = {1\0 \pi \eta} e^{- (\vcsm r - \vcsm r^{\prime\prime} 
     )^2 / \eta} f^{ab} (\ov{z}) \qquad
\ee 
und nach Auswertung bitte noch immer. Man blicke gleich 
jetzt voraus auf \eq{11.26} und teste selbst. 

Es erweist sich als n\"utzlich, eine Stammfunktion $S$
von $\s$ zu kennen$\,$: 
\be{11.19}
 \6 S (\vc r ) = \s (\vc r ) \;\; , \;\; \hbox{\ft Ansatz} 
 \; S = {z\0 r^2} h(r) \; , \; \ldots \quad 
  \Rightarrow \quad  S(\vc r ) = {1\0 \pi\, \ov{z} } 
  \( 1 - e^{-r^2/\eta} \) \quad , \quad
\ee 
wobei die Integrationskonstante zur $h$--Dgl auf $C=1$
festgelegt wurde. Mittels $S$ und nach partieller 
Integration bekommt $\cl G$ zwei Terme (die Indices 
$a$, $b$ lassen wir vor\"ubergehend weg)$\,$:
\be{11.20}
  \cl G_{r r^{\prime\prime}} = \int^\prime G_{r r^\prime} 
        \,\lb \;\6^{\,\prime} S(\vc r^\prime - 
   \vc r^{\prime\prime} ) \;\rb\, f (\ov{z}^\prime ) 
   \, = \, \cl G_1 + \cl G_2  \quad . \quad
\ee 
Hierin steht $\cl G_1$ f\"ur $\int^\prime \6^\prime \lk 
G S f\rk$ und $\cl G_2$ f\"ur $ -\int^\prime \lk \6^\prime 
G \rk S f\,$. Folglich ist
\bea{11.21}
 \cl G_2 &=& S(\vc r - \vc r^{\prime\prime} ) f(\ov{z})
  = {1\0\pi} {1\0 \ov{z}- \ov{z}^{\prime\prime} }
   \( 1 - e^{(\vcsm r - \vcsm r^{\prime\prime} )^2 / 
   \eta} \)  f(\ov{z}) \nonu \\
  &\Rightarrow& \; G_{r r^{\prime\prime}}   \( 1 - 
   e^{(\vcsm r - \vcsm r^{\prime\prime} )^2 / \eta} \)
   f(\ov{z}) \quad , \quad
\eea 
wobei die Ersetzung durch $G$ erlaubt war, weil die
runde Klammer am $G$--Pol verschwindet (darum die Wahl
$C=1$). Bleibt $\cl G_1\,$. Wir schreiben $\cl G_1 = 
\cl G_{1\, c} + G_{1\, e}\,$ und lassen die Indizes auf 
den $C\!=\!1$--Term bzw. den $e$--Term in $S$ verweisen. 
In $\cl G_{1\, c}$ wirkt $\6^\prime$ nur auf $G$ und gibt 
$\6^\prime G_{r r^\prime} = - \d_{r r^\prime}\,$:
\be{11.22}
  \cl G_{1c} = \, - \,{1\0 \pi} \; {1\0 \ov{z}  - 
  \ov{z}^{\prime\prime} } \; f(\ov{z} ) \quad . \quad
\ee 
Zum Term $\cl G_{1\, e}$,
\be{11.23}
  \cl G_{1 e} = - \int\! d^2 r^\prime \;\, \6^{\,\prime} 
  \lk {1\0\pi} \, { z-z^\prime \0 (z-z^\prime )\,
   (\ov{z} - \ov{z}^\prime ) + \e^2 } \;\, {1\0\pi}\,
   {1 \0 \ov{z}^\prime - \ov{z}^{\prime\prime} } \;
   e^{- (\vcsm r^\prime - \vcsm r^{\prime\prime} )^2 / \eta}
   \; f(\ov{z}^\prime ) \rk \quad , 
\ee 
ist zu begreifen, da\ss\ hierin der limes $\e\to 0$ 
ausgef\"uhrt werden darf (Details seien dem Leser 
\"uberlassen). Jetzt wird die Sache h\"ubsch. Nach 
Verschiebung $\vc r^\prime \to \vc r^\prime + 
\vc r^{\prime\prime}$ und in den Variablen
$$ 
  x^\prime = {1\02} (u-iv) \;\; , \;\; y^\prime 
  = {1\02} (-iu+v)  \;\; , \;\; \hbox{Jacobi--Det.} 
  = {1\02}  \quad , \quad \Big(\; \ov{z}^\prime = u 
    \;\; , \;\; z^\prime  = -iv \; \Big) \qquad  $$
folgt n\"amlich
\be{11.24}
   \cl G_{1\, e} = - {i\0 2} \, {1\0 \pi} \int \! 
   du^\prime  \; { f(u^\prime + u^{\prime\prime})
   \0 \; u-u^{\prime\prime} - u^\prime \;} \;\, 
   {1 \0 \pi\, u^\prime} \int \! dv^\prime \; \6_{v^\prime}
   \; e^{i\, {u^\prime\0\eta}\,v^\prime }\,
   \; = \; {1\0\pi} \, {1\0 \ov{z} - \ov{z}^{\prime\prime}} 
   \; f(\ov{z}^{\prime\prime}) \quad . \quad
\ee 
Die rechte Seite ensteht per Ausf\"uhren von $\6_{v^\prime}
\,$, $\d(u^\prime)$ aus der $v^\prime$--Integration und 
reuiger R\"uckkehr zu $u^{\prime\prime} = \ov{z}^{\prime
\prime}\,$. Addition von \eq{11.22} und \eq{11.24} gibt
\be{11.25}
 \cl G_1 = \cl G_{1\,e} + \cl G_{1\,c} =
  {1\0 \pi} \, {1\0 \ov{z} -  \ov{z}^{\prime\prime} }
  \lk f( \ov{z}^{\prime\prime}) - f( \ov{z} ) \rk
  \;\; \Rightarrow\;\; G_{r r^{\prime\prime}} 
 \lk f( \ov{z}^{\prime\prime}) - f( \ov{z} ) \rk
\ee 
und zusammen mit \eq{11.21} das Resultat
\be{11.26}
  \cl G^{ab}_{r r^{\prime\prime}} = G_{r r^{\prime\prime}}
  \( \d^{ab} - e^{-(\vcsm r - \vcsm r^{\prime\prime} )^2 
  / \eta} \; f^{ab} (\ov{z} ) \;\) \quad . \quad
\ee 
\eq{11.26} ist \ek{3.8}. Wir haben zwar von der konkreten
$\s$--Gestalt Gebrauch gemacht, aber jede Wette, da\ss\
die Prozedur (incl. n\"achster Unterabschnitt) von 
$\s$--Details nicht wirklich abh\"angt (der Leser bekommt 
allm\"ahlich immer mehr zu tun).  
\pagebreak[2]
     \nz[.5cm]  \hspace*{2cm} 
{\bf 11.4 \quad Der limes \boldmath$\vc r^\prime \to \vc r$}
     \nz[.1cm] 
\nopagebreak[3]
Jetzt \ --- \ nach Regularisierung \ --- \ ist der
Koinzidenz--limes problemlos ausf\"uhrbar. Wir blicken auf
$\cl G_2$, \eq{11.21}, und erkennen, da\ss\ die
runde Klammer viel schneller verschwindet, n\"amlich $\sim 
(z-z^{\prime\prime})\,(\ov{z} - \ov{z}^{\prime\prime})\,$,
als der Nenner. Kurz, $\cl G_2$ geht gegen {\bf Null}. Zu 
$\cl G_1$ zeigt der mittlere Ausdruck in \eq{11.25} einen 
Differentialquotienten$\,$:
\bea{11.27}
  \cl G_{r r}^{ab} \; = \;\cl G_{1 \; r r}^{ab} &=& - 
    \, {1\0\pi} \, 
  \apf{\ov{\6}} \lk K^{-1} (z, \apf{\ov{z}} ) \, 
  K(z,\ov{z}) \rki^{ab} = {1\0\pi} \lk H^{-1} \ov{\6} 
  H \rki^{ab} \nonu \\ 
  & & \hspace*{-3.7cm} {} =  
  {1\0\pi} \lk M^{-1} M^{\dagger -1} \Big( (\ov{\6} 
  M^\dagger ) M + M^\dagger  \ov{\6} M \Big) \rki^{ab}
  = {-1\0\pi} \lk M^{-1} \Big( A^\dagger - (\ov{\6} M)
  M^{-1} \Big) M \rki^{ab} . \qquad  
\eea 
\eq{11.27} ist \ek{3.10}. In \eq{9.9} haben wir nun
nur noch $G$ durch $\cl G$ zu ersetzen$\,$:
\be{11.28}
  \Big[ \; D^{-1}_{\rm reg} \; \Big]^{ab}_{rr} 
   = M^{ac}\, \cl G^{cd}_{rr} \, (M^{-1})^{db} 
   =  -\,{1\0 \pi} \,\lk   A^\dagger 
   - (\ov{\6} M) M^{-1}  \rki^{ab} \quad , 
   \quad \hbox{q.e.d.} \quad 
\ee 
und \anfu hurra$\,$!\anfo, denn das ist das 
Wunschresultat \eq{9.14}.


\sec{Hamiltonian, mass gap and CFT}

Irgendwo \ --- \ weit hinter uns \ --- \ lag schon einmal 
die klassische Hamilton--Dichte auf dem Papier, 
\eq{2.8} bis \eq{2.10}$\;$:
\be{12.1}
  \cl L = {1\0 2 e^2} \p A{\!}^a_j \p A{\!}^a_j - \cl V
          \quad , \quad 
    \P^a_j = {1\0 e^2} \p A{\!}^a_j \quad , \quad
  \cl H = {e^2\02} \P^a_j \P^a_j + \cl V \quad   
\ee 
mit $\,\cl V \gll {1\0 2 e^2} B^a B^a\,$ und $\,B^a = 
\6_1 A^a_2 - \6_2 A^a_1 + f^{abc} A^b_1 A^c_2\,$. 
Bei Abwesenheit von Eichfreiheit  \ \hbox{w \"u r d e} \
nun Quantenmechanik durch $\P^a_j \;\Rightarrow\; (1/i)
\,\d_{A^a_j}\,$ einzul\"auten sein und
\be{12.2}
 \cl H = \cl T + \cl V \qquad \hbox{mit} \qquad
  \cl T \, = - {e^2\02} \d_{A^a_j} \d_{A^a_j} 
  \; = \; - {e^2\02} \d^{a*} \d^a \quad\quad
\ee 
geben \ --- \ wie schon in \eq{11.10} behauptet. 
Es gibt eine recht billige Antwort darauf, wie sich 
$\cl T$ auf den Raum $\cl C$ beschr\"anken  l\"a\ss t
(und obiges \anfu w\"urde\anfo ausr\"aumen 
l\"a\ss t)$\,$: wende $\cl T$ nur auf Funktionale von 
$H$ an ! \ KKN's Bem\"uhungen um den {\sl Laplacian on 
$\cl C$} laufen darauf hinaus, jenen Anteil von $\d^a$ 
(oder von $p^a$) abzuspalten und wegzulassen, welcher 
Umeich--Anteile ver\"andert. Sind solche Anteile erst 
gar nicht da, dann ist $\cl T$ \ \hbox{a u t o m a t 
i s c h} \ geeignet reduziert.
\pagebreak[2]
     \nz[.5cm]  \hspace*{2cm} 
{\bf 12.1 \quad \boldmath$ T\; J^a \; = m \; J^a $}
     \nz[.1cm] 
\nopagebreak[3]
Wir wenden $\cl T$ auf spezielle $\psi[H]$ an. Welche$\,$?
Einmal mehr lassen wir uns von KKN leiten, n\"amlich vom 
Text vor \ek{2.28}). Dort findet sich eine ebenso 
interessante wie (vorerst) mysteri\"ose Behauptung 
(beruhend auf Konformer Feldtheorie $\glr$ CFT). Es seien, 
so die Behauptung, nur Funktionale $\psi[H]\,$ normierbar, 
welche \"uber die \anfu Str\"ome\anfo
\bea{12.3}
  J^a (\vc r) &=& {2N\0\pi} \; \Sp \Big( \, T^a \,(\6H) 
  \, H^{-1}\, \Big) \nonu \\
  &=& {N\0\pi}\, 2 \, 
  \Sp \( T^a \lk (\6 M^\dagger) M^{\dagger -1}
  +  M^\dagger iT^c A^c M^{\dagger-1} \rk \)
\eea 
von $H$ abh\"angen. Unter diesen wiederum betrachten 
wir (wie Karabali, Nair in \cite{vor}) speziell
\be{12.4}
  \psi_{\rm sp} \lk H \rk \;\gll\; \int c^a (\vc r)
  J^a (\vc r ) \nonu \\
\ee 
(mit beliebigen c--Zahl--Funktionen $c^a\,$) und wenden 
${\bf T} = \int \cl T\,$ auf \eq{12.4} an. Die Details 
$(*)$, $(**)$ werden nachgestellt, denn die Sache ist 
allzu aufregend$\,$:
\bea{12.5}
 {\bf T} \;\psi_{\rm sp} &=& - {e^2\02} \int \d^{a*}_r \, 
    \d^a_r \, \int^\prime c_{r^\prime}^d \, 
     J^d_{r^\prime} \nonu \\
  &=& - {e^2\02} \int \,\d_r^{a*} \; {N\0\pi}\, i\, c^d_r 
        (M_r^\dagger)^{da}  \; = \;
     - i \; {e^2N\0 2\pi} \int c^d \lk \d^{a*}_{r^\prime}
      (M_r^\dagger)^{da} \rki_{\vcsm r^\prime \to \vcsm r} 
        \nonu \\  
  &=& - i \; {e^2N\0 2\pi} \int c^d \lk  
      (M_{r^\prime}^\dagger)^{db} f^{bae} \( 
       D^{* -1}_{r r^\prime} \)^{ae} \rki_{\vcsm r^\prime 
       \to \vcsm r}          \qquad\qquad (*) \nonu \\
  &=&  - i \; {e^2N\0 2\pi} \int c^d \, (M^\dagger)^{db}
       \;{iN\0\pi}\, 2\,\Sp \(\, T^b \lk M^{\dagger -1} 
       \6 M^\dagger + (\6 M) M^{-1} \rk \) 
                        \qquad (**) \qquad \nonu \\  
  &=&  {e^2N\0 2\pi} \int c^d \lk 
       {N\0\pi} \, 2 \, \Sp \( T^d (\6 H ) H^{-1} \) \rk   
       \nonu \\  
  &=& {e^2 N \0 2\pi} \int c^d J^d \;\; = \;\; 
       m  \;\; \psi_{\rm sp} \qquad , \qquad
       m \gll {e^2 N \0 2\pi}  \quad . \quad
\eea 
Eigenfunktionen von ${\bf T}$ sind gefunden. Und wegen
der Beliebigkeit der Gewichte $c^a(\vc r )$ ist der 
Eigenwert $m$ unendlichfach entartet.

Der Schritt zur zweiten Zeile brauchte $\d^a_r 
J^d_{r^\prime}\,$, und dies kann direkt aus \eq{12.3} 
abgelesen werden. Die Zeile $(*)$ zeigt, da\ss\ hier genau 
der gleiche Koinzidenzlimes auftaucht wie schon in \eq{9.6}. 
Mehr noch, die Faktoren $f^{^{\ldots}} D^{*-1}$ in der 
eckigen Klammer \ s i n d \ $\,\d^b \,\G\,$ und k\"onnen 
\ --- \ regularisiert und im limes \ --- \ durch \eq{9.20} 
ersetzt werden. Dies erkl\"art die Zeile $(**)$. Von 
dieser zur n\"achsten ist nur noch eine {\sl concatenation} 
vorzunehmen. Es bleibt die Frage, wie man von der zweiten 
zur Zeile $(*)$ gelangt. Der Inhalt der eckigen Klammern 
sei gleich, wird behauptet. $\,\d^a_{r^\prime} \hat{M}_r\,$ 
\ --- \ c.c. nehmen wir sp\"ater \ --- \ mu\ss\ sich aus 
$\, 0 = \(\6 + \hat{A}\)\, \hat{M}\,$ ergr\"unden lassen, 
d.h. gleich aus der adjungierten Version \eq{9.12} 
(Erinnerung$\,$: $A^{ab}=A^\bullet f^{a\bullet b}\,$). 
Wir wenden $\d^a_{r^\prime}$ an, benutzen \eq{7.13} und 
erhalten 
\be{12.6}
  \int^{\prime\prime} D^{eb}_{r r^{\prime\prime}} 
  \; \d^a_{r^\prime} M_{r^{\prime\prime}}^{bd}
   = \d_{r r^\prime} f^{aeb} M_{r^\prime}^{bd} 
      \quad \Rightarrow \quad  
  \d_{r^\prime}^a M_r^{cd} = (D^{-1}_{r r^\prime})^{ce} 
     f^{aeb} M_{r^\prime}^{bd} \quad . \quad
\ee 
Hier kann nun $a=c$ gesetzt und \"uber $a$ summiert 
werden. Mittels $M^{db\, *} = (M^\dagger)^{bd}\,$ 
schreibt sich das gew\"unschte c.c. leicht daneben$\,$:
\be{12.7}
  \d_{r^\prime}^a M_r^{ad} = (D^{-1}_{r r^\prime})^{ae} 
     f^{aeb} M_{r^\prime}^{bd} \quad , \quad
  \d^{a*}_{r^\prime} (M_r^\dagger)^{da} =
  (M_{r^\prime}^\dagger)^{db} f^{bae} \( 
   D^{* -1}_{r r^\prime} \)^{ae} \quad , \quad
\ee 
was zu zeigen war. Wer da aber gern eigene Wege geht, kann
leicht in einem Morast versinken. Er gelangt z.B. anstelle 
von \eq{12.6} zu einer anderen, recht \"ahnlichen Beziehung, 
n\"amlich 
\be{12.8} 
  \d^a_{r^\prime} M_r^{cd} = (D^{-1}_{r r^\prime})^{ea}
  f^{ecb} M_r^{bd} \quad . \quad
\ee 
Bei $c=a$ und Summation \"uber $a$ ergibt sich \eq{12.7} 
nur fast, denn $M^{bd}$ tr\"agt jetzt den Index $r$ (statt 
$r^\prime$). Also verflucht er \eq{12.5} und sucht 
stundenlang nach seinem oder unserem Fehler \ --- \ ohne 
Erfolg. Alles ist richtig$\,$! \ \eq{12.6} und \eq{12.8}
lassen sich mittels \eq{9.9} auseinander herleiten. In der 
speziellen Bildung \eq{12.7} \ \hbox{d a r f} \ man den 
$\vc r$--Index an $M$ wechseln.
\pagebreak[2]
     \nz[.5cm]  \hspace*{2cm} 
{\bf 12.2 \quad \boldmath$S[H]$ ist konform invariant}
     \nz[.1cm] 
\nopagebreak[3]
Schon seit \S~2 leben wir im 2D Euklidischen. Das ist 
der spezielle Fall, in welchem die Forderung nach 
konformer Invarianz besonders viel verlangt 
\cite{gins,fuchs,ket,schell}, weil in der globalen 
Koordinatentransformation
\bea{12.9}
    z \;\to\; f(z) \quad & &, \;\; \hbox{d.h.} \;\;\;
    z_{\rm neu} = u(x,y)-iv(x,y) \;\ueb{!}{=}\; f(x-iy) \\ 
    & & \quad \Rightarrow \quad u_{\prime x} = v_{\prime y} 
    \quad , \quad  u_{\prime y} = - v_{\prime x} \nonu
\eea 
die Funktion $f(z)$ beliebig bleibt. Da\ss\ dabei (a) 
\ --- \ dem Wort \anfu konform\anfo entsprechend \ --- \ 
der Winkel zwischen zwei infinitesimalen Vektoren $d \vc r$, 
$d \vc \rho$ der gleiche bleibt und (b) die flache Metrik 
{\scriptsize $\( \matrix{ 1 & 0 \cr 0 & 1 \cr} \) $} in 
$ f^\prime f^{\prime *} ${\scriptsize $ \(
\matrix{ 1 & 0 \cr 0 & 1 \cr} \) $} \"ubergeht \cite{ket}, 
kann man sich gem\"utlich selbst aufkl\"aren.  

\eq{12.9} ist eine reine Koordinatentransformation.
Die Werte der Felder $H(z, \ov{z})$ \"andern sich
nicht. Infinitesimal ($\e\to0 \,$, $\, g(z)$ beliebig) 
hei\ss t das
\bea{12.10}
 z\to z^\prime = z + \e\, g(z) \quad , \quad
 H^\prime (z^\prime , \ov{z}^\prime) &=& H(z,\ov{z})
   = H(z^\prime - \e g , \ov{z}^\prime - \e \ov{g}) 
   \hspace*{2cm} \nonu \\
 &=& H(z^\prime, \ov{z}^\prime ) 
     - \e g \6 H - \e \ov{g} \, \ov{\6} H \quad . \quad
\eea 
Zum Nachweis der konformen Invarianz von $ S[H] $ haben wir
zu zeigen\footnote{\
     In der ersten Zeile \eq{12.11} wurde rechts die 
     Variable $\vc r^\prime$ in $\vc r$ umbenannt. Bei 
     Subtraktion von $S$ vergi\ss t man daraufhin 
     Randterm--Unterschiede. Das ist der Grund daf\"ur, 
     da\ss\ \anfu Noether 1\anfo (Nachtmann, Flie\ss bach, 
     die Jacobi--Determinante der Transformation  explizit 
     bedenkend) nur selten eine totale Ableitung in der 
     Invarianzbedingung ben\"otigt, insbesondere nicht bei 
     Translationen. \anfu Noether 2\anfo w\"alzt hingegen 
     auf Felder ab und kann aus Noe 1 mit der genannten 
     Umbenennung hergeleitet werden. Auf Noe 2 sind 
     s\"amtliche Experten des Hauses (Brandt, Dragon, 
     Lechtenfeld, Reuter) eingeschworen.}, 
da\ss\ 
\bea{12.11}
 S^\prime - S &=& \int d^2 r^\prime \; \cl L 
 \lb H^\prime (z^\prime , \ov{z}^\prime )\, , \, \6^\prime 
  \, , \, \ov{\6}^\prime \, \rb - S
  = \int d^2 r \; \cl L \lb H+ \d H \, , \, \6 
  \, , \, \ov{\6}\, \rb - S \nonu \\
  &=& S\lk H+\d H\rk - S\lk H\rk = \d S \quad 
  \hbox{\ft zu speziell}   \quad 
  \d H = - \e g \6 H - \e \ov{g} \ov{\6} H \qquad \quad
\eea 
verschwindet.
Nat\"urlich d\"urfen sich $S^\prime$ und $S$ auch um eine 
Konstante unterscheiden, resultierend aus Randtermen
bei partieller Integration. Solche Beitr\"age seien hier
ignoriert. Zur Variation von $S$ \ --- \ unter welcher 
Vorgabe $\d H$ auch immer \ --- \ haben wir die fertige 
Formel \eq{9.33}$\;$:
$$  
    \d S = {1\0 \pi } \int \Sp \Big( \;\ph  \;\6 
           \ov{X} \,\Big) \quad  \hbox{mit} \quad
  \lb \ph , X , \ov{X} \rb = H^{-1} 
  \lb \d , \6 , \ov{\6} \rb H \quad . \quad \eqno{(9.33)} 
$$  
Jetzt ist nur noch $\d H$ aus \eq{12.11} hier einzusetzen. 
\bea{12.12}
  \d_{\rm konfo}\;\, S &=& - {\e\0\pi} \int 
     \bigg[ g  \,\Sp \( X  \, \6 \ov{X} \) \; + \: 
        \ov{g} \,\Sp \( \ov{X} \, \6 \ov{X} \) \bigg]  
        \nonu \\
  &=& - {\e\0\pi} \int \bigg[ g\,\Sp \( X  
      \lk \ov{\6} X + \ov{X} X  - X \ov{X} \rk \) \; + \; 
      \ov{g} \, \Sp \( \ov{X} \6 \ov{X} \) \bigg]  
      \;\;\; \nonu \\
  &=& - {\e\02\pi} \int \bigg[ \ov{\6}\; g \;
       \Sp \( X X \)  \; + \;  \6\; \ov{g} \;
       \Sp \( \ov{X}\,\ov{X} \) \bigg] \;\; 
      = \;\; 0 \quad , \quad \hbox{q.e.d.} \quad                 
\eea 
In der zweiten Zeile wurde \eq{9.34}, d.h. $\ov{\6} X 
+ \ov{X}X = \6\ov{X} + X \ov{X}\,$, benutzt, und
$\,\ov{\6} g = \6 \ov{g} = 0\,$ in der dritten. 

Die Bewegungsgleichung der $S$--Theorie folgt aus $\d S=0$ 
zu unabh\"angiger Variation  der $N^2-1$ $H$--Elemente. 
Man kann sie aus \eq{9.33} ablesen$\,$:
\bea{12.13}
  \6 \;\ov{X} = 0 \qquad & & \hbox{oder} \qquad 
    \6 \, J^\dagger = 0 \quad \hbox{mit} \quad 
    J^\dagger \gll H^{-1} \ov{\6} H \nonu \\
   & & \hbox{sowie} \qquad 
    \ov{\6}\, J = 0 \quad \hbox{mit} \quad 
     J \gll (\6 H)  H^{-1} \quad . \quad 
\eea 
Der Matrix--Strom $J$ kann nat\"urlich in $J= (J_1 + i 
J_2) / 2$ zerlegt werden mit $J_\ell = (\6_\ell H) H^{-1}$. 
Wie \eq{10.1} zeigt, kann die Wirkung $S[H]$ restlos durch 
diese $J_\ell$ ausgedr\"uckt werden. Jene Funktionen 
$J^a(\vc r )$ in \eq{12.3} m\"ochten nun bitte (bis auf 
Vorfaktor) nichts weiter sein als die Koeffizienten in 
der Entwicklung $\,J = J^a T^a\,$, woraus $\,J^a = 2\,
\Sp \( T^a J \)\,$ folgen w\"urde. Um so aufspannen zu 
d\"urfen, hat $J = (\6 H) H^{-1}$ spurfrei zu sein. Hierzu 
hatte O. Lechtenfeld die rechte Idee. Mit der auf $\det 
(H) = 1 $ beruhenden Darstellung $\, H = e^{\vcsm \s 
\vcsm T}\;$, $\s^a$ reell$\,$, vgl. \eq{5.8} und \eq{8.14}, 
ergibt sich  
\be{12.14}
  \Sp \Big( (\6 H)H^{-1} \Big) = \Sp \( \int_0^1 ds \; 
   e^{s \vcsm \s \vcsm T} (\6 \s^a )T^a 
   e^{- s \vcsm \s \vcsm T} \) =  (\6 \s^a ) 
   \;\Sp \Big( T^a \Big) = 0 \qquad 
\ee 
tats\"achlich.
\pagebreak[2]
     \nz[.5cm]  \hspace*{2cm} 
{\bf 12.3 \quad Ende}
     \nz[.1cm] 
\nopagebreak[3]
Unverbesserlichen Fu\ss g\"angern war schon auf Seite 1 
ein fr\"uhes Ende prophezeit worden. Was hier alsbald 
abbricht, ist nat\"urlich nur das \anfu Logbuch\anfo dieser 
Expedition. Wir befinden uns mitten im zerkl\"ufteten 
Gel\"ande. Die Luft wird d\"unner. Es folgen nur noch ein 
paar Notizen zur momentanen Situation.

Die Masse $m$ ist m\"oglicherweise die, welche quadriert 
im Gluon--Propagator steht. Aber das physikalische 
(holomorph invariante) Spektrum d\"urfte \ --- \ sofern 
wir KKN recht vertstehen \ --- \ als ersten angeregten 
Zustand eine Zwei--$J$--Bildung ({\sl glue ball}) haben$\,$: 
{\sl mass gap} $= 2m$. \ Offene Frage ist vorerst auch, 
weshalb der Potentialterm $\cl V$ nur eine untergeordnete 
(Entartung aufhebende) Rolle spielt. Der Wilson--loop 
\ek{2.32} d\"urfte den Einstieg in die 
{\sl confinement}--Region \cite{nach} bieten. KKN's 
Paragraphen 4 ({\sl An expression for {\bf T} in terms of 
currents}), 6 ({\sl Construction of eigenstates of {\bf T}}) 
und 7 ({\sl Corrections due to the potential term}) werden 
schon noch irgendwie zu bezwingen sein. Wohlan \ --- \
{\ft und 
Helau$\,$: Aschermittwoch 1999}$\,$.  

\vskip .2cm\noindent \leftskip .5cm {\small\sl
Im Text ist sporadisch von allerlei Leuten die Rede, 
welche mal hier, mal da weitergeholfen haben. Ein 
herzliches Dankesch\"on! Die Leute sind F. Brandt, 
N. Dragon, O. Lechtenfeld, S. Ketov, J. Reinbach, 
M. Reuter, Y. Schr\"oder und J. Schulze. I am also 
indebted to D. Karabali and V.P. Nair for one or 
another helpful discussion during the TFT'98 in August 
at Regensburg.}  

{\small

}
\end{document}